\documentclass[11pt]{article}
\usepackage[T1]{fontenc}
\usepackage{array,longtable}
\usepackage{xcolor}
\usepackage{fancyhdr}
\usepackage{graphicx}
\usepackage{float}
\usepackage{algorithmic}
\usepackage[american]{babel}
\usepackage{csquotes}
\usepackage{enumitem} 
\usepackage{subcaption} 
\usepackage{multirow, booktabs}
\usepackage{setspace} 
\usepackage{algorithm}
\usepackage{algorithmic}
\usepackage{makecell}

\usepackage[backend=biber,citestyle=authoryear,sortcites=true,natbib,sorting=nyt,style=apa]{biblatex}
\DeclareLanguageMapping{american}{american-apa}
\bibliography{bibtex-MonoDecomp.bib}
\usepackage[
  top=2.5cm, bottom=2.5cm,
  left=2.6cm, right=2.6cm,
  headsep=3mm,
  headheight=15pt]{geometry}
\usepackage{newpxtext}
\usepackage{tikz}
\usetikzlibrary{shapes.geometric}
\usetikzlibrary{through} 
\usepackage{hyperref}
\hypersetup{
    colorlinks=true,
    linkcolor=blue,
    filecolor=magenta,      
    urlcolor=cyan,
    }
\usepackage{amsmath, amssymb}
\DeclareMathOperator*{\argmax}{arg\,max}
\DeclareMathOperator*{\argmin}{arg\,min}
\DeclareMathOperator*{\tr}{tr}

\DeclareMathOperator*{\MSE}{MSE}
\DeclareMathOperator*{\Var}{Var}
\DeclareMathOperator*{\sign}{sign}
\DeclareMathOperator{\Bias}{Bias} 
\newcommand\IR{\mathrm{I\!R}}

\newcommand\cH{{\mathcal H}}

\newcommand\bA{{\mathbf A}}
\newcommand\bB{{\mathbf B}}

\newcommand\bD{{\mathbf D}}

\newcommand\bG{{\mathbf G}}
\newcommand\bH{{\mathbf H}}
\newcommand\bI{{\mathbf I}}

\newcommand\bK{{\mathbf K}}

\newcommand\bQ{{\mathbf Q}}

\newcommand\bU{{\mathbf U}}
\newcommand\bV{{\mathbf V}}
\newcommand\bW{{\mathbf W}}

\newcommand\bZ{{\mathbf Z}}

\newcommand{\bOmega}{{\boldsymbol\Omega}}
\newcommand{\bSigma}{{\boldsymbol\Sigma}}
\newcommand{\one}{{\boldsymbol 1}}
\newcommand\zero{\boldsymbol{0}}

\newcommand\bbE{{\mathbb E}}

\newcommand\ls{\mathrm{ls}}
\newcommand\sspl{\mathrm{ss}}
\newcommand\up{\mathrm{up}}
\newcommand\down{\mathrm{down}}
\newcommand{\MSFE}{\mathrm{MSFE}}
\newcommand{\MSPE}{\mathrm{MSPE}}
\newcommand\subto{\mathrm{s.t.}}

\newcommand\bfa{{\mathbf a}}
\newcommand\bfb{{\mathbf b}}

\newcommand\bff{{\mathbf f}}

\newcommand\bfy{{\mathbf y}}

\newcommand{\supp}{\href{bka-supp.pdf}{Supplementary Material}}

\usepackage{amsthm}
\ifthenelse{\isundefined{\definition}}{\newtheorem{definition}{Definition}}{}
\ifthenelse{\isundefined{\fact}}{\newtheorem{fact}{Fact}}{}
\ifthenelse{\isundefined{\theorem}}{\newtheorem{theorem}{Theorem}}{}
\ifthenelse{\isundefined{\proposition}}{\newtheorem{proposition}{Proposition}}{}
\ifthenelse{\isundefined{\corollary}}{\newtheorem{corollary}{Corollary}}{}
\ifthenelse{\isundefined{\lemma}}{\newtheorem{lemma}{Lemma}}{}
\ifthenelse{\isundefined{\remark}}{\newtheorem{remark}{Remark}}{}
\ifthenelse{\isundefined{\assumption}}{\newtheorem{assumption}{Assumption}}{}

\usepackage[affil-it]{authblk}
\title{Decomposition with Monotone B-splines: Fitting and Testing}
\author[1,2]{Lijun Wang%
  \thanks{Email: \texttt{ljwang@link.cuhk.edu.hk}, \texttt{lijun.wang@yale.edu}}
}
\affil[1]{Department of Statistics, The Chinese University of Hong Kong, Hong Kong SAR, China}
\affil[2]{Department of Biostatistics, Yale University, New Haven, Connecticut, USA}
\author[1]{Xiaodan Fan%
  \thanks{Email: \texttt{xfan@cuhk.edu.hk}}
}
\author[2]{Hongyu Zhao%
    \thanks{Email: \texttt{hongyu.zhao@yale.edu}}
}
\author[3]{Jun S. Liu%
\thanks{Email: \texttt{jliu@stat.harvard.edu}}
}
\affil[3]{Department of Statistics, Harvard University, Cambridge, Massachusetts, USA}
\date{}
\begin{document}
\maketitle

\begin{abstract}

A univariate continuous function can always be decomposed as the sum of a non-increasing function and a non-decreasing one. 
Based on this property, we propose a non-parametric regression method that combines two spline-fitted monotone curves.
We demonstrate by extensive simulations that, compared to standard spline-fitting methods,  the proposed approach is particularly advantageous in high-noise scenarios. 
Several theoretical guarantees are established for the proposed approach. Additionally, we present statistics to test the monotonicity of a function based on monotone decomposition, which can better control Type I error and achieve comparable (if not always higher) power compared to existing methods. Finally, we apply the proposed fitting and testing approaches to analyze the single-cell pseudotime trajectory datasets, identifying significant biological insights for non-monotonically expressed genes through Gene Ontology enrichment analysis. The source code implementing the methodology and producing all results is accessible at \texttt{\url{https://github.com/szcf-weiya/MonotoneDecomposition.jl}}.
    
    
    
    

\end{abstract}
\noindent%
\vfill
\newpage

\section{Introduction}

Suppose we have $n$ pairs of observations 
$(x_i, y_i), i=1,\ldots,n$, with $x_i\in \IR^d, y_i\in \IR$, independent and identically distributed (i.i.d.) according to an unknown probability distribution $P(X, Y)$. Various methods exist for estimating the conditional expectation function $f(x) = \bbE(Y|X=x)$, ranging from simple linear regressions (including ridge and lasso) to more sophisticated nonlinear techniques. 
Spline is one of the most popular methods, particularly when $d=1$. 
%
Unlike existing methods, we aim to estimate the monotonic components of $f(x)$ and then use their sum as an estimator for $f(x)$. This is because any general function can be decomposed into the sum of an increasing function $f_\up(x)$ and a decreasing function $f_\down(x)$ (a more formal proof is given in the \supp{} for self-contained).

The monotone decomposition idea has been  exploited  by \textcite{chipmanMBARTMultidimensionalMonotone2022a} in their recent algorithm,
where the monotone decomposition is incorporated into fitting monotone Bayesian additive regression trees (mBART). They found that fitting by monotone decomposition with mBART outperforms the corresponding BART algorithm proposed earlier in
\textcite{chipmanBARTBayesianAdditive2010}.
We  here focus on the case of $d=1$, and adopt B-spline basis functions to represent the monotone components, 
$$
f_{\up}(x) = \sum_{j=1}^{J_u}\gamma_j^uB^u_j(x) + \varepsilon_u\,,\qquad f_{\down}(x) = \sum_{j=1}^{J_d}\gamma_j^dB^d_j(x)+\varepsilon_d\,,
$$
where the superscripts and subscripts ``$u$'' and ``$d$'' 
indicate for \emph{up} (increasing) and \emph{down} (decreasing), respectively.
The monotonicity of each monotone component is ensured by the monotonicity of the coefficients \parencite{wangMonotoneCubicBSplines2023a}, i.e.,
$$
\gamma_1^u\le \gamma_2^u\le\cdots\le \gamma^u_{J_u}; \quad \gamma_1^d\ge \gamma_2^d\ge\cdots\ge\gamma_{J_d}^d\,.
$$

This paper is organized as follows. In Section \ref{sec:md_cs}, we formulate monotone decomposition with cubic splines and establish properties of the solution, particularly for monotone functions. 
In Section \ref{sec:md_ss}, we propose the monotone decomposition with smoothing splines and establish similar properties and theoretical guarantees. In Section \ref{sec:md_sim}, we present simulation results to demonstrate how fitting via monotone decomposition can outperform the corresponding unconstrained fitting, particularly in high-noise scenarios. In Section \ref{sec:test_of_mono}, we propose statistics for testing monotonicity and, in Section \ref{sec:test_sim}, we demonstrate the power of the proposed method via simulations. In Section~\ref{sec:app}, we present the results of our analysis on single-cell pseudo-time trajectory datasets using the fitting and testing techniques based on monotone decomposition. Finally, we discuss the limitations and potential future work in Section \ref{sec:md_discuss}.

\section{Monotone Decomposition with Cubic Splines}\label{sec:md_cs}

Cubic splines are the most popular polynomial splines for practitioners. Presumably, cubic splines are the lowest-order splines for which the knot-discontinuity is not visible to the human eye, and there is scarcely any good reason to go beyond cubic splines \parencite{hastieElementsStatisticalLearning2009}. On the other hand, although there are many equivalent bases for representing a spline function, the B-spline basis system developed by \textcite{deboorPracticalGuideSplines1978} is attractive numerically \parencite{ramsayFunctionalDataAnalysis2005}. Thus, we take the order-4 B-spline basis to represent cubic splines, under which the problem reduces to solving the following optimization problem:
\begin{equation}
    \begin{split}
\min_{\gamma^u,\gamma^d}&\;\Vert \bfy-\bB_u\gamma^u - \bB_d\gamma^d\Vert_2\,, \\
\text{s.t. }&\gamma_1^u\le \gamma_2^u\le\cdots\le\gamma_J^u;\; \gamma_1^d\ge \gamma_2^d\ge\cdots\ge\gamma_J^d\,, \\
    \end{split}
    \label{eq:prob_JuJd}
\end{equation}
where $\bfy =(y_1,\ldots,y_n)$ is an $n$-vector of the responses (note that we use \emph{round brackets} to denote column vectors), $(\bB_u)_{ik} = B^u_k(x_i), k=1,\ldots,J_u, (\bB_d)_{i\ell} = B^d_\ell(x_i), \ell=1,\ldots,J_d$ are the matrices constructed by evaluating the B-spline basis at $\{x_i\}_{i=1}^n$, and $\gamma^u=(\gamma_1^u,\ldots,\gamma_J^u), \gamma^d=(\gamma_1^d,\ldots,\gamma_J^d)$ are the coefficient vectors.

For simplicity, we consider $J_u=J_d=J$. Note that the knots for determining the B-spline basis are conventionally on the quantiles of $x$'s, then the B-spline basis functions are also the same, $B_k^u=B_\ell^d$, so the above problem \eqref{eq:prob_JuJd} becomes
\begin{equation}
    \begin{split}
\min_{\gamma^u,\gamma^d}&\;\Vert \bfy-\bB(\gamma^u + \gamma^d)\Vert_2\\
\text{s.t. }&\gamma_1^u\le \gamma_2^u\le\cdots\le\gamma_J^u;\; \gamma_1^d\ge \gamma_2^d\ge\cdots\ge\gamma_J^d\,,
    \end{split}
    \label{eq:prob_J}
\end{equation}
where $\bB_{ij}=B_j(x_i), j=1,\ldots,J$.

First of all, Proposition \ref{prop:nonconstraint} establishes the equivalence between problem \eqref{eq:prob_J} with the corresponding unconstrained B-spline fitting.
\begin{proposition}
\label{prop:nonconstraint}
Regardless of the component solutions $\hat\gamma^u, \hat\gamma^d$ to problem \eqref{eq:prob_J}, the overall solution $\hat\gamma^u+\hat\gamma^d$ is equivalent to the unconstrained B-spline fitting, i.e., the least squares solution,
\begin{equation}
    \hat\gamma^\ls = \argmin_\gamma \Vert \bfy-\bB\gamma\Vert_2 = (\bB^T\bB)^{-1}\bB^T\bfy\,.
    \label{eq:prob_nonconstraint}
\end{equation}
Specifically,
$\hat\gamma^u + \hat\gamma^d = \hat\gamma^\ls\,.$
\end{proposition}

Note that the monotone components in \eqref{eq:prob_J} are not identifiable, since
$$
\gamma^u + \gamma^d = \gamma^u + \delta + \gamma^d - \delta\,,
$$
where $\delta$ is an arbitrary increasing sequence, $\delta_1\le \cdots\le \delta_J$. In other words, the decomposition for a general function is not unique since
$$
f_{\up}(x) + f_{\down}(x) = f_{\up}(x) + h(x) + f_{\down}(x) - h(x)\,,
$$
where $h(x)$ is an arbitrary increasing function. 

In order to have a unique decomposition, we consider the \emph{closest} decomposition in some sense, such as the difference between two monotone components being the smallest in the $L_2$-norm. Thus, 
we consider imposing some discrepancy constraint on problem \eqref{eq:prob_J}, as detailed in the following subsections, to help obtain a unique solution.

\subsection{Discrepancy Constraint: A Motivating Example}

Consider the simple function $y=x^3,x\in[-1,1]$, which may be decomposed as 
$$
x^3 = \{x^3 + h(x)\} + \{0-h(x)\}\triangleq f_{\up}(x) + f_{\down}(x)\,,
$$
where $h(x)$ is an increasing function. 
If we set $h(0) = 0$, then it is easy to show that the magnitude of the difference between the two monotone components is lower-bounded by $|x^3|$, i.e.,
\begin{equation}\label{eq:lower_bound}
    \vert x^3 \vert \leq \vert x^3 + 2h(x)\vert.
\end{equation}
Since it is unreasonable to have a decreasing component for a strictly increasing function, the ideal decomposition should correspond to $h(x)=0$. 
Equation \eqref{eq:lower_bound} suggests to us that such an  ideal decomposition may be obtained by requiring the two monotone components to differ the least.

In light of this observation, we introduce the following {\it discrepancy constraint} for the two monotone components in the decomposition:
\begin{equation}\label{eq:discrepancy_constraint}
    \Vert f_{\up} - f_{\down}\Vert_2 \le s,
\end{equation}
where $s\ge 0$ is a tuning parameter. The role of parameter $s$ can be summarized as follows,
\begin{itemize}
    \item if $s\rightarrow 0$, then the solution is $\gamma^u=\gamma^d=c\one_J$, and hence $\hat f_{\up} = \hat f_{\down} = c\bB\one_J = c\one_n$ are constant functions;
    \item if $s\rightarrow \infty$, then the problem reduces to be equivalent to the unconstrained B-spline fitting;
    \item a moderate $s$ imposes some regularization, which is preferable for a better fitting.
\end{itemize}

\subsection{General Functions}

With the  discrepancy constraint in \eqref{eq:discrepancy_constraint}, we can restate problem \eqref{eq:prob_J} as
\begin{equation}
    \begin{split}
\min_{\gamma^u,\gamma^d}&\;\Vert \bfy-\bB(\gamma^u + \gamma^d)\Vert_2\\
\text{s.t. }&\gamma_1^u\le \gamma_2^u\le\cdots\le\gamma_J^u;\; \gamma_1^d\ge \gamma_2^d\ge\cdots\ge\gamma_J^d\\
&\Vert\bB(\gamma^u - \gamma^d)\Vert_2\le s\,.
    \end{split}
    \label{eq:prob_J_s}
\end{equation}
Defining $\Upsilon \triangleq \{\gamma = (\gamma^u, \gamma^d) \in\IR^{2J}: \gamma_1^u\le \gamma_2^u\le\cdots\le\gamma_J^u;\; \gamma_1^d\ge \gamma_2^d\ge\cdots\ge\gamma_J^d\}$ and
$$
\bZ \triangleq \begin{bmatrix}
\bB & \bB
\end{bmatrix},\quad
\bW \triangleq \begin{bmatrix}
\bB^T \\
-\bB^T
\end{bmatrix}
\begin{bmatrix}
\bB & -\bB
\end{bmatrix},
$$
we further rewrite problem \eqref{eq:prob_J_s} as
\begin{equation}
    \min_{\gamma\in \Upsilon}\;\Vert \bfy-\bZ\gamma\Vert_2^2\qquad
        \subto \ \  \gamma^T\bW\gamma \le s^2\,.
        \label{eq:prob_md_cs}
\end{equation}
It is more convenient to consider its Lagrangian form
\begin{align}
    \min_{\gamma\in\Upsilon} \left[\Vert \bfy-\bZ\gamma\Vert_2^2+\mu(\gamma^T\bW\gamma -s^2) \right]
    =\min_{\gamma\in\Upsilon} \Vert \bfy-\bZ\gamma\Vert_2^2+\mu\gamma^T\bW\gamma\,,\label{eq:lag_md_cs}
\end{align}
where $\mu\ge 0$ is the Lagrange multiplier. By Lagrangian duality, there is a one-to-one correspondence between the constrained problem \eqref{eq:prob_md_cs} and the Lagrangian form \eqref{eq:lag_md_cs}: for each value of $s$ in the range where the constraint $\gamma^T\bW\gamma\le s^2$ is active, there is a corresponding value of $\mu$ that yields the same solution from the Lagrangian form \eqref{eq:lag_md_cs}. Conversely, the solution $\hat\gamma_\mu$ to problem \eqref{eq:lag_md_cs} solves the bound problem \eqref{eq:prob_md_cs} with $s^2 = \hat\gamma^T_\mu\bW\hat\gamma_\mu$.
Some basic properties of the solution to problem \eqref{eq:lag_md_cs} are summarized in Proposition \ref{thm:md_cs_sol_general}.
\begin{proposition}
\label{thm:md_cs_sol_general}
Let $\hat\gamma=(\hat\gamma^u, \hat\gamma^d)$ be the monotone decomposition to problem \eqref{eq:lag_md_cs} (or problem~\eqref{eq:lag_md_ss} discussed later). 
\begin{enumerate}[label=(\roman*)]
    \item There must be ties in the solution $\hat\gamma^u$ or $\hat\gamma^d$, i.e., there exists $i$ or $j$ such that
$\hat\gamma^u_i = \hat\gamma^u_{i+1}$ or $\hat\gamma^d_j = \hat\gamma^d_{j+1}$.
    \item The mean values of two monotone components are equal, $\one^T\bB\hat\gamma^u = \one^T\bB\hat\gamma^d$.
\end{enumerate}
\end{proposition}

\subsection{Monotone Functions}

To delve deeper into the properties of the solution to problem \eqref{eq:lag_md_cs}, this section discusses the monotone decomposition of monotone functions. Without loss of generality, we consider increasing functions.

\begin{proposition}\label{thm:md_cs_increase}
Let $\hat\gamma = (\hat\gamma^u, \hat\gamma^d)$ be the monotone decomposition to problem \eqref{eq:lag_md_cs}. Suppose $\hat\gamma^u+\hat\gamma^d$ is increasing, then
\begin{enumerate}[label=(\roman*)]
    \item $\hat\gamma^d$ is a vector with identical elements, i.e., $\hat\gamma^d = c\one$, where the constant $c=\frac{\one^T\bB\hat\gamma^u}{n}$; 
    \item \label{thm:md_cs_increase_item2} if there is no ties in $\hat\gamma^u$, i.e., 
    $\hat\gamma_1^u< \hat\gamma_2^u<\ldots<\hat\gamma^u_J$, then
\begin{equation}
\hat\gamma^u = \frac{1}{\mu+1}\hat\gamma^\ls + \frac{\mu-1}{\mu+1}c\one\,,\label{eq:md_cs_no_ties}
\end{equation}
where the unconstrained B-spline solution $\hat\gamma^\ls$ is given in Equation \eqref{eq:prob_nonconstraint};
\item if $\hat\gamma^u_1 < \cdots < \hat\gamma^u_{k_1}=\cdots=\hat\gamma^u_{k_2}<\cdots <\hat\gamma^u_{k_{2m-1}} =\cdots = \hat\gamma^u_{k_{2m}}< \cdots< \hat\gamma_J^u$, where $1\le k_1 \le k_2\le \cdots\le k_{2m-1}\le k_{2m}\le J$, then 
\begin{equation}
\hat\gamma^u = \frac{1}{\mu+1}\bG^T(\bG\bB^T\bB\bG^T)^{-1}\bG\bB^T\bfy +\frac{\mu-1}{\mu+1}c\one\,,\label{eq:md_cs_g}
\end{equation}
where $\bG$ is a block diagonal matrix with elements
$$
\{\bI_{k_1-1}, \one^T_{k_2-k_1+1}, \ldots, \bI_{k_{2m-1}-k_{2m-2}-1}, \one^T_{k_{2m}-k_{2m-1}+1}, \bI_{J-k_{2m}}\}\,.
$$
The above result \ref{thm:md_cs_increase_item2} can be viewed as a special case when  $k_1=k_2=J$.
\end{enumerate}
\end{proposition}
Intuitively, solution \eqref{eq:md_cs_no_ties} can be viewed as a  \emph{shrinkage with offset} applied to the unconstrained B-spline fitting $\hat\gamma^\ls$, where $\frac{1}{\mu+1}$ is the shrinkage factor, and $\frac{\mu-1}{\mu+1}c\one$ is the offset. Even with the general matrix $\bG$, solution \eqref{eq:md_cs_g} also exhibits a similar \emph{shrinkage with offset} pattern.




\subsection{MSE Comparisons}
To quantify the performance of fitting by monotone decomposition, consider the model
\begin{equation}
y = f(x) + \varepsilon\,,\quad \varepsilon\sim N(0, \sigma^2)\,.\label{eq:md_model}
\end{equation}
Define the mean squared error of the fitness,
\begin{align*}
\MSE(\hat \bfy) = \bbE \Vert\bff-\bB(\hat\gamma^u+\hat\gamma^d)\Vert^2_2 \,,\quad
\MSE(\hat \bfy^\ls) = \bbE \Vert\bff-\bB\hat\gamma^\ls\Vert^2_2\,,
\end{align*}
where $\bff=[f(x_1),\ldots, f(x_n)]^T$, and the expectations are taken over $\bfy \sim N(\bff, \sigma^2\bI)$.
Proposition \ref{thm:md_cs_mse} shows that the fitting with monotone decomposition can achieve better performance, particularly in high-noise scenarios, when the underlying function is monotone.

\begin{proposition}\label{thm:md_cs_mse}
Suppose the monotone decomposition $\hat\gamma=(\hat\gamma^u,\hat\gamma^d)$ satisfies that $\hat\gamma^u+\hat\gamma^d$ is increasing. Let $\bG$ be a $g\times J$ matrix defined in Proposition~\ref{thm:md_cs_increase} such that $\bG^T\hat\gamma^u$ is the sub-vector with unique elements. If 
$$
\sigma^2 > \frac{-C_1(J-\bbE g)+C_2(\bbE g+1)+\sqrt{\Delta}}{2[(J-\bbE g)(\bbE g+1)+(\bbE g-1)^2]}\,,
$$
where 
\begin{align*}
    C_1 &= \bff^T\bA \bff-\frac{(\one^T_n\bff)^2}{n} \ge 0\,,\quad
    C_2 = \bff^T(\bI - \bA)\bff \ge 0\\
    \bA &= \bbE[\bB\bG^T(\bG\bB^T\bB\bG^T)^{-1}\bG\bB^T]\\
    \Delta &=\left[C_1(J-\bbE g)+C_2(\bbE g+1)\right]^2+4C_1C_2(\bbE g-1)^2\,,
\end{align*}
and the expectations are taken over $\bfy$ since $\bG$ (and hence $g$) depends on $\bfy$, then there exists monotone decomposition such that $\MSE(\hat \bfy) < \MSE(\hat \bfy^\ls)$.

Particularly, 
if we assume there is no ties in $\hat\gamma^u$, i.e.,$\bG\equiv\bI$ for different $\bfy$, then there always exists a monotone decomposition such that $\MSE(\hat \bfy) < \MSE(\hat \bfy^\ls)$ regardless of the noise level.
\end{proposition}
The lower bound of $\sigma^2$ in Proposition \ref{thm:md_cs_mse} might not be easy to evaluate. Nonetheless, the pivotal implication is that the monotone decomposition fitting can achieve better performance when the noise level is large enough. Extensive simulations in Section \ref{sec:md_sim} agree with this argument. Moreover, although Proposition \ref{thm:md_cs_mse} is specifically established for monotone functions, the simulations show that the monotone decomposition fitting with cubic splines can also outperform the corresponding unconstrained cubic splines applied to random functions, particularly in high-noise scenarios.

\section{Monotone Decomposition with Smoothing Splines}\label{sec:md_ss}

When dealing with cubic splines, it is typically necessary to ascertain both the number of basis functions, denoted as $J$, and the optimal placement of knots. In contrast, smoothing splines take a different approach by employing all unique data points as knots, thus bypassing the need for an optimization process to determine the knot placement and the number of knots required for B-spline basis functions.

With B-spline basis functions, the smoothing spline $f(x)=\sum_{j=1}^J\gamma_jB_j(x)$ can be estimated as follows,
\begin{equation}
    \hat\gamma^\sspl = \argmin \Vert \bfy-\bB\gamma\Vert_2^2 + \lambda \gamma^T\bOmega\gamma\,,
\end{equation}
where $\{\bOmega\}_{jk}=\int B_j''(s)B_k''(s)ds$ is called the roughness penalty matrix and $\lambda > 0$ is the penalty parameter. For this reason, smoothing splines are also referred to as penalized splines.

Imposing the roughness penalty $\gamma^T\bOmega\gamma = (\gamma^u+\gamma^d)^T\bOmega(\gamma^u+\gamma^d) = \gamma^T\bSigma\gamma$ on problem \eqref{eq:lag_md_cs}, where $\bSigma \triangleq 
\begin{bmatrix}
\bOmega & \bOmega\\
\bOmega & \bOmega
\end{bmatrix}$, we have the Lagrangian form of monotone decomposition with smoothing splines,
\begin{equation}
    \min_{\gamma\in \Upsilon}\Vert \bfy-\bZ\gamma\Vert_2^2 + \mu\gamma^T\bW\gamma + \lambda\gamma^T\bSigma\gamma\,.
    \label{eq:lag_md_ss}
\end{equation}

For general functions, the properties in Proposition~\ref{thm:md_cs_sol_general} also hold for the monotone decomposition with smoothing splines.


\subsection{Monotone Functions}

Likewise, we delve deeper into the characteristics of monotone decomposition with smoothing splines on monotone functions. The solutions demonstrate analogous  \emph{shrinkage with offset} patterns, akin to those observed in Proposition \ref{thm:md_cs_increase} for monotone decomposition with cubic splines, and the results are articulated in the following Proposition \ref{thm:md_ss_sol}.

\begin{proposition}\label{thm:md_ss_sol}
Let $\hat\gamma=(\hat\gamma^u, \hat\gamma^d)$ be the monotone decomposition to problem \eqref{eq:lag_md_ss}. Suppose $\hat\gamma^u+\hat\gamma^d$ is increasing, then
\begin{enumerate}[label=(\roman*)]
    \item $\hat\gamma^d$ is a vector with identical elements, i.e., $\hat\gamma^d = c\one$, where the constant $c=\frac{\one^T\bB\hat\gamma^u}{n}$; 
    \item \label{thm:ss_item2} if there is no ties in $\hat\gamma^u$, i.e., the inequalities hold strictly, $\hat\gamma_1^u< \hat\gamma_2^u<\ldots<\hat\gamma^u_J$, then
\begin{equation}\label{eq:md_ss_no_ties}
    \begin{split}
\hat\gamma^u 
&=\frac{1}{1+\mu}\hat\gamma^\sspl\left({\frac{\lambda}{1+\mu}}\right) - c((1+\mu) \bB^T\bB + \lambda\bOmega)^{-1}((1-\mu)\bB^T\bB + \lambda\bOmega)\one_J
    \end{split}
\end{equation}
where $\hat\gamma^\sspl\left({\frac{\lambda}{1+\mu}}\right)$ is the solution to smoothing spline with penalty parameter $\frac{\lambda}{1+\mu}$,
$$
\hat\gamma^\sspl\left({\frac{\lambda}{1+\mu}}\right) = \left(\bB^T\bB + \frac{\lambda}{1+\mu}\bOmega\right)^{-1}\bB^T\bfy\,.
$$
\item if $\hat\gamma^u_1 < \cdots < \hat\gamma^u_{k_1}=\cdots=\hat\gamma^u_{k_2}<\cdots <\hat\gamma^u_{k_{2m-1}} =\cdots = \hat\gamma^u_{k_{2m}}< \hat\gamma_J^u$, where $1\le k_1 \le k_2\le \cdots\le k_{2m-1}\le k_{2m}\le J$, then 
\begin{align*}
\hat\gamma^u &= \bG^T((1+\mu) \bG\bB^T\bB\bG^T + \lambda\bG\bOmega\bG^T)^{-1}\bG\bB^T\bfy - \\
&\qquad
c\bG^T((1+\mu) \bG\bB^T\bB\bG^T + \lambda\bG\bOmega\bG^T)^{-1}((1-\mu)\bG\bB^T\bB\bG^T + \lambda\bG\bOmega\bG^T)\one_g\,,
\end{align*}
where $\bG$ is defined in Proposition~\ref{thm:md_cs_increase}.
The above result \ref{thm:ss_item2} can be viewed as a special case when  $k_1=k_2=J$.
\end{enumerate}
\end{proposition}

For the no-tie solution \eqref{eq:md_ss_no_ties}, the \emph{shrinkage} is on the ridge solution $\hat\gamma^\sspl({\frac{\lambda}{1+\mu}})$, but different from the constant \emph{offset} in Equation \eqref{eq:md_cs_no_ties}, the offset $c((1+\mu) \bB^T\bB + \lambda\bOmega)^{-1}((1-\mu)\bB^T\bB + \lambda\bOmega)\one_J$ depends on $\bB$ and $\bOmega$.


\subsection{MSE Comparisons}

Based on model \eqref{eq:md_model}, for a comparative analysis of the fitting performance between monotone decomposition with smoothing splines and their smoothing splines counterparts, we further define the mean squared error for smoothing splines,
\begin{align*}
\MSE(\hat \bfy^\sspl(\lambda)) &= \bbE \Vert\bff-\bB\hat\gamma^\sspl(\lambda)\Vert^2_2\,,
\end{align*}
where $\hat\gamma^\sspl(\lambda)$ is the solution to smoothing splines with penalty parameter $\lambda$.

Proposition \ref{thm:md_ss_mse} shows that employing monotone decomposition with smoothing splines can result in a superior mean squared error (MSE) compared to smoothing splines in the context of monotone functions, particularly when the noise level is sufficiently high. While the condition \eqref{eq:md_ss_mse_cond} outlined in Proposition \ref{thm:md_ss_mse} may appear intricate, the simulations presented in the next section empirically substantiate this assertion. Furthermore, although the proposition is specifically formulated for monotone functions, the simulations show that the monotone decomposition applied to general functions can still achieve better performance in high-noise scenarios.

\begin{proposition}\label{thm:md_ss_mse}
Consider the smoothing spline with penalty parameter $\lambda_0$.
Let $\hat\gamma^\sspl(\lambda_0)$ be the coefficient vector and denote its hat matrix by $\bQ = \bB(\bB^T\bB+\lambda_0\bOmega)^{-1}\bB^T$. If
\begin{equation}
\sigma^2 > \frac{\bff^T\bQ(\bI-\frac{\one_n\one_n^T\bQ}{n})(\bI-\bQ)\bff}{\tr[(\bI-\frac{\one_n\one_n^T\bQ}{n})\bQ^2]}\label{eq:md_ss_mse_cond}\,,
\end{equation}
and suppose the monotone decomposition $\hat\gamma^u+\hat\gamma^d$ is increasing with no ties, then there exists a monotone decomposition with parameters $\lambda = \lambda_0/k, \mu = 1/k-1$, where $k\in(0,1)$, that achieves smaller mean squared error than $\MSE(\hat \bfy^\sspl(\lambda_0))$. 

\end{proposition}







\section{Simulations for Fitting}\label{sec:md_sim}

In this section, we compare the performance of monotone decomposition using simulated examples. We generate data from function $f$ with standard Gaussian noises,
$$
y = f(x) + N(0, \sigma^2)\,,\; x\in[-1,1]\,.
$$
To cover a diverse range of functional forms, we consider the following different types of functions.
\begin{itemize}
\item monotone functions: (i) polynomial function: $y=x^3$; (ii) exponential function: $y=\exp(x)$; (iii) sigmoid function: $y = \frac{1}{1+\exp(-5x)}$. 
\item general functions: (i) unimodal: $y=x^2$; (ii)
random functions, where the kernel can be Squared Exponential (SE), Rational Quadratic (RQ), Mat\'{e}rn (Mat) and Periodic \parencite{rasmussenGaussianProcessesMachine2006}. The numerical values appended to the kernel names in Tables~\ref{tab:bspl_vs_mbss_mu_J} and \ref{tab:ss_vs_md_lambda_mu} are the kernel parameters. For example, ``Mat12'' refers to the Mat\'{e}rn kernel with parameter $\nu=1/2$. The detailed procedure for generating a random function and a visualization of those curves can be found in the \supp.
\end{itemize}

To compare the performance of different methods, we adopt the mean squared fitting error (MSFE), i.e., the residual sum of squares (RSS) divided by sample size, and the mean squared prediction error (MSPE),
$$
\MSFE = \frac 1n\sum_{i=1}^n\left(y_i-\hat f(x_i)\right)^2\,,\qquad \MSPE=\frac 1N\sum_{i=1}^N\left(\hat f(t_i) - f(t_i)\right)^2\,,
$$
where $t_i = x_{(1)} + (i-1)\cdot\frac{x_{(N)}-x_{(1)}}{N},i=1,\ldots,N$ are equally spaced within the same range of $x$. Based on $R=100$ replications, we estimate the mean MSFE and MSPE, together with their respective standard errors. 
To judge how significant the differences of MSPE between the fitting by monotone decomposition and the corresponding spline fitting, we consider
$$
\Delta = \MSPE(\text{Monotone decomposition}) - \MSPE(\text{Spline fitting})\,,
$$
and denote the differences for each experiment as $\delta_i,i=1,\ldots,R$. We report the $p$-value for the one-sided $t$-test $H_0: \Delta=0$ versus $H_1^<:\Delta < 0$ when $\sum_{i=1}^n\delta_i < 0$ (or $H_1^>:\Delta > 0$ when $\sum_{i=1}^n\delta_i > 0$).
Besides, we also count the proportion for the fitting by monotone decomposition that achieves better performance, $\text{prop} \triangleq \frac{\sum_{i=1}^R\#\{\delta_i < 0\}}{R}\,.$

\subsection{Cubic Splines}

\begin{figure}[H]
    \centering
    \begin{subfigure}{0.5\textwidth}
    \includegraphics[width=0.9\textwidth]{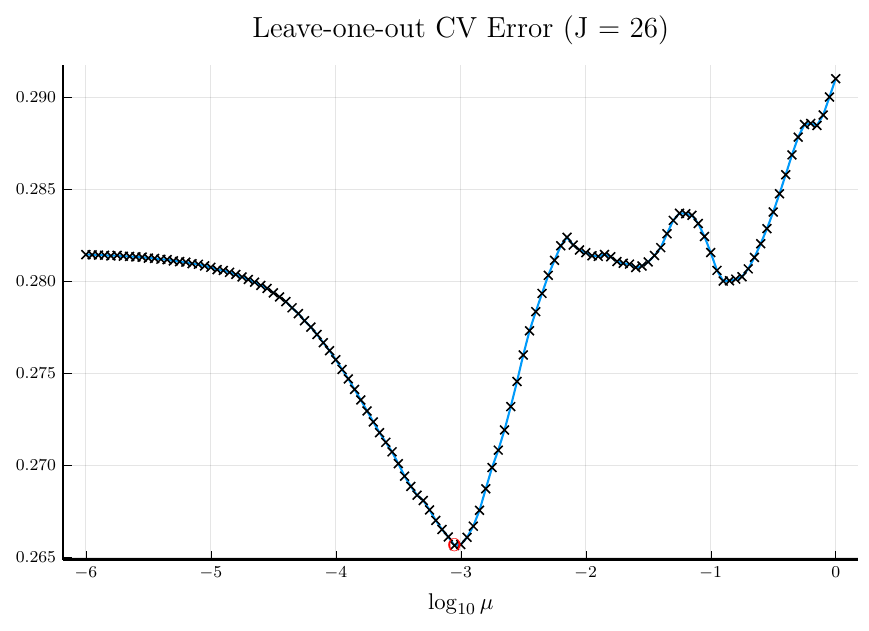}
    \end{subfigure}%
    \begin{subfigure}{0.5\textwidth}
    \includegraphics[width=\textwidth]{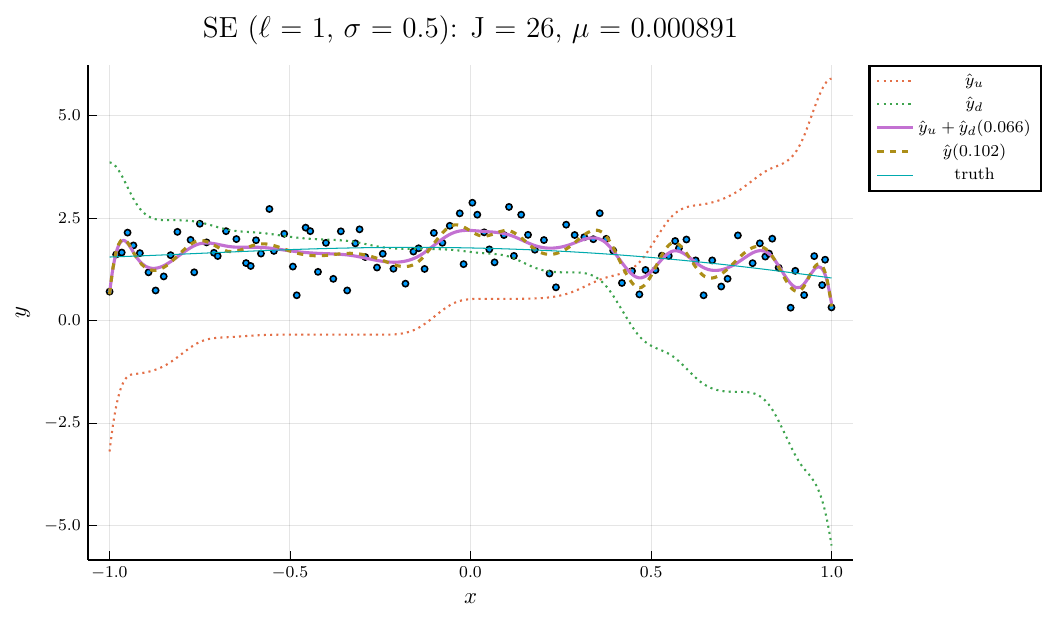}
    \end{subfigure}
    \caption[Monotone Decomposition with Cubic Splines using CV-tuned $\mu$]{Demo for monotone decomposition with cubic splines. The left panel shows the leave-one-out cross-validation error curve for each candidate $\mu$ when $J$ is CV-tuned for the cubic spline. The right panel shows the corresponding fitting curves, together with the truth and the noised observations. The values in the parentheses are the mean squared prediction error.}
    \label{fig:demo_bspl}
\end{figure}

\begin{figure}[H]
    \centering
    \begin{subfigure}{0.5\textwidth}
    \includegraphics[width=\textwidth]{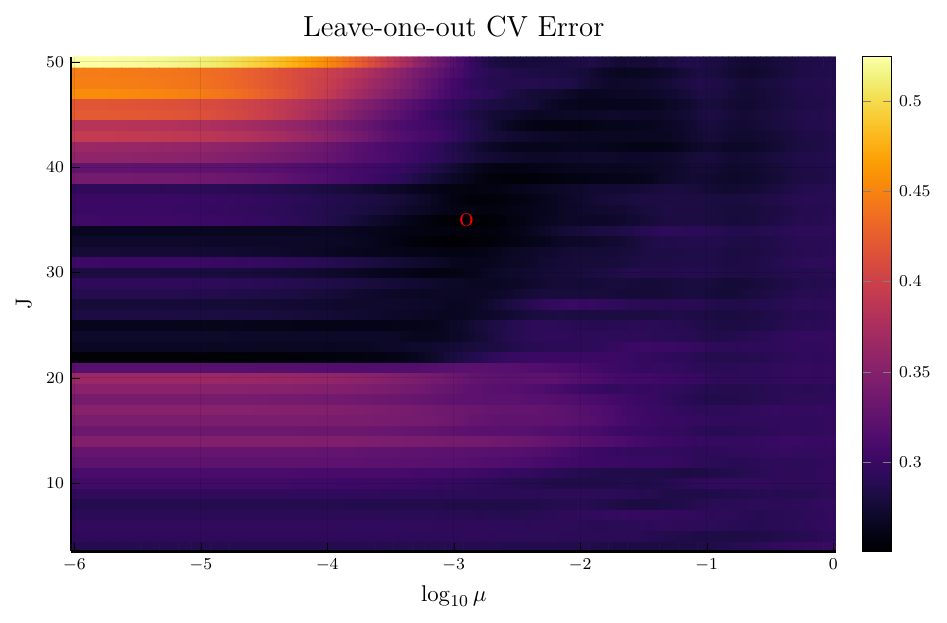}
    \end{subfigure}%
    \begin{subfigure}{0.5\textwidth}
    \includegraphics[width=\textwidth]{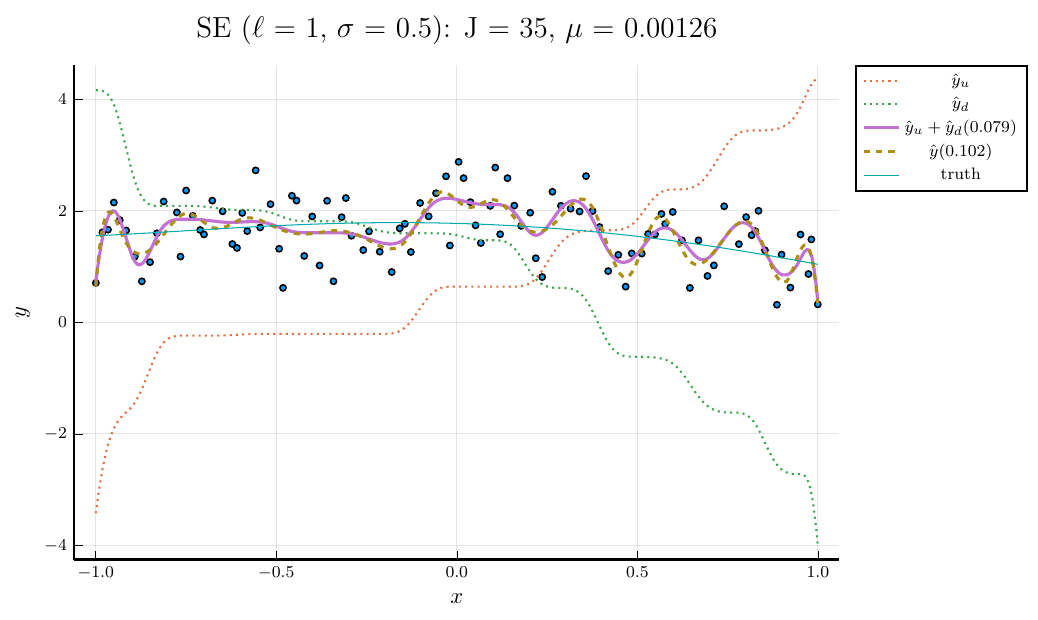}
    \end{subfigure}
    \caption[Monotone Decomposition with Cubic Splines using CV-tuned $\mu$ and $J$]{Demo for monotone decomposition with cubic splines. The left panel shows the leave-one-out cross-validation error curve for each candidate pair $(\mu,J)$. The right panel shows the corresponding fitting curves, together with the truth and noised observations. The values in the parentheses are the mean squared prediction error.}
    \label{fig:demo_bspl2}
\end{figure}

Firstly, we consider the monotone decomposition with cubic splines. There are two tuning parameters: the number of basis functions $J$, and the Lagrange multiplier $\mu$ for the discrepancy between the two components. We adopt two strategies to optimize these parameters:
    \paragraph{Tune $\mu$ with fixed $J$:}
    We pick $J$,  the tuning parameter for cubic splines, to be a minimizer of the cross-validation (CV) error, and then perform the monotone decomposition with cubic splines using the same $J$ while tuning the parameter $\mu$. Figure \ref{fig:demo_bspl} shows an example, with $J = 26$ selected by CV.
The left panel shows the leave-one-out CV error plotted against $\mu$. The cubic spline fitting and the monotone decomposition fitting are displayed in the right panel, where the former achieves 0.102 MSPE, while the latter improves the MSPE to 0.066.
    \paragraph{Tune $\mu$ and $J$ simultaneously:}
    Instead of fixing $J$ in the monotone decomposition procedure, we also use cross-validation to choose it, together with $\mu$. Figure \ref{fig:demo_bspl2} displays an example of this process, where the left panel shows the CV error for each parameter pair $(\mu, J)$, and the right panel compares the fitting given the parameters that minimize the CV error to cubic spline fitting, whose parameter $J$ is separately tuned by CV.





Note that the right panels of Figures \ref{fig:demo_bspl} and \ref{fig:demo_bspl2} depict the same training data. Although the MSPE of 0.079 by tuning $(\mu, J)$ simultaneously is slightly larger than the MSPE of 0.066 by tuning $\mu$ with fixed $J$, the monotone decomposition method achieves better performance than the cubic spline fitting, which has an MSPE of 0.102.

To provide comprehensive comparisons, we conducted 100 repetitions for 12 types of curves under different noise levels. The results by tuning $\mu$ and $J$ simultaneously are summarized in Table \ref{tab:bspl_vs_mbss_mu_J}, and the results by tuning $\mu$ with fixed $J$ can be found in the \supp. For some curves with small noises (e.g., $\sigma = 0.2$), such as $y=x^3$, the decomposition method performs slightly worse than cubic spline fitting. Nevertheless, the monotone decomposition always outperforms cubic spline fitting in higher noise (e.g., $\sigma=1.0$) scenarios, regardless of the optimization strategies. 

\begin{table}[H]
    \centering
    \caption[MSE for Monotone Decomposition with Cubic Splines using CV-tuned ($\mu, J$)]{Results for comparing the cubic splines and the fitting by monotone decomposition with CV-tuned $(\mu, J)$. The values are averages over 100 replications, with the standard error of the average in parentheses. The bold values highlight the smaller mean squared prediction error. The complete table with finer noise levels can be found in the \supp.}
    \label{tab:bspl_vs_mbss_mu_J}
    \resizebox{1.0\textwidth}{!}{%
        \begin{tabular}{ccccccll}
\toprule
\multirow{2}{*}{curve} & \multirow{2}{*}{$\sigma$}&\multicolumn{2}{c}{MSFE}&\multicolumn{2}{c}{MSPE}& \multirow{2}{*}{p-value}& \multirow{2}{*}{prop.}\tabularnewline
\cmidrule(lr){3-4}
\cmidrule(lr){5-6}
&&CubicSpline&MonoDecomp&CubicSpline&MonoDecomp\tabularnewline
\midrule
\multirow{1}{*}{$x^2$}
&1.0& 9.71e+00 (7.3e-02)& 9.73e+00 (7.3e-02)& 7.50e+00 (3.1e-01)& \textbf{6.93e+00} (2.6e-01)& 5.91e-03 (**)& 0.59\tabularnewline
\midrule
\multirow{1}{*}{$x^3$}
&1.0& 9.65e+00 (7.4e-02)& 9.60e+00 (7.2e-02)& 7.63e+00 (3.5e-01)& \textbf{7.17e+00} (2.7e-01)& 2.21e-02 (*)& 0.6\tabularnewline
\midrule
\multirow{1}{*}{$\exp(x)$}
&1.0& 9.57e+00 (5.9e-02)& 9.49e+00 (7.3e-02)& 7.56e+00 (2.8e-01)& \textbf{7.08e+00} (3.1e-01)& 3.59e-02 (*)& 0.57\tabularnewline
\midrule
\multirow{1}{*}{sigmoid}
&1.0& 9.51e+00 (8.5e-02)& 9.50e+00 (8.7e-02)& 7.33e+00 (3.2e-01)& \textbf{6.61e+00} (2.6e-01)& 1.45e-03 (**)& 0.56\tabularnewline
\midrule
\multirow{1}{*}{SE-1}
&1.0& 9.55e+00 (7.0e-02)& 9.51e+00 (7.6e-02)& 7.29e+00 (3.2e-01)& \textbf{6.62e+00} (2.7e-01)& 5.51e-03 (**)& 0.63\tabularnewline
\midrule
\multirow{1}{*}{SE-0.1}
&1.0& 9.29e+00 (8.5e-02)& 9.20e+00 (9.3e-02)& 1.44e+01 (2.6e-01)& \textbf{1.38e+01} (2.4e-01)& 5.93e-04 (***)& 0.7\tabularnewline
\midrule
\multirow{1}{*}{Mat12-1}
&1.0& 9.79e+00 (8.9e-02)& 9.73e+00 (9.1e-02)& 1.26e+01 (2.9e-01)& \textbf{1.17e+01} (2.4e-01)& 9.31e-07 (***)& 0.68\tabularnewline
\midrule
\multirow{1}{*}{Mat12-0.1}
&1.0& 1.04e+01 (1.3e-01)& 1.04e+01 (1.3e-01)& 2.07e+01 (2.4e-01)& \textbf{2.01e+01} (2.4e-01)& 1.85e-04 (***)& 0.73\tabularnewline
\midrule
\multirow{1}{*}{Mat32-1}
&1.0& 9.62e+00 (8.2e-02)& 9.61e+00 (8.3e-02)& 9.01e+00 (3.5e-01)& \textbf{8.00e+00} (2.6e-01)& 1.46e-04 (***)& 0.56\tabularnewline
\midrule
\multirow{1}{*}{Mat32-0.1}
&1.0& 9.62e+00 (1.1e-01)& 9.53e+00 (9.7e-02)& 1.68e+01 (2.5e-01)& \textbf{1.58e+01} (2.1e-01)& 4.67e-10 (***)& 0.72\tabularnewline
\midrule
\multirow{1}{*}{RQ-0.1-0.5}
&1.0& 9.55e+00 (1.1e-01)& 9.50e+00 (1.2e-01)& 1.50e+01 (2.4e-01)& \textbf{1.44e+01} (2.6e-01)& 5.01e-03 (**)& 0.68\tabularnewline
\midrule
\multirow{1}{*}{Periodic-0.1-4}
&1.0& 9.41e+00 (1.2e-01)& 9.31e+00 (1.1e-01)& 1.72e+01 (3.0e-01)& \textbf{1.60e+01} (2.5e-01)& 2.03e-10 (***)& 0.63\tabularnewline
\bottomrule
\end{tabular}

    }
\end{table}

\subsection{Smoothing Splines}

This section compares the fitting performance of monotone decomposition with smoothing splines to the fitting of the smoothing splines. There are two tuning parameters: the penalty parameter $\lambda$ and the Lagrange multiplier $\mu$ for the discrepancy. We consider three strategies to optimize these parameters:
\begin{itemize}
    \item Tune $\lambda$ for smoothing splines first, then tune $\mu$ for the monotone decomposition with smoothing splines using the tuned $\lambda$;
    \item According to Proposition \ref{thm:md_ss_mse}, tune $\lambda$ for smoothing splines first, then tune the shrinkage factor $k=\frac{1}{1+\mu}$ for monotone decomposition with smoothing splines using penalty parameter $\lambda/k$;
    \item Simultaneously tune $\lambda$ and $\mu$ for monotone decomposition with smoothing splines.
\end{itemize}
All strategies use cross-validation (CV) to determine the parameters. Specifically, we pick the tuning parameters that minimize the CV error. 

\begin{figure}
    \centering
    \begin{subfigure}{0.5\textwidth}
    \includegraphics[width=\textwidth]{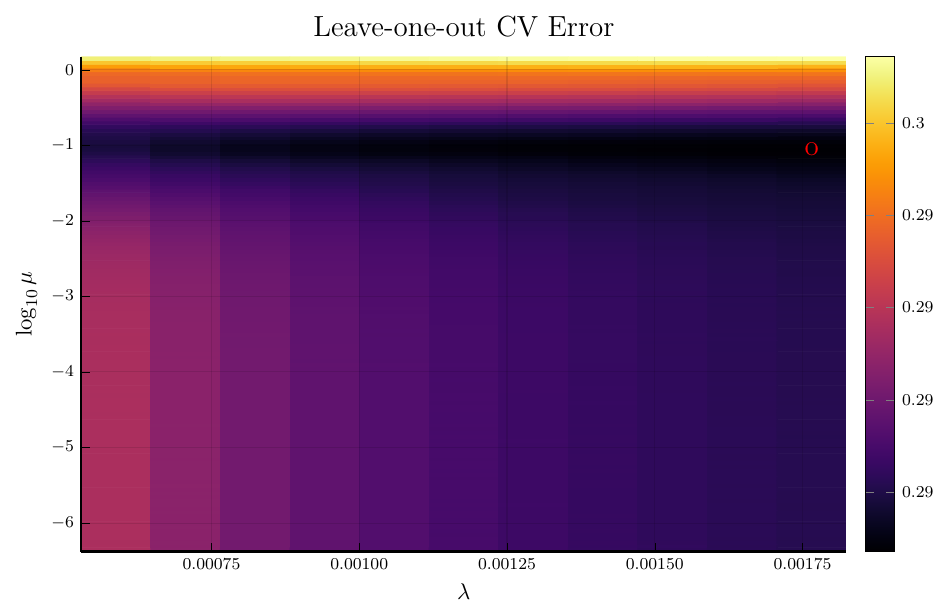}
    \end{subfigure}%
    \begin{subfigure}{0.5\textwidth}
    \includegraphics[width=\textwidth]{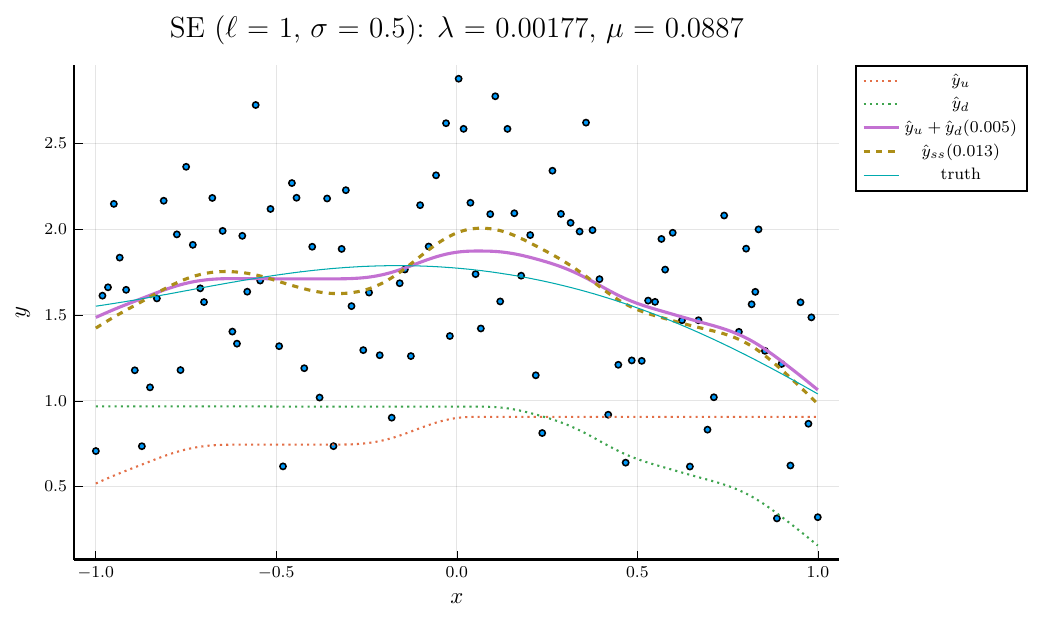}
    \end{subfigure}
    \caption[Monotone Decomposition with Smoothing Splines using CV-tuned $\mu$ and $\lambda$]{Demo for monotone decomposition with smoothing splines. The left panel shows the leave-one-out cross-validation error heatmap for each candidate pair $(\mu, \lambda)$. The right panel shows the corresponding fitting curves, together with the truth and the noised observations. The values in the parentheses are the mean squared prediction errors.}
    \label{fig:demo_ss2}
\end{figure}

For brevity, we only present the results using the third strategy in this section. Results based on the other two strategies can be found in the \supp. Figure \ref{fig:demo_ss2} illustrates the simultaneous tuning of $(\mu,\lambda)$ using the same toy example presented in Figures~\ref{fig:demo_bspl} and \ref{fig:demo_bspl2}. The smoothness penalty in smoothing splines results in fitted curves that are notably less wiggly compared to the curves fitted by cubic splines without such a penalty.

The results of 100 repetitive experiments are summarized in Table \ref{tab:ss_vs_md_lambda_mu}. We observe that the monotone decomposition consistently outperforms the corresponding smoothing splines.
Moreover, the monotone decomposition with CV-tuned $(\lambda, \mu)$ is more likely to obtain better performance compared to the other two strategies. Regardless of the optimization strategies employed, all results consistently show that the monotone decomposition fitting can achieve a good performance, especially in high-noise scenarios. 



\begin{table}[H]
    \centering
    \caption[MSE for Monotone Decomposition with Smoothing Splines using CV-tuned ($\mu,\lambda$)]{Results for comparing the smoothing splines with CV-tuned $\lambda$ and the fitting by monotone decomposition with CV-tuned $(\lambda, \mu)$. The values are averages over 100 replications, with the standard error of the average in parentheses. The bold values highlight the smaller mean squared prediction error. The complete table with finer noise levels can be found in the \supp.}
    \label{tab:ss_vs_md_lambda_mu}
    \resizebox{\textwidth}{!}{%
        \begin{tabular}{ccccccll}
\toprule
\multirow{2}{*}{curve} & \multirow{2}{*}{$\sigma$}&\multicolumn{2}{c}{MSFE}&\multicolumn{2}{c}{MSPE}& \multirow{2}{*}{p-value}& \multirow{2}{*}{prop.}\tabularnewline
\cmidrule(lr){3-4}
\cmidrule(lr){5-6}
&&SmoothSpline&MonoDecomp&SmoothSpline&MonoDecomp\tabularnewline
\midrule
\multirow{1}{*}{$x^2$}
&1.0& 9.65e+00 (8.9e-02)& 9.69e+00 (8.5e-02)& 6.44e+00 (2.9e-01)& \textbf{6.35e+00} (2.5e-01)& 1.68e-01& 0.49\tabularnewline
\midrule
\multirow{1}{*}{$x^3$}
&1.0& 9.75e+00 (8.2e-02)& 9.77e+00 (8.2e-02)& 6.47e+00 (1.8e-01)& \textbf{6.21e+00} (1.7e-01)& 1.18e-04 (***)& 0.65\tabularnewline
\midrule
\multirow{1}{*}{$\exp(x)$}
&1.0& 9.74e+00 (8.4e-02)& 9.75e+00 (8.4e-02)& 5.94e+00 (2.3e-01)& \textbf{5.82e+00} (2.1e-01)& 3.12e-02 (*)& 0.58\tabularnewline
\midrule
\multirow{1}{*}{sigmoid}
&1.0& 9.60e+00 (7.8e-02)& 9.64e+00 (7.5e-02)& 5.99e+00 (2.9e-01)& \textbf{5.68e+00} (2.4e-01)& 4.47e-04 (***)& 0.67\tabularnewline
\midrule
\multirow{1}{*}{SE-1}
&1.0& 9.67e+00 (8.3e-02)& 9.70e+00 (8.1e-02)& 6.32e+00 (2.8e-01)& \textbf{6.11e+00} (2.5e-01)& 3.86e-03 (**)& 0.6\tabularnewline
\midrule
\multirow{1}{*}{SE-0.1}
&1.0& 8.92e+00 (9.2e-02)& 8.99e+00 (9.0e-02)& 1.23e+01 (2.1e-01)& \textbf{1.23e+01} (2.1e-01)& 4.32e-01& 0.57\tabularnewline
\midrule
\multirow{1}{*}{Mat12-1}
&1.0& 9.65e+00 (9.5e-02)& 9.69e+00 (9.2e-02)& 1.15e+01 (2.3e-01)& \textbf{1.13e+01} (2.1e-01)& 3.40e-04 (***)& 0.56\tabularnewline
\midrule
\multirow{1}{*}{Mat12-0.1}
&1.0& 9.76e+00 (1.2e-01)& 9.92e+00 (1.1e-01)& 1.90e+01 (2.2e-01)& \textbf{1.90e+01} (2.2e-01)& 2.06e-01& 0.58\tabularnewline
\midrule
\multirow{1}{*}{Mat32-1}
&1.0& 9.59e+00 (8.5e-02)& 9.64e+00 (7.5e-02)& 7.20e+00 (2.4e-01)& \textbf{7.04e+00} (2.0e-01)& 3.65e-02 (*)& 0.49\tabularnewline
\midrule
\multirow{1}{*}{Mat32-0.1}
&1.0& 8.96e+00 (1.1e-01)& 9.14e+00 (1.0e-01)& 1.48e+01 (2.2e-01)& \textbf{1.47e+01} (2.1e-01)& 1.44e-01& 0.54\tabularnewline
\midrule
\multirow{1}{*}{RQ-0.1-0.5}
&1.0& 9.10e+00 (1.1e-01)& 9.22e+00 (1.1e-01)& 1.27e+01 (2.4e-01)& \textbf{1.25e+01} (2.2e-01)& 1.21e-03 (**)& 0.6\tabularnewline
\midrule
\multirow{1}{*}{Periodic-0.1-4}
&1.0& 8.69e+00 (1.0e-01)& 8.89e+00 (9.1e-02)& 1.44e+01 (2.1e-01)& \textbf{1.43e+01} (1.9e-01)& 3.03e-01& 0.49\tabularnewline
\bottomrule
\end{tabular}

    }
\end{table}

\section{Test of Monotonicity}\label{sec:test_of_mono}

Once obtaining two monotone components through monotone decomposition, in addition to utilizing the sum of these two components as a fitting method, we can also derive statistics for the monotonicity testing.
Consider the model $Y=f(X)+\epsilon$, where $X$ is a scalar covariate, $Y$ is a scalar dependent random variable, $\epsilon$ is the noise satisfying $\bbE[\epsilon\mid X]=0$, and $f$ is an unknown function. Testing of monotonicity aims to test
$$
H_0: f \text{ is monotone} \quad H_1: f\text{ is not monotone}\,,
$$
where ``monotone'' can be specifically (strictly) ``increasing'' or ``decreasing''.

\subsection{Related Work}\label{sec:test_related}

There are many existing approaches for testing monotonicity.
\textcite{bowmanTestingMonotonicityRegression1998} constructed a test based on critical bandwidth. They fitted a local linear regression and determined the smallest bandwidth value such that the estimate becomes monotone. This critical bandwidth is then used as a test statistic, and the $p$-value is calculated by the bootstrap method. \textcite{hallTestingMonotonicityRegression2000} pointed out the shortcoming of the test when the true function has flat and nearly flat spots, and they proposed a test that estimates local slopes and approximates the distribution of the weighted minimum. 
\textcite{ghosalTestingMonotonicityRegression2000} proposed test statistics that are functionals of a U-process, which is based on the signs of $(Y_{i+k}-Y_i)(X_{i+k}-X_i)$. They approximated the limiting distribution by Gaussian processes and then derived the critical values for an asymptotic significance level $\alpha$. \textcite{chetverikovTestingRegressionMonotonicity2019} used the similar U-statistics, but he introduced a weighting function $Q(X_i, X_j)$ and proposed a statistic based on $(Y_i-Y_j)\sign(X_j-X_i)Q(X_i, X_j)$, where $Q$ is chosen from a large set of potentially useful weighting functions to maximize the statistic. \textcite{wangTestingMonotonicityConvexity2011} used quadratic regression splines to fit the data, took the minimum of the slopes at the knots as the test statistic, and then estimated the null distribution of such a statistic by performing constrained quadratic regression splines. 

\subsection{Test by Monotone Decomposition}\label{sec:test_method}

Suppose we have obtained the monotone components. Propositions \ref{thm:md_cs_increase} and \ref{thm:md_ss_sol} imply that the coefficients for one component would be constant if the function is monotone. Thus, we can test the monotonicity of a function by testing whether the coefficients of monotone components are constant. The equivalences are summarized in Table \ref{tab:H0s}.
\begin{table}[H]
    \centering
    \caption[Equivalent Hypotheses]{Equivalent hypotheses.}
    \label{tab:H0s}
    \begin{tabular}{c|c}
    \toprule
    Original Hypothesis & Hypothesis in terms of Monotone Decomposition\\
    \hline
        $\cH_0^u: f \text{ is increasing}$ &  $H_0^u:\gamma^d = c\one\,.$ \\
        $\cH_0^d: f \text{ is decreasing}$ & $H_0^d: \gamma^u = c\one\,.$\\
        $\cH_0^{su}: f \text{ is strictly increasing}$ & $H_0^{su}: \gamma^d = c\one, \gamma^u\neq c\one\,.$\\
        $\cH_0^{sd}: f \text{ is strictly decreasing}$ & $H_0^{sd}: \gamma^u =c\one, \gamma^d\neq c\one\,.$\\
        \midrule
        $\cH_0^m: f \text{ is monotone}$ & \makecell{$H_0^m: \text{one monotone component is constant}$,\tabularnewline i.e., $\min(\gamma^u,\gamma^d)=c\one\,.$}\\
        $\cH_0^{sm}: f \text{ is strictly monotone }$ & \makecell{$H_0^{sm}: \text{one and only one monotone component is constant}$,\tabularnewline i.e., $\min(\gamma^u,\gamma^d)=c\one, \max(\gamma^u,\gamma^d)\neq c\one\,.$}\\
    \bottomrule
    \end{tabular}
\end{table}
In Table \ref{tab:H0s}, the minimum (maximum) of two vectors $a, b\in\IR^J$ are defined as:
\begin{align*}
\min(a, b) \triangleq \argmin_{x\in \{a, b\}} V(x), \ \ \ \ \ 
\max(a, b) &\triangleq \argmax_{x\in \{a, b\}} V(x) ,
\end{align*}
where $V$ is the sample variance on the elements of a vector. To test $H_0^u, H_0^d$ and $H_0^m$, consider the test statistics
\begin{align*}
T_u =V(\hat\gamma^d), \ \ 
T_d =V(\hat\gamma^u), \ \ 
T_m =V(\min(\hat\gamma^u, \hat\gamma^d)).
\end{align*}
Note that the null hypothesis will be rejected if the test statistic $T$ is large enough. Specifically, given a significance level $\alpha$, the respective null hypothesis would be rejected if $T_u\ge t_{u, 1-\alpha}, T_d\ge t_{d, 1-\alpha}$ or $T_m \ge t_{m, 1-\alpha}$, where $t_{u,1-\alpha}, t_{d, 1-\alpha}$ and $t_{m, 1-\alpha}$ denote the critical values of the distributions of $T_u, T_d, T_m$ under the respective null hypotheses, respectively. The distributions of $T_u, T_d, T_m$ under their null hypotheses can be characterized by the bootstrap samples. Note that the $\epsilon$ and $\hat\epsilon$ can be heterogeneous, so we take the wild bootstrap \parencite{davidsonWildBootstrapTamed2008}.
Without loss of generality, we focus on the test of increasing functions, and the procedure for testing $H_0^u$ is outlined in Algorithm \ref{alg:mono_test}.

\begin{algorithm}
    \caption{Test of Monotonicity (Increasing): $H^u_0$}
    \label{alg:mono_test}
    \begin{algorithmic}[1]
    \REQUIRE Significance level $\alpha$, number of bootstrap samples $R$.
    \STATE Fit $\{x_i, y_i\}_{i=1}^n$ by monotone decomposition, either with cubic splines or smoothing splines. Denote the increasing component as $\hat y^u$, let $\hat c = \frac{\one^T\hat y^u}{n}$, and denote the fitting method as $\hat m$. 
    \STATE Calculate the test statistic $T$.
    \STATE Calculate the errors, $\epsilon_i=y_i-\hat y_i$.
    \FOR{$r$ from 1 to $R$}
    \STATE Sample $\eta_i\sim N(0,1)$ and construct $\epsilon^\star_i = \eta_i\epsilon_i$. [Wild Bootstrap]
    \STATE Construct bootstrap samples $y^\star_i = \hat y^u_i + \hat c + \epsilon^\star_i, i=1,\ldots,n$.
    \STATE Apply $\hat m$ on $\{x_i,y_i^\star\}_{i=1}^n$, and calculate the bootstrap statistic $\delta_r$
    \ENDFOR
    \STATE The critical value $t_{1-\alpha} $ is the $1-\alpha$ quantile of $\{\delta_r\}_{r=1}^R$, then reject the null hypothesis if 
    $$
    T > t_{1-\alpha}
    $$
    Alternatively, construct $p$-value $\frac 1R\sum_{r=1}^R\#\{\delta_r > T\}$, and reject if $p < \alpha$.
    \end{algorithmic}
\end{algorithm}

\section{Simulations for Testing}\label{sec:test_sim}

Firstly, we want to check whether the methods can control the Type I error. 
Specifically, consider five monotone functions, $x, x^3, x^{1/3}, e^x$ and $\frac{1}{1+e^{-x}}$. We conducted 100 simulations and 
calculated the proportion of rejecting the null hypothesis. Ideally, the rejection proportion should be less than 0.05 if we pick the commonly used significance level $\alpha=0.05$. The results are reported in Table \ref{tab:mono_test_mono}.

\begin{table}
    \centering
    \caption[Simulated Size on Monotone Curves]{Simulated size (the proportion of rejecting the null hypothesis) of monotone curves under different noise levels and different sample sizes given the significance level $\alpha = 0.05$.}
    \label{tab:mono_test_mono}
     \resizebox{0.75\textwidth}{!}{\begin{tabular}{ccccccccccc}
\toprule
\multirow{2}{*}{Methods} & \multirow{2}{*}{Curves}&\multicolumn{3}{c}{$\sigma = 0.001$}&\multicolumn{3}{c}{$\sigma = 0.01$}&\multicolumn{3}{c}{$\sigma = 0.1$}\tabularnewline
\cmidrule(lr){3-5}
\cmidrule(lr){6-8}
\cmidrule(lr){9-11}
&&n = 50&100&200&n = 50&100&200&n = 50&100&200\tabularnewline
\midrule
\midrule
\multirow{5}{*}{\textcite{wangTestingMonotonicityConvexity2011}}&$x$& 0.0& 0.0& 0.0& 0.0& 0.0& 0.0& 0.0& 0.01& 0.0\tabularnewline
&$x^3$& 0.53& 0.84& 1.0& 0.05& 0.08& 0.08& 0.02& 0.06& 0.04\tabularnewline
&$x^{1/3}$& 0.0& 0.0& 0.0& 0.0& 0.0& 0.0& 0.03& 0.02& 0.01\tabularnewline
&$e^x$& 0.0& 0.0& 0.0& 0.0& 0.0& 0.0& 0.0& 0.0& 0.0\tabularnewline
&$1/(1+e^{-x})$& 0.0& 0.0& 0.0& 0.01& 0.0& 0.0& 0.04& 0.06& 0.05\tabularnewline
\midrule
\multirow{5}{*}{\textcite{ghosalTestingMonotonicityRegression2000}}&$x$& 0.0& 0.0& 0.0& 0.0& 0.0& 0.0& 0.0& 0.0& 0.0\tabularnewline
&$x^3$& 0.0& 0.0& 0.0& 0.0& 0.0& 0.0& 0.0& 0.0& 0.0\tabularnewline
&$x^{1/3}$& 0.0& 0.0& 0.0& 0.0& 0.0& 0.0& 0.0& 0.0& 0.0\tabularnewline
&$e^x$& 0.0& 0.0& 0.0& 0.0& 0.0& 0.0& 0.0& 0.0& 0.0\tabularnewline
&$1/(1+e^{-x})$& 0.0& 0.0& 0.0& 0.0& 0.0& 0.0& 0.0& 0.0& 0.0\tabularnewline
\midrule
\multirow{5}{*}{\textcite{bowmanTestingMonotonicityRegression1998}}&$x$& 0.0& 0.0& 0.0& 0.02& 0.0& 0.01& 0.01& 0.0& 0.01\tabularnewline
&$x^3$& 0.0& 0.0& 0.0& 0.2& 0.19& 0.18& 0.26& 0.2& 0.15\tabularnewline
&$x^{1/3}$& 0.0& 0.0& 0.0& 0.0& 0.0& 0.0& 0.05& 0.02& 0.01\tabularnewline
&$e^x$& 0.0& 0.0& 0.0& 0.0& 0.0& 0.0& 0.04& 0.02& 0.0\tabularnewline
&$1/(1+e^{-x})$& 0.0& 0.0& 0.0& 0.02& 0.0& 0.02& 0.02& 0.01& 0.0\tabularnewline
\midrule
\multirow{5}{*}{MDCS}&$x$& 0.01& 0.0& 0.12& 0.0& 0.0& 0.0& 0.0& 0.0& 0.0\tabularnewline
&$x^3$& 0.0& 0.0& 0.0& 0.01& 0.0& 0.0& 0.0& 0.0& 0.0\tabularnewline
&$x^{1/3}$& 0.0& 0.0& 0.0& 0.0& 0.0& 0.0& 0.0& 0.0& 0.0\tabularnewline
&$e^x$& 0.02& 0.11& 0.03& 0.0& 0.0& 0.0& 0.0& 0.0& 0.0\tabularnewline
&$1/(1+e^{-x})$& 0.0& 0.0& 0.0& 0.0& 0.0& 0.0& 0.0& 0.0& 0.0\tabularnewline
\midrule
\multirow{5}{*}{MDSS}&$x$& 0.0& 0.0& 0.0& 0.0& 0.0& 0.0& 0.0& 0.0& 0.0\tabularnewline
&$x^3$& 0.1& 0.08& 0.08& 0.08& 0.1& 0.07& 0.09& 0.05& 0.03\tabularnewline
&$x^{1/3}$& 0.0& 0.02& 0.02& 0.01& 0.03& 0.07& 0.05& 0.06& 0.05\tabularnewline
&$e^x$& 0.04& 0.05& 0.02& 0.08& 0.07& 0.05& 0.04& 0.08& 0.08\tabularnewline
&$1/(1+e^{-x})$& 0.03& 0.03& 0.03& 0.0& 0.01& 0.0& 0.0& 0.0& 0.0\tabularnewline
\bottomrule
\end{tabular}

     }
\end{table}

\textcite{ghosalTestingMonotonicityRegression2000} always accepts the null hypothesis. \textcite{wangTestingMonotonicityConvexity2011} fails to control the Type I error when the noise level is small on curve $x^3$, and \textcite{bowmanTestingMonotonicityRegression1998} cannot control the Type I error when the noise level is large on curve $x^3$. In contrast, our proposed methods, monotone decomposition with cubic splines (MDCS) and monotone decomposition with smoothing splines (MDSS), demonstrate strong Type I error control in the majority of cases, even though the rejection rates are slightly elevated in a few instances.

\begin{figure}[H]
    \centering
    \begin{subfigure}{0.5\textwidth}
    \includegraphics[width=\textwidth]{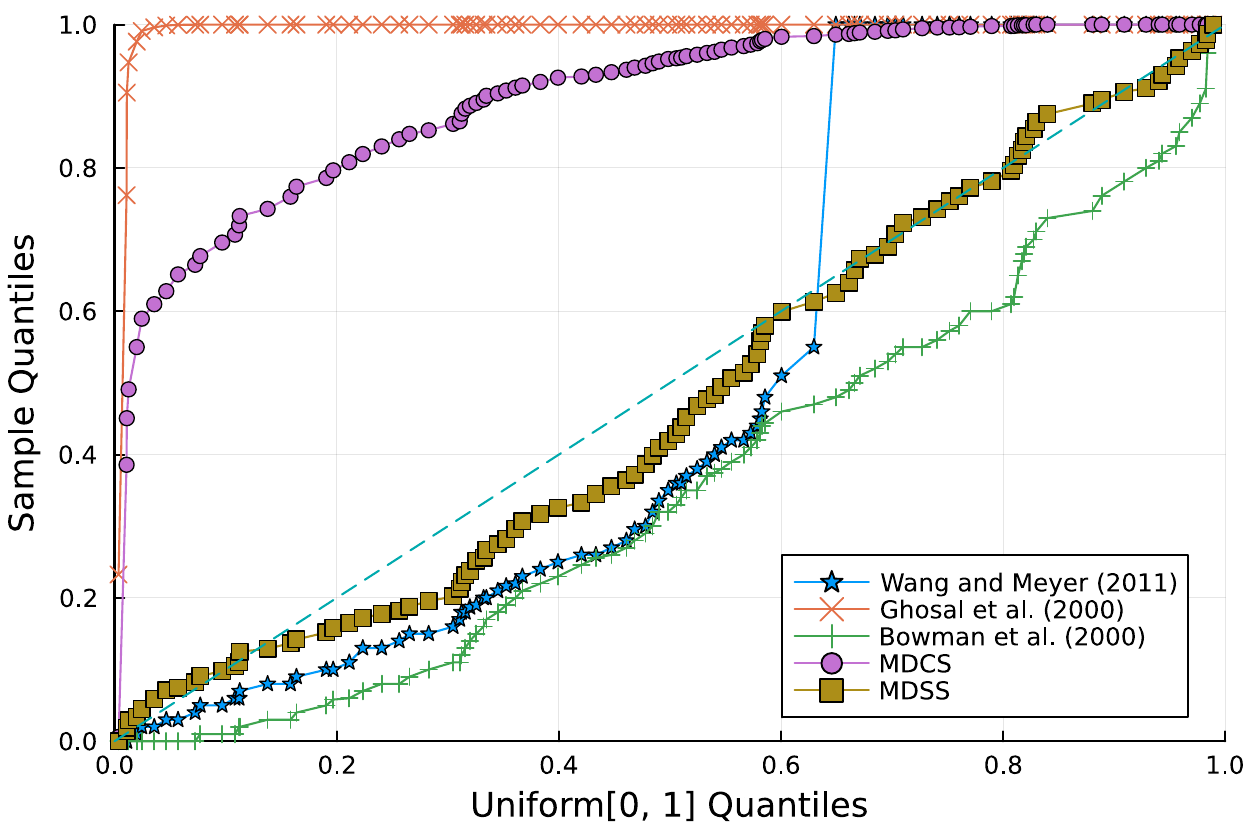}
    \caption{$x^3$}
    \end{subfigure}%
    \begin{subfigure}{0.5\textwidth}
    \includegraphics[width=\textwidth]{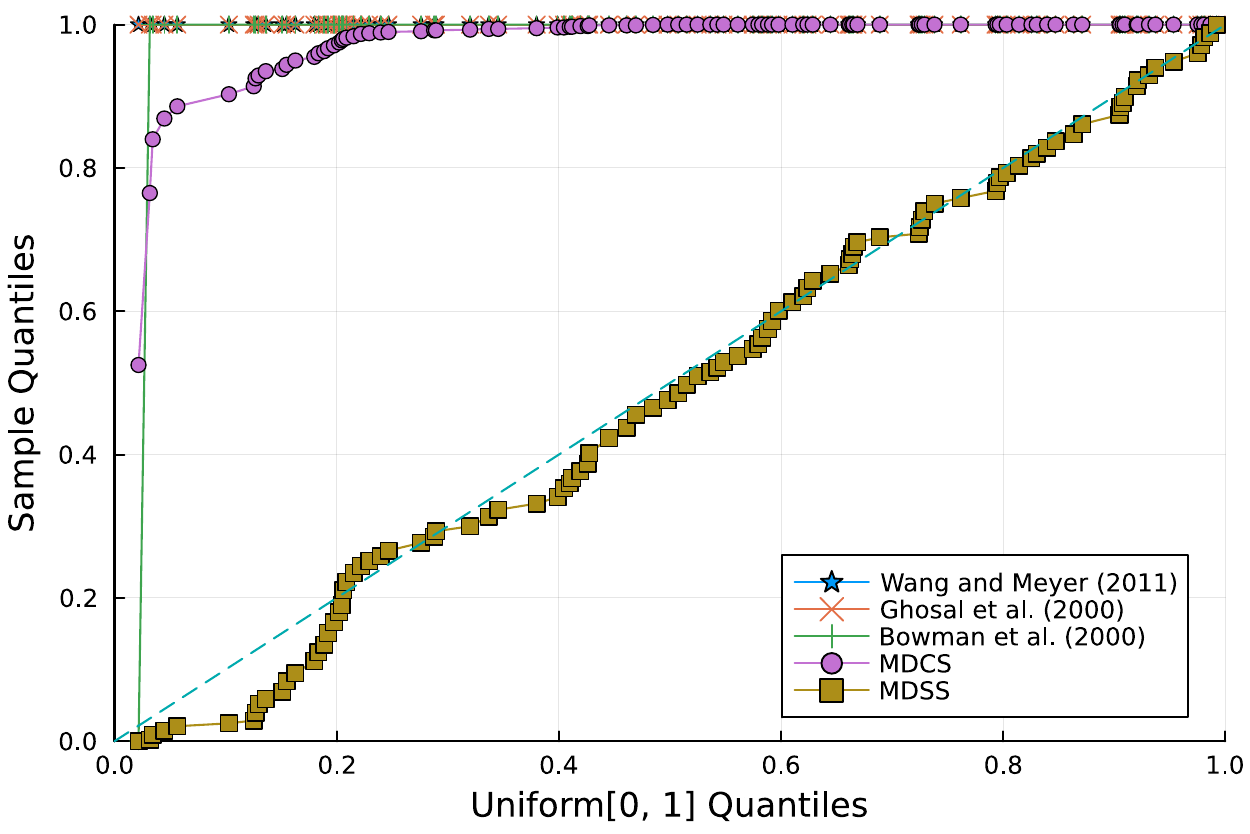}
    \caption{$e^x$}
    \end{subfigure}

    \caption{Uniform QQ plots of 1000 $p$-values for five approaches on two curves with $n=200,\sigma=0.01$, respectively.}
    \label{fig:pval_dist}
\end{figure}

Furthermore, the $p$-value should follow $\text{Uniform}[0, 1]$ under the null hypothesis. To check the distribution of $p$-value for each approach, Figure~\ref{fig:pval_dist} displays the uniform QQ plots of 1000 $p$-values for $x^3$ and $e^x$ with sample size $n=200$ and noise level $\sigma=0.01$, respectively. The uniform QQ plots of $p$-values for all five curves with different noise levels can be found in the \supp. Notably, our proposed MDSS aligns pretty well with the diagonal line in the QQ plots, indicating the closest resemblance to the uniform distribution.

Next, we compare the simulated size and power under the settings of two competitors. The first one is the simulation setting in \textcite{bowmanTestingMonotonicityRegression1998},
$$
f(x, a) = 1+x-a\exp\left(-\frac{(x-0.5)^2}{2\cdot 0.1^2}\right)\,,\quad y = f(x, a) + N(0, \sigma^2)\,,
$$
where $a=0$, and 0.15 generate strongly and just monotonic curves, respectively; $a=0.25$, and 0.45 produce mildly and strongly non-monotonic shapes, respectively. Also, we consider the simulation setting in \textcite{ghosalTestingMonotonicityRegression2000}, $y=m_i(x)+N(0,\sigma^2)$, where
  \begin{align*}
    m_1(x) &= 0\,,\quad m_2(x) = x(1-x)\,,\quad m_3(x) = x + 0.415\exp(-50x^2)\,,\\
    m_4(x) &= \begin{cases}
      10(x-0.5)^3 - \exp(-100(x-0.25)^2) & \text{ if } x < 0.5\\
      0.1(x-0.5) - \exp(-100(x-0.25)^2) & \text{ otherwise}
    \end{cases}\,,
  \end{align*}
and $m_i(x), i\ge 2$ are non-monotone curves. A visualization of those curves can be found in the \supp.

For each combination of parameters on each curve, 100 simulations are carried out, using a bootstrap simulation size of 100. The proportions of rejecting the null hypothesis, i.e., the simulated size and power, are reported.
The complete results are displayed in Table \ref{tab:mono_test_bowman_and_ghosal}, from which we have the following observations:

\begin{itemize}
    \item \textcite{ghosalTestingMonotonicityRegression2000} fails to achieve high power on \textcite{bowmanTestingMonotonicityRegression1998}'s curve $a=0.25$, but our proposed method can achieve as high power as \textcite{bowmanTestingMonotonicityRegression1998}.
    \item \textcite{bowmanTestingMonotonicityRegression1998} loses power on \textcite{ghosalTestingMonotonicityRegression2000}'s curve $m_4$, but our proposed method can obtain as high power as \textcite{ghosalTestingMonotonicityRegression2000}.
    \item For most curves, the behaviors of MDCS are similar to MDSS. But MDCS has better control over Type I errors while losing some power.
\end{itemize}

In summary, our proposed method can achieve comparable (and even better) power as other methods while controlling the Type I error.

\begin{table}
    \centering
    \caption[Simulated Power on \textcite{bowmanTestingMonotonicityRegression1998}'s Curves]{Simulated size and power on curves from \textcite{bowmanTestingMonotonicityRegression1998} and \textcite{ghosalTestingMonotonicityRegression2000} based on 100 repetitive experiments.}
    \label{tab:mono_test_bowman_and_ghosal}    
    \resizebox{\textwidth}{!}{\aboverulesep = 0pt
\belowrulesep = 0pt
\begin{tabular}{ccccccccccc| cccccccccc}
\toprule
\multirow{3}{*}{Methods} & \multicolumn{10}{c|}{\textcite{bowmanTestingMonotonicityRegression1998}} & \multicolumn{10}{c}{\textcite{ghosalTestingMonotonicityRegression2000}} \tabularnewline
\cmidrule(lr){2-11}
\cmidrule(lr){12-21}
&\multirow{2}{*}{Curves}&\multicolumn{3}{c}{$\sigma = 0.001$}&\multicolumn{3}{c}{$\sigma = 0.01$}&\multicolumn{3}{c|}{$\sigma = 0.1$}&\multirow{2}{*}{Curves}&\multicolumn{3}{c}{$\sigma = 0.001$}&\multicolumn{3}{c}{$\sigma = 0.01$}&\multicolumn{3}{c}{$\sigma = 0.1$}\tabularnewline
\cmidrule(lr){3-5}
\cmidrule(lr){6-8}
\cmidrule(lr){9-11}
\cmidrule(lr){13-15}
\cmidrule(lr){16-18}
\cmidrule(lr){19-21}
&&n = 50&100&200&n = 50&100&200&n = 50&100&200 &&n = 50&100&200&n = 50&100&200&n = 50&100&200\tabularnewline
\midrule
\midrule
\multirow{4}{*}{\textcite{wangTestingMonotonicityConvexity2011}}&a = 0.0& 0.0& 0.0& 0.0& 0.0& 0.0& 0.0& 0.01& 0.01& 0.01&m1& 0.1& 0.04& 0.04& 0.06& 0.02& 0.06& 0.07& 0.07& 0.1\tabularnewline
&a = 0.15& 0.0& 0.0& 0.0& 0.07& 0.04& 0.06& 0.03& 0.03& 0.01&m2& 1.0& 1.0& 1.0& 1.0& 1.0& 1.0& 0.09& 0.27& 0.45\tabularnewline
&a = 0.25& 1.0& 1.0& 1.0& 1.0& 1.0& 1.0& 0.13& 0.24& 0.64&m3& 1.0& 1.0& 1.0& 1.0& 1.0& 1.0& 0.34& 0.56& 0.88\tabularnewline
&a = 0.45& 1.0& 1.0& 1.0& 1.0& 1.0& 1.0& 0.56& 0.95& 1.0&m4& 0.36& 0.57& 0.98& 0.39& 0.57& 0.98& 0.23& 0.38& 0.78\tabularnewline
\midrule
\multirow{4}{*}{\textcite{ghosalTestingMonotonicityRegression2000}}&a = 0.0& 0.0& 0.0& 0.0& 0.0& 0.0& 0.0& 0.0& 0.0& 0.0 &m1& 0.01& 0.0& 0.01& 0.01& 0.04& 0.02& 0.02& 0.0& 0.01\tabularnewline
&a = 0.15& 0.0& 0.0& 0.0& 0.0& 0.0& 0.0& 0.0& 0.0& 0.0 &m2& 1.0& 1.0& 1.0& 1.0& 1.0& 1.0& 0.37& 0.7& 0.94\tabularnewline
&a = 0.25& 0.0& 0.0& 1.0& 0.0& 0.0& 0.31& 0.0& 0.0& 0.0  &m3& 0.97& 1.0& 1.0& 0.94& 1.0& 1.0& 0.1& 0.33& 0.87\tabularnewline
&a = 0.45& 1.0& 1.0& 1.0& 1.0& 1.0& 1.0& 0.06& 0.71& 0.99&m4& 0.01& 0.19& 0.53& 0.02& 0.13& 0.53& 0.0& 0.05& 0.34\tabularnewline
\midrule
\multirow{4}{*}{\textcite{bowmanTestingMonotonicityRegression1998}}&a = 0.0& 0.0& 0.0& 0.0& 0.02& 0.01& 0.0& 0.01& 0.01& 0.03&m1& 0.0& 0.0& 0.0& 0.0& 0.0& 0.0& 0.0& 0.0& 0.0\tabularnewline
&a = 0.15& 0.0& 0.0& 0.0& 0.0& 0.03& 0.0& 0.01& 0.02& 0.04&m2& 0.0& 0.0& 0.0& 0.0& 0.0& 0.0& 0.0& 0.0& 0.0\tabularnewline
&a = 0.25& 1.0& 1.0& 1.0& 1.0& 1.0& 1.0& 0.09& 0.16& 0.44&m3& 1.0& 1.0& 1.0& 1.0& 1.0& 1.0& 0.9& 1.0& 1.0\tabularnewline
&a = 0.45& 1.0& 1.0& 1.0& 1.0& 1.0& 1.0& 0.3& 0.72& 0.98&m4& 0.0& 0.0& 0.0& 0.01& 0.02& 0.0& 0.34& 0.33& 0.28\tabularnewline
\midrule
\multirow{4}{*}{MDCS}&a = 0.0& 0.01& 0.0& 0.07& 0.0& 0.0& 0.0& 0.0& 0.0& 0.0&m1& 0.03& 0.02& 0.02& 0.0& 0.01& 0.0& 0.0& 0.0& 0.0\tabularnewline
&a = 0.15& 0.05& 0.02& 0.04& 0.01& 0.0& 0.0& 0.0& 0.0& 0.0&m2& 0.98& 0.99& 0.95& 1.0& 1.0& 0.94& 0.18& 0.35& 0.65\tabularnewline
&a = 0.25& 0.97& 1.0& 1.0& 0.74& 0.92& 0.95& 0.0& 0.0& 0.0&m3& 0.89& 0.93& 1.0& 0.9& 0.97& 0.99& 0.16& 0.22& 0.29\tabularnewline
&a = 0.45& 1.0& 1.0& 1.0& 0.99& 1.0& 1.0& 0.01& 0.08& 0.17&m4& 0.89& 0.91& 0.94& 0.88& 0.92& 0.99& 0.51& 0.68& 0.81\tabularnewline
\midrule
\multirow{4}{*}{MDSS}&a = 0.0& 0.02& 0.07& 0.06& 0.02& 0.08& 0.03& 0.02& 0.04& 0.04&m1& 0.05& 0.07& 0.04& 0.06& 0.04& 0.03& 0.05& 0.07& 0.05\tabularnewline
&a = 0.15& 0.05& 0.03& 0.05& 0.03& 0.06& 0.05& 0.02& 0.05& 0.06&m2& 1.0& 1.0& 1.0& 1.0& 1.0& 1.0& 0.79& 0.96& 1.0\tabularnewline
&a = 0.25& 0.99& 1.0& 1.0& 0.98& 1.0& 1.0& 0.03& 0.04& 0.03&m3& 1.0& 1.0& 1.0& 1.0& 1.0& 1.0& 0.67& 0.83& 0.97\tabularnewline
&a = 0.45& 1.0& 1.0& 1.0& 1.0& 1.0& 1.0& 0.46& 0.82& 0.98&m4& 0.98& 1.0& 1.0& 0.99& 1.0& 1.0& 0.84& 0.99& 1.0\tabularnewline
\bottomrule
\end{tabular}
}
\end{table}

\section{Application: Monotonicity Test for scRNA-seq Trajectory Inference}\label{sec:app}

Single-cell transcriptome sequencing (scRNA-seq) is a powerful technique that allows researchers to profile transcript abundance at the resolution of individual cells. Trajectory inference aims first to allocate cells to lineages and then order them based on pseudotimes within these lineages. Based on trajectory inference, researchers can discover differentially expressed genes within lineages, such as \textcite{vandenbergeTrajectorybasedDifferentialExpression2020}'s \texttt{tradeSeq}, \textcite{songPseudotimeDEInferenceDifferential2021}'s \texttt{PseudotimeDE}, and \textcite{houStatisticalFrameworkDifferential2023}'s \texttt{Lamian}. 
These methods mostly focus on the differential genes by checking whether the trajectory is constant along the pseudotime. Once a gene is identified as differentially expressed, researchers may further check whether its expression exhibits a monotone pattern. A non-decreasing expression pattern indicates that the corresponding gene is turning on and needed thereafter along the cell lineage. A decreasing expression pattern indicates that the corresponding gene is needed less and less along the pseudotime. On the other hand, a non-monotone expression pattern indicates that the corresponding gene is part of a more complex dynamics.
Such detailed dynamics may illuminate the critical regulatory mechanism of cell differentiation along the corresponding lineage.   

As an analogy to the term \emph{differentially expressed (DE)} gene when the null hypothesis that the expression of the gene along the trajectory is constant is rejected, we call a \emph{non-monotonically expressed (nME)} gene when the null hypothesis that the expression is monotonic is rejected. We adopt \texttt{tradeSeq} to identify DE genes, and the monotonicity test via monotone decomposition with cubic spline (MDCS) to find nME genes. Both DE genes and nME genes are selected using the Benjamini–Hochberg (BH) procedure to control the false discovery rate (FDR) with cutoff $\alpha = 0.05$.

To explore the biological functions of DE genes and nME genes, we examined
the Gene Ontology (GO), which is a relational database of terms (concepts) used to describe gene functions, and  
conducted enrichment analysis \parencite{boyleGOTermFinderOpen2004}.

Suppose there are $N$  genes in the reference gene list, among which $n$ genes are in our analyzed gene set. For a GO term of interest, suppose there are $M$ and $k$ genes within the reference gene list and our analyzed gene set, respectively, that are annotated to have the GO term.
The $p$-value for the one-sided Fisher's exact test of the null hypothesis that the GO term is not enriched in the analyzed gene set can be calculated based on the hypergeometric distribution:
$$
p = 1 - \sum_{i=0}^{k-1}\frac{\binom{M}{i}\binom{N-M}{k-i}}{\binom{N}{n}}\,.
$$
We repeat this test for multiple GO terms of interest and correct for multiple comparisons via
the BH procedure to control the FDR at the cutoff $\alpha = 0.05$.

\subsection{nME genes can identify significant GO terms when DE genes fail}

\begin{figure}[h]
    \centering
    \begin{subfigure}{0.25\textwidth}
        \centering
        \includegraphics[width=\textwidth]{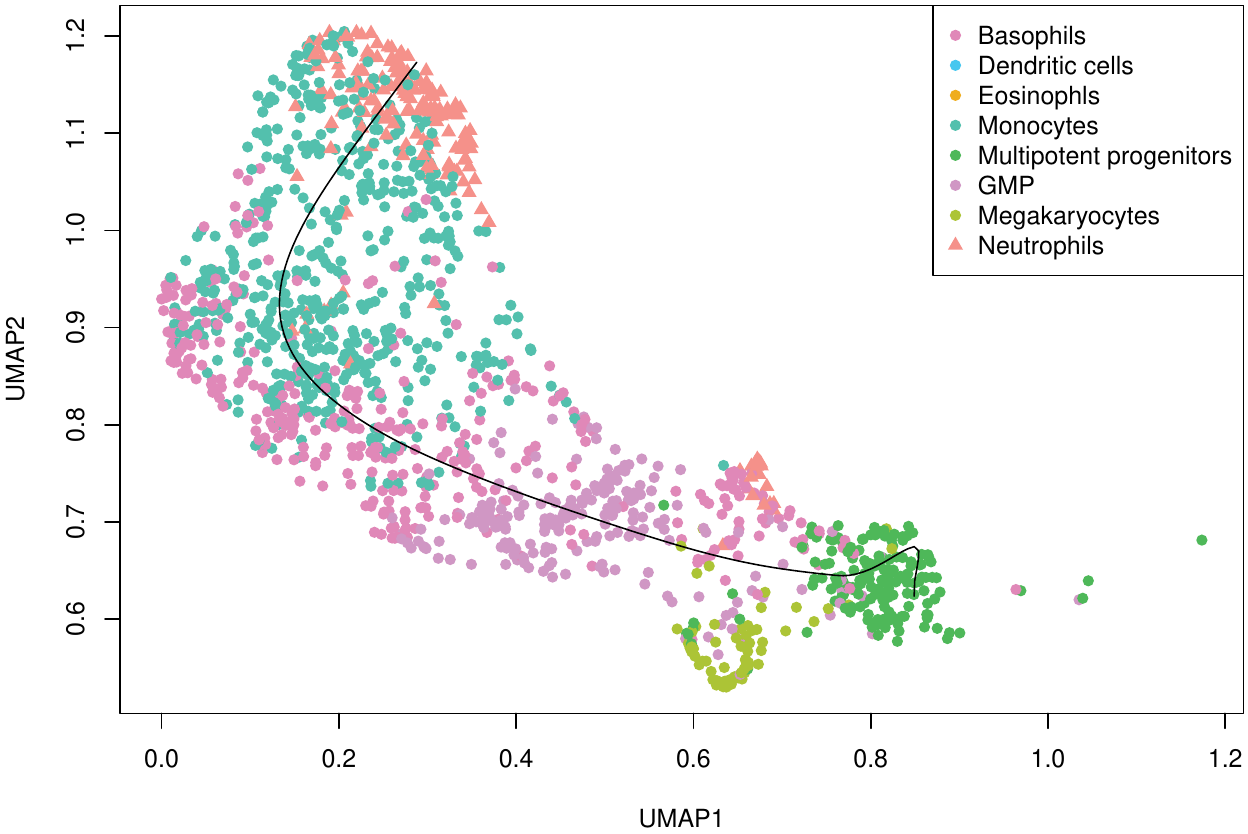}
        \caption{}
        \label{fig:paul1_umap}
    \end{subfigure}%
    \begin{subfigure}{0.25\textwidth}
    \centering
        \includegraphics[width=\textwidth]{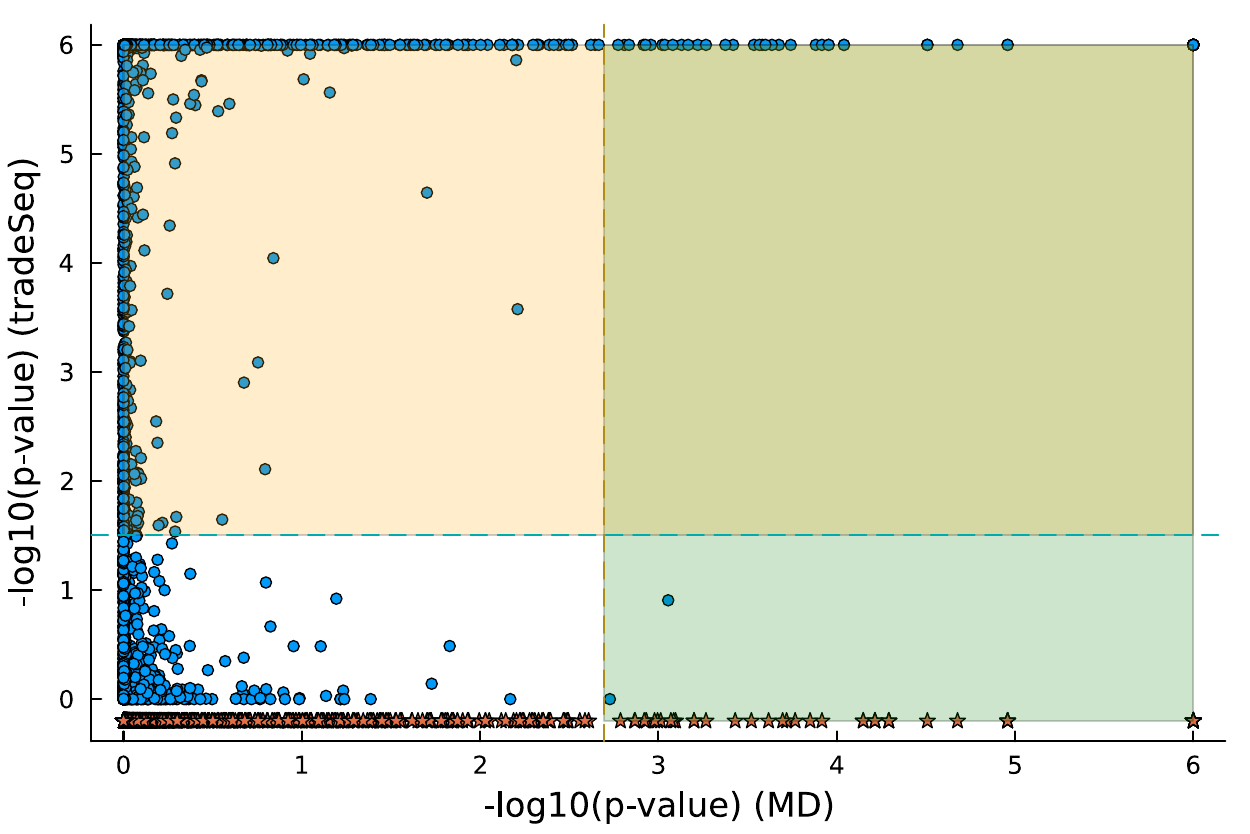}
        \caption{}
        \label{fig:paul1_pvals}
    \end{subfigure}%
    \begin{subfigure}{0.25\textwidth}
        \includegraphics[width=\textwidth]{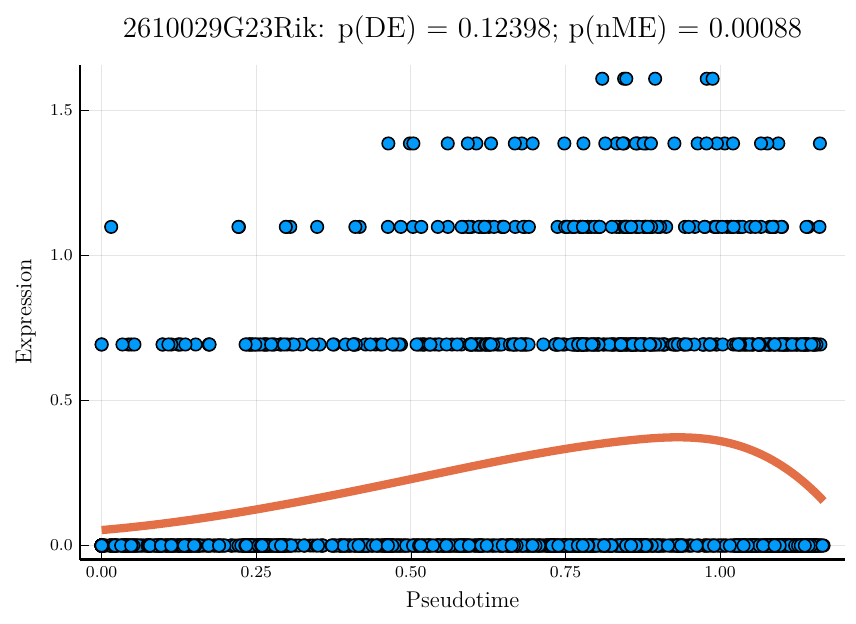}
        \caption{}
        \label{fig:paul1_45}
    \end{subfigure}%
    \begin{subfigure}{0.25\textwidth}
        \centering
        \includegraphics[width=\textwidth]{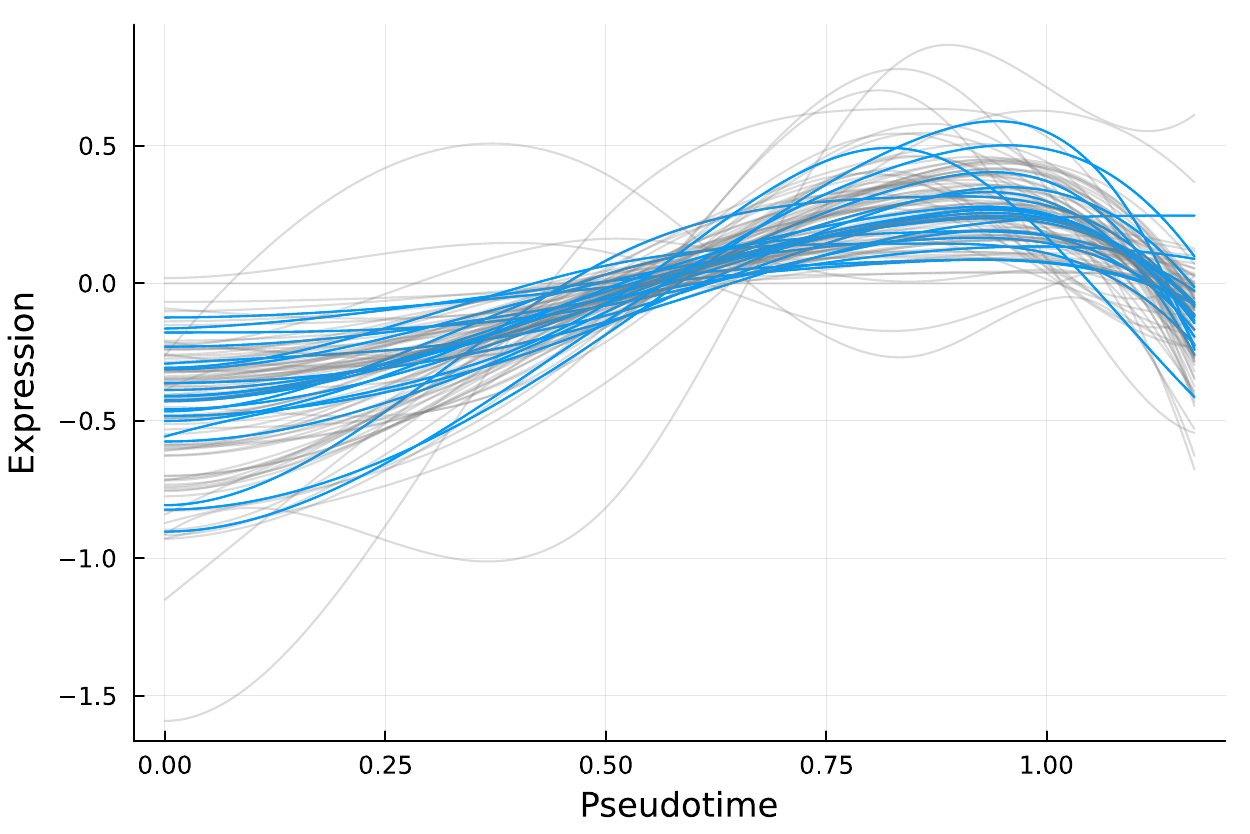}
        \caption{}
        \label{fig:paul1_curve}
    \end{subfigure}

    \caption{\emph{(a):} Two-dimensional representation of the data using UMAP. Different cell types are denoted with different colors (and shapes). The pseudotime axis is displayed, which starts from the \emph{Multipotent progenitors} cell type at the bottom-right to the \emph{Neutrophils} cell type at the top-left. \emph{(b)}: the scatter plot of the paired $p$-values $(p_{\text{nME}}, p_{\text{DE}})$, where the star symbols denote \texttt{NA} values when \texttt{tradeSeq} failed. \emph{(c)}: one example gene located in the bottom right green block of \emph{(b)}, which is identified as an nME gene but not a DE gene. \emph{(d)}: Trajectory curves of 109 nME genes. The curves of 22 genes annotated in the GO term ``translation'' are highlighted.}
    \label{fig:paul1_test}
\end{figure}

\begin{figure}[h]
    \centering
    \begin{subfigure}{0.25\textwidth}
    \centering
        \includegraphics[width=\textwidth]{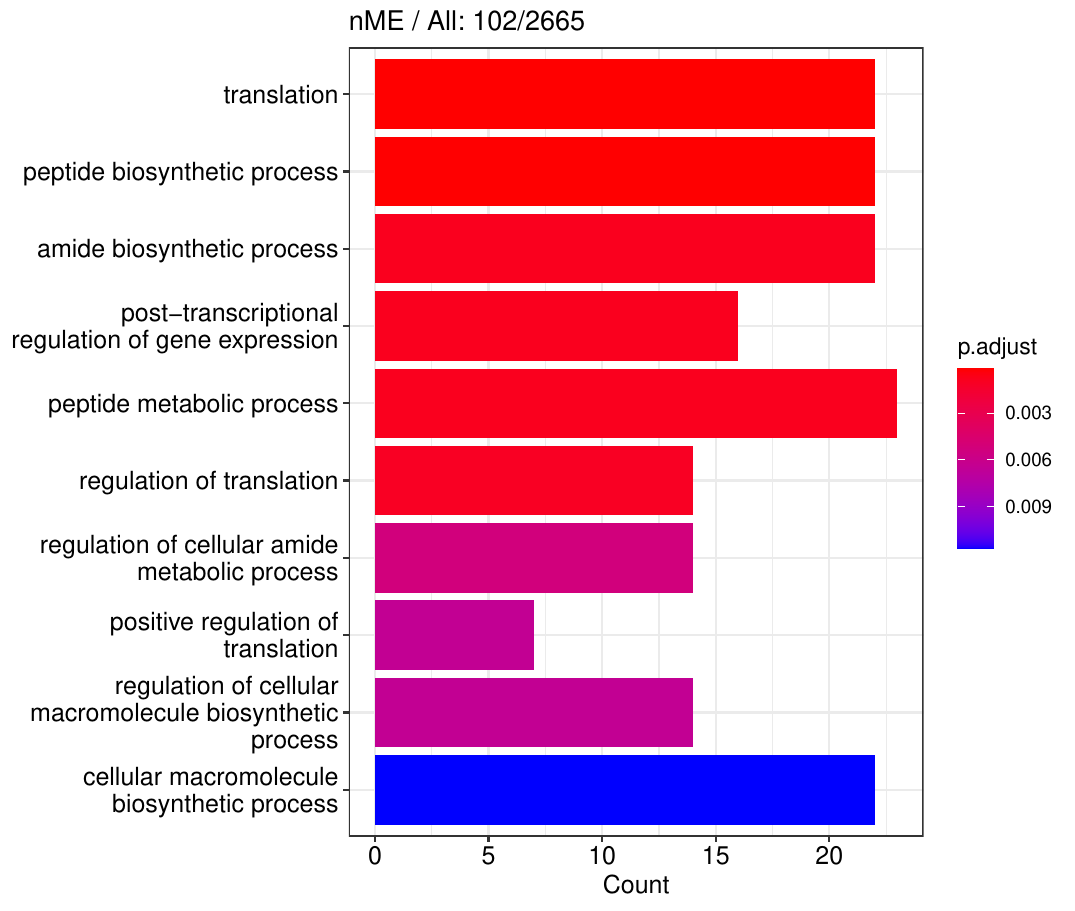}
        \caption{}
        \label{fig:paul_1}
    \end{subfigure}%
    \begin{subfigure}{0.25\textwidth}
    \centering
        \includegraphics[width=\textwidth]{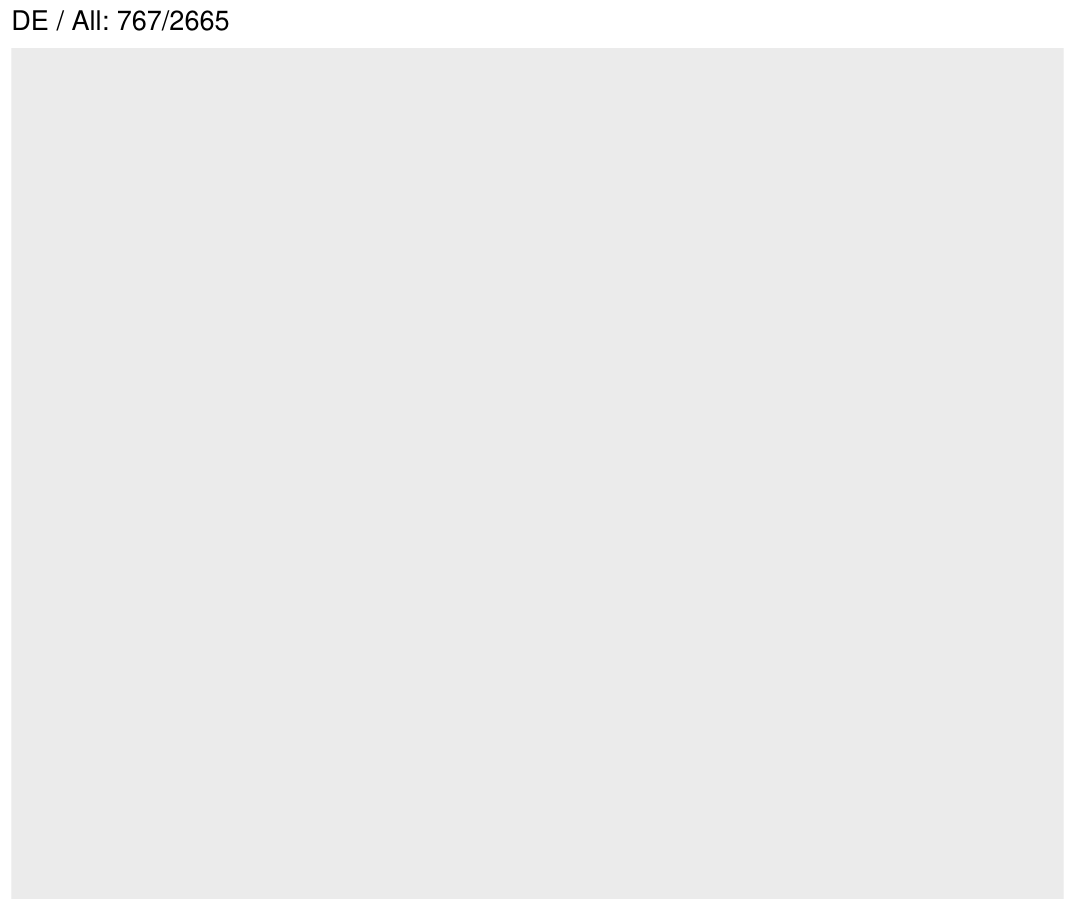}
        \caption{}
        \label{fig:paul_2}
    \end{subfigure}%
    \begin{subfigure}{0.25\textwidth}
    \centering
        \includegraphics[width=\textwidth]{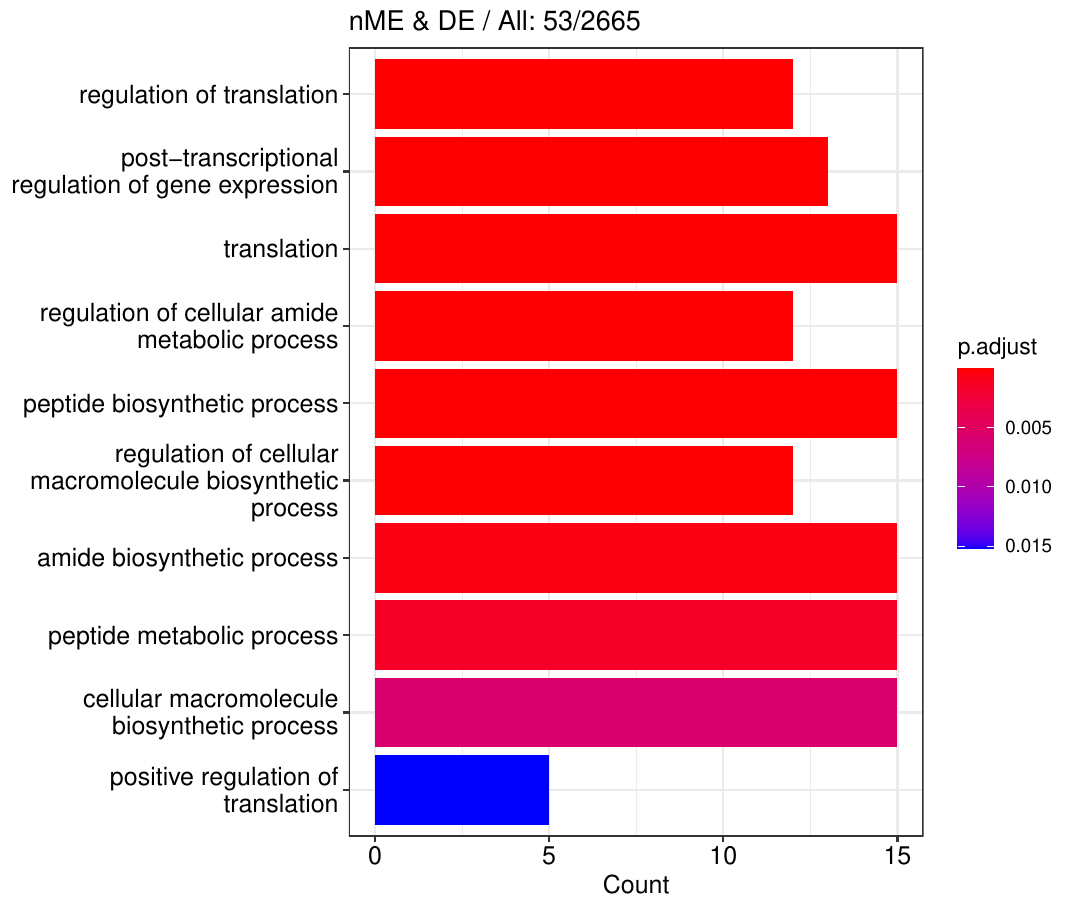}
        \caption{}
        \label{fig:paul_9}
    \end{subfigure}%
    \begin{subfigure}{0.25\textwidth}
    \centering
        \includegraphics[width=\textwidth]{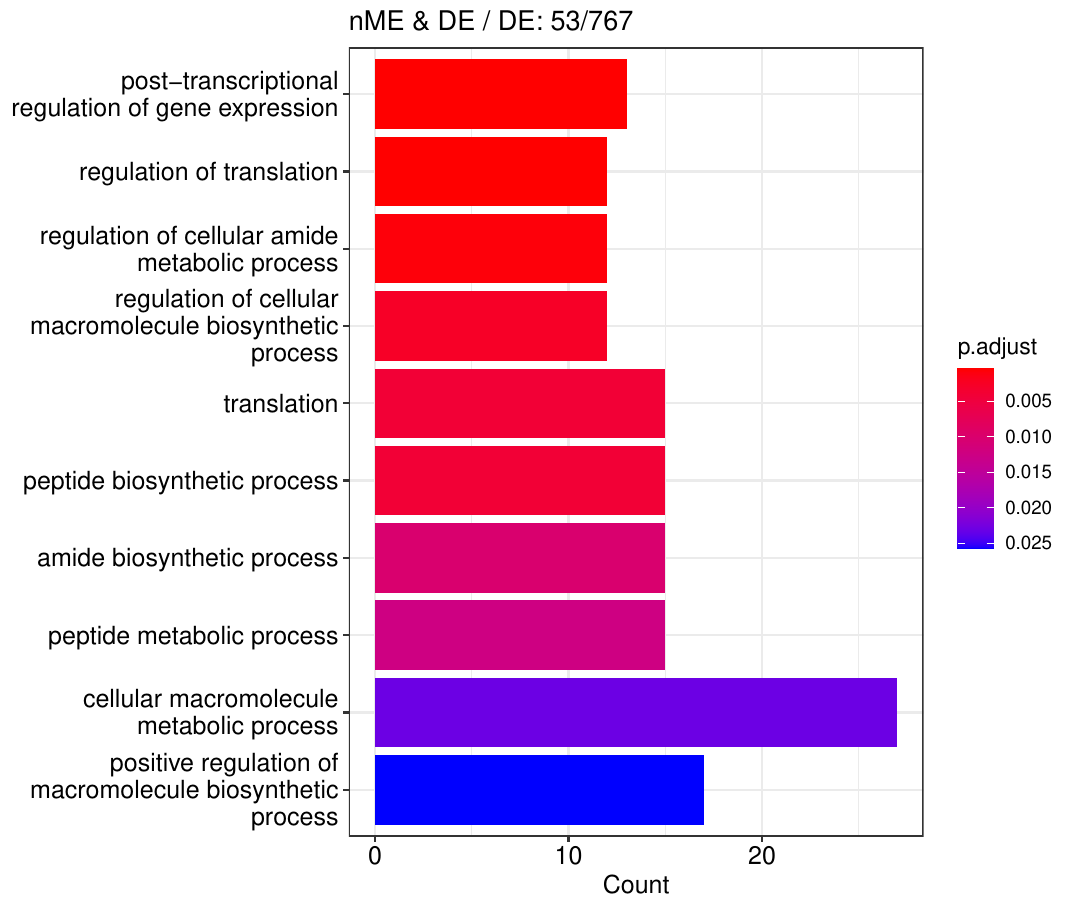}
        \caption{}
        \label{fig:paul_12}
    \end{subfigure}

    \caption{GO enrichment analysis of four gene sets (\emph{(a)}: nME gene set; \emph{(b)}: DE gene set; \emph{(c)}: the intersection of nME and DE genes; \emph{(d)}: the same gene set as \emph{(c)} but the reference list is the DE genes) with BH FDR cutoff $\alpha = 0.05$. The numbers of genes are noted in the title of each subfigure. The enriched GO terms are sorted by the adjusted $p$-values. \emph{(b)} is left empty deliberately since no significant GO terms are selected.}
    \label{fig:paul}
\end{figure}

We studied the leukocyte lineage of the mouse bone marrow data set \parencite{paulTranscriptionalHeterogeneityLineage2015}, which consists of the expression measurements of 3004 genes at 1474 pseudotime points. Figure~\ref{fig:paul1_umap} shows the reduced two-dimensional representation of the data using uniform manifold approximation and projection (UMAP) \parencite{mcinnesUMAPUniformManifold2018}. Eight cell types are denoted with different colors and shapes. The solid curve is the pseudotime axis, which starts from the cell type \emph{Multipotent progenitors} at the bottomright and ends at the cell type \emph{Neutrophils} at the topleft. Note that although a monotone pattern is a special DE pattern, we do \emph{not} perform the monotonicity test in a two-step manner, i.e., firstly find DE genes and then perform the monotonicity test among those found DE genes. Instead, for each gene, we test whether it is a DE gene or an nME gene independently.
Figure~\ref{fig:paul1_pvals} displays the paired $p$-values $(p_{\text{nME}}, p_{\text{DE}})$ in the logarithmic scale, where the dash lines denote the cutoff determined by the BH procedure. As a result, we identified 109 nME genes (it is 102 after GO analysis since 7 genes are not mapped in the GO database) and 767 DE genes, of which 53 genes are in common. These numbers are also noted in the titles of GO bar plots in Figure~\ref{fig:paul}. 
Figure~\ref{fig:paul1_45} illustrates the fitted trajectory for gene 2610029G23Rik, which is identified as an nME gene but not a DE gene, i.e., it lies in the bottom right green block of Figure~\ref{fig:paul1_pvals}.

Figure~\ref{fig:paul} displays GO enrichment analysis on the DE genes and nME genes by R package \texttt{clusterProfiler} \parencite{yuClusterProfilerUniversalEnrichment2023}. Figures~\ref{fig:paul_1}, \ref{fig:paul_2} and \ref{fig:paul_9} take the whole 3004 genes as the reference gene list, but note that, because some genes are not mapped in the GO database, there are only 2665 genes after filtering. We cannot find significant GO terms for the DE gene list, as shown in Figure~\ref{fig:paul_2}, which is left blank deliberately due to no significant results. In contrast, we can identify several significant GO terms for the nME gene list. Figures~\ref{fig:paul_9} and \ref{fig:paul_12} display the intersection of the DE gene list and the nME gene list with respect to the whole gene list or the DE gene list, respectively. Both can identify significant GO terms. 
One possible reason for DE genes failing to identify significant GO terms is that the range of DE genes might be too broad,
thus different sub-categories of DE genes (the non-monotonic pattern is a special sub-category) might contribute to different GO terms, but the increased sample size due to incorporating unrelated genes might reduce the significance for determining the significant GO terms. Another potential reason is that the \texttt{tradeSeq} test is not robust enough. As shown by the star symbol in Figure~\ref{fig:paul1_pvals}, there are many \texttt{NA} values returned by \texttt{tradeSeq}, which is a known issue discussed in their GitHub repository\footnote{\url{https://github.com/statOmics/tradeSeq/issues/209}}. 

We can further check whether the pattern of trajectory curves in the enriched GO terms agrees with the biological mechanism. Take the first GO term ``translation'' as an example. Figure~\ref{fig:paul1_curve} displays the trajectory curves of 109 nME genes along the pseudotime. Among these 109 nME genes, the curves of 22 genes annotated in the GO term ``translation'' are highlighted. These genes exhibit a coherent pattern, characterized by an initial increase in expression followed by a subsequent decrease. If we cast the pseudotime axis in Figure~\ref{fig:paul1_curve} back to Figure~\ref{fig:paul1_umap}, the curve pattern implies that the gene expression increases when the cell develops from \emph{Multipotent progenitors} to \emph{Monocytes}, and roughly after the gene expression reaching the peak, the cell evolves to \emph{Neutrophils}. This behavior is consistent with the biological fact that specialized cell types (here \emph{Neutrophils}) might reduce rates of translation (and hence protein synthesis), since their structure and function are relatively stable.



\subsection{nME genes can fine-locate GO terms when DE genes identify too many terms}\label{sec:fine-locate}

In some scenarios, although GO enrichment analysis can identify significant GO terms given the DE genes, nME genes can further fine-locate GO terms and focus on a small but significant set of GO terms. We analyzed the cholangiocyte lineage from the mouse liver data studied in \textcite{ghazanfarInvestigatingHigherorderInteractions2020} to demonstrate such an advantage of nME genes. In the dataset, there are 6038 genes and 308 pseudotime points.

\begin{figure}[h]
    \centering

    \begin{subfigure}{0.5\textwidth}
        \includegraphics[width=\textwidth]{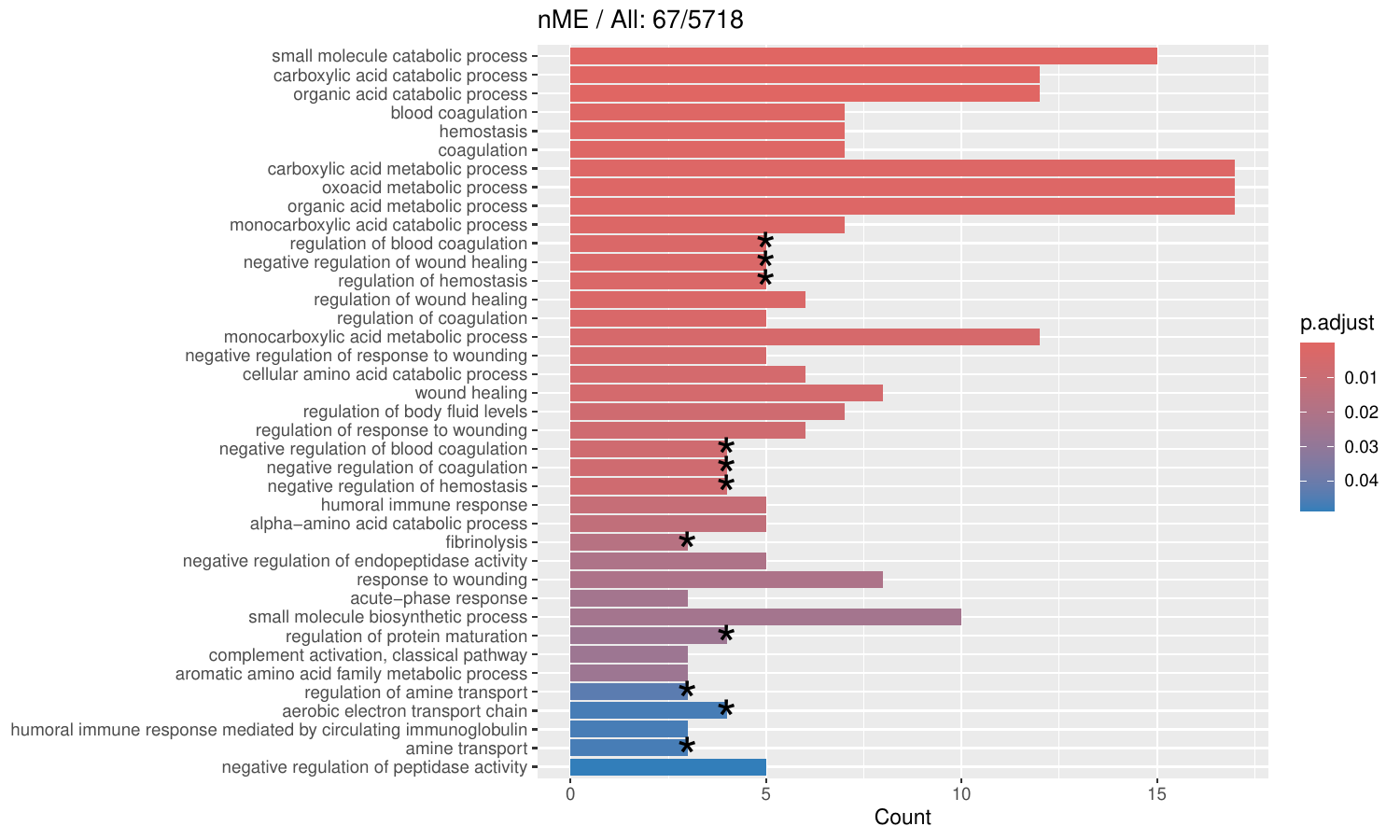}
        \caption{}
        \label{fig:liver_go}
    \end{subfigure}%
    \begin{subfigure}{0.5\textwidth}
        \includegraphics[width=\textwidth]{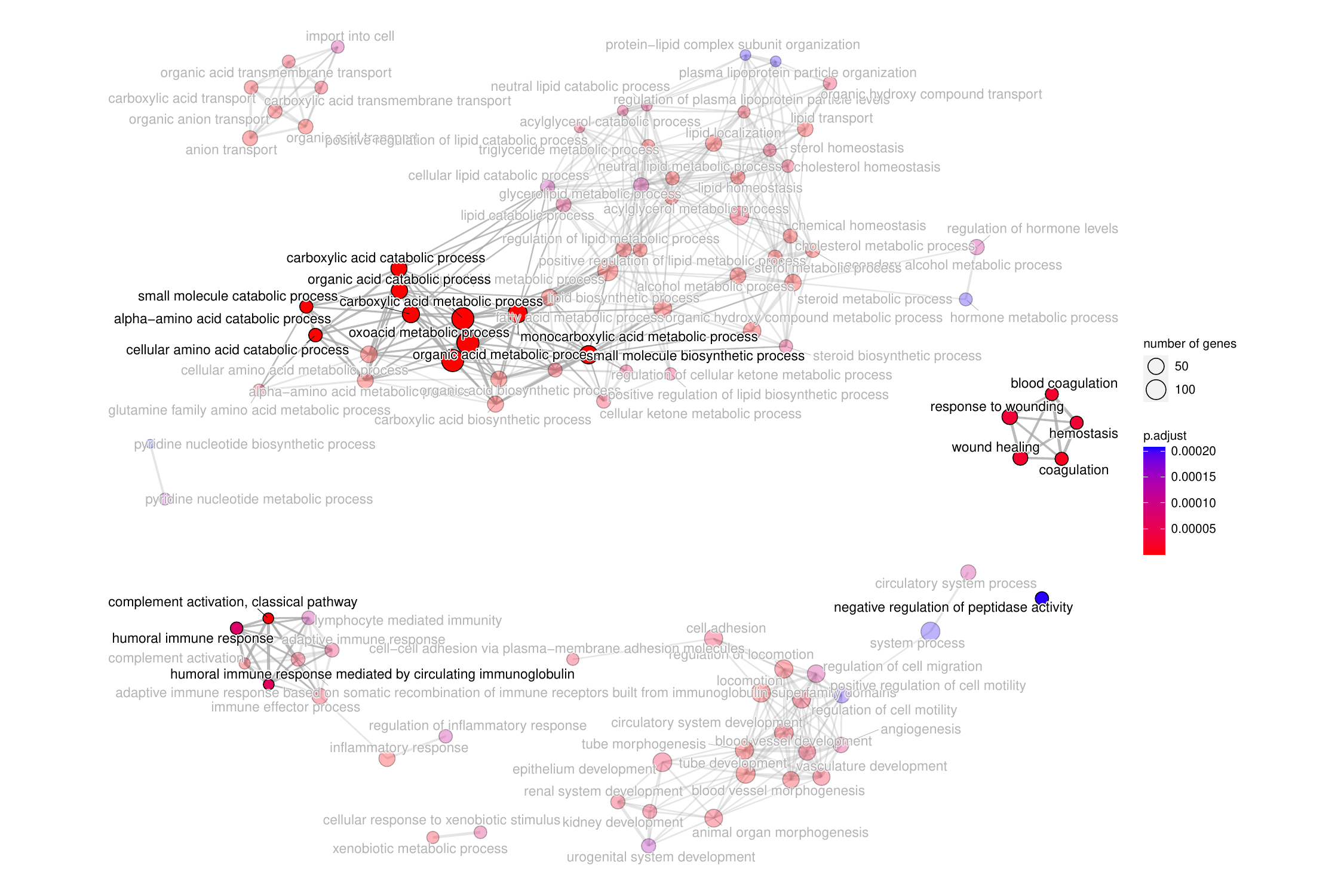}
        \caption{}
        \label{fig:liver_graph}
    \end{subfigure}
    
    \caption{GO enrichment analysis for the DE genes and nME genes on the cholangiocyte lineage of the liver data. 
    \emph{(a)} GO terms sorted by adjusted $p$-values enriched by the nME genes. \emph{(b)} Enrichment map of the GO terms by the DE genes, where the common GO terms shared by the nME genes are highlighted.}
    \label{fig:liver_go_all}
\end{figure}

\texttt{tradeSeq} identifies 767 DE genes (it is 801 before filtering due to unmapping in GO), and the monotonicity test identifies 67 nME genes (it is 69 before filtering due to unmapping in GO), of which 45 genes are in common. For the 767 DE genes, we identify 439 significant GO terms, whereas for the 67 nME genes, we find 39 significant GO terms. Between the two sets, 28 GO terms are in common. 
Figure~\ref{fig:liver_go} displays all GO terms returned by the 39 nME genes, where the star symbol indicates GO terms not shared by the DE genes. Figure~\ref{fig:liver_graph} shows the enrichment map constructed by the GO terms from DE genes, in which the common GO terms shared by the nME genes are highlighted. In the enrichment map, an edge connects two GO terms if there are overlapped genes, and hence, mutually overlapping GO terms tend to cluster together, making it easy to identify functional modules. It is clear that the shared GO terms mainly focus on two clusters: one is isolated from others on the right side and forms a pentagon shape with the keyword ``coagulation''; another cluster is located at the left corner of the biggest cluster, related to ``catabolic process''. In other words, nME genes can help fine-locate GO terms, which might help save time for researchers without checking too many GO terms from DE genes.

\begin{figure}[h]
    \centering
    \begin{subfigure}{0.25\textwidth}
        \includegraphics[width=\textwidth]{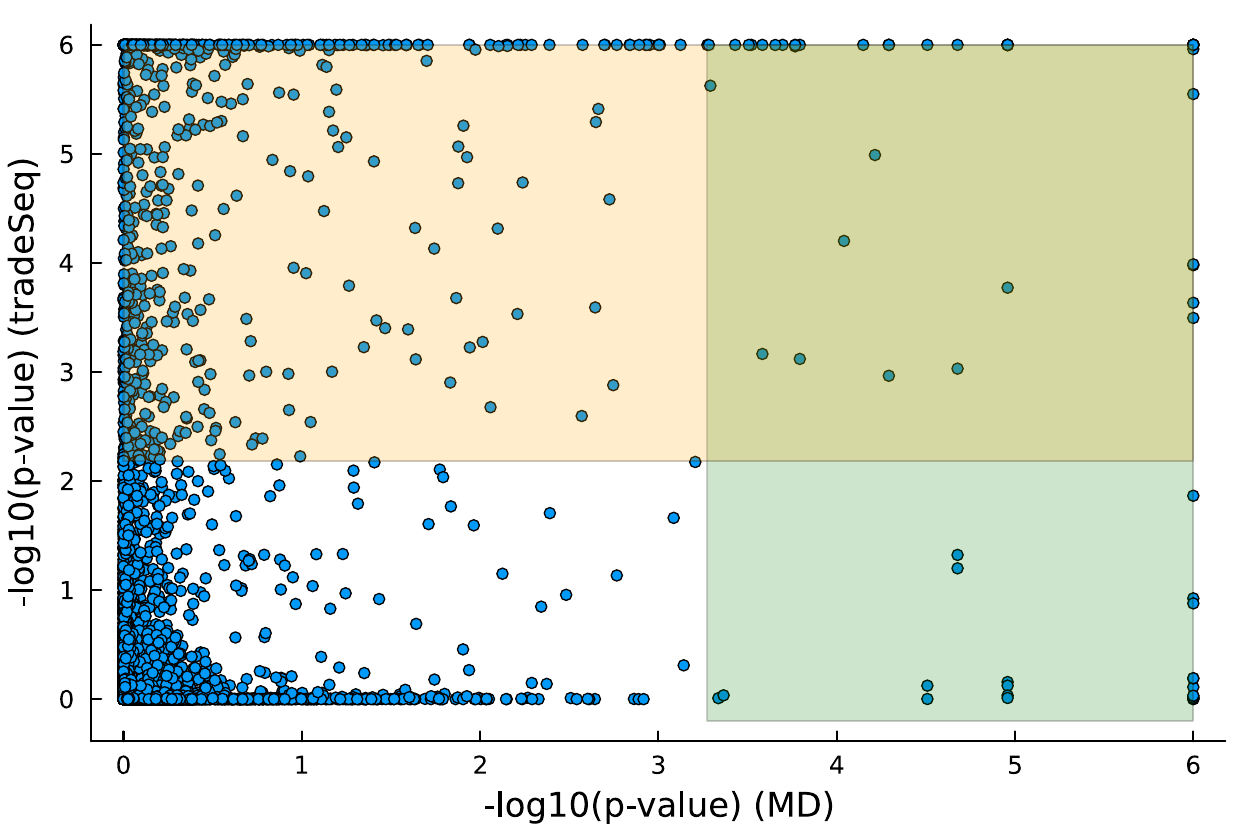}
        \caption{}
        \label{fig:liver2_pval_grid}
    \end{subfigure}%
    \begin{subfigure}{0.25\textwidth}
        \includegraphics[width=\textwidth]{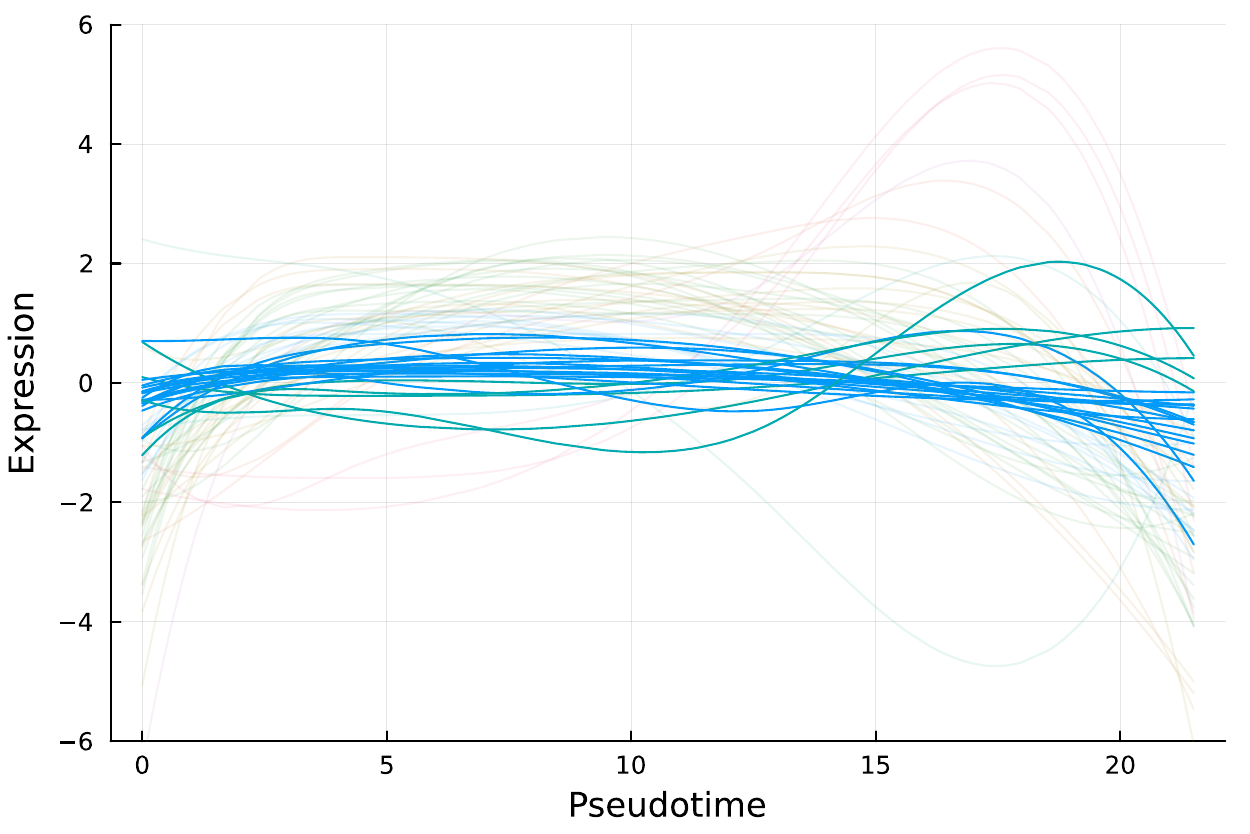}
        \caption{}
        \label{fig:liver2_curves}
    \end{subfigure}%
    \begin{subfigure}{0.25\textwidth}
        \includegraphics[width=\textwidth]{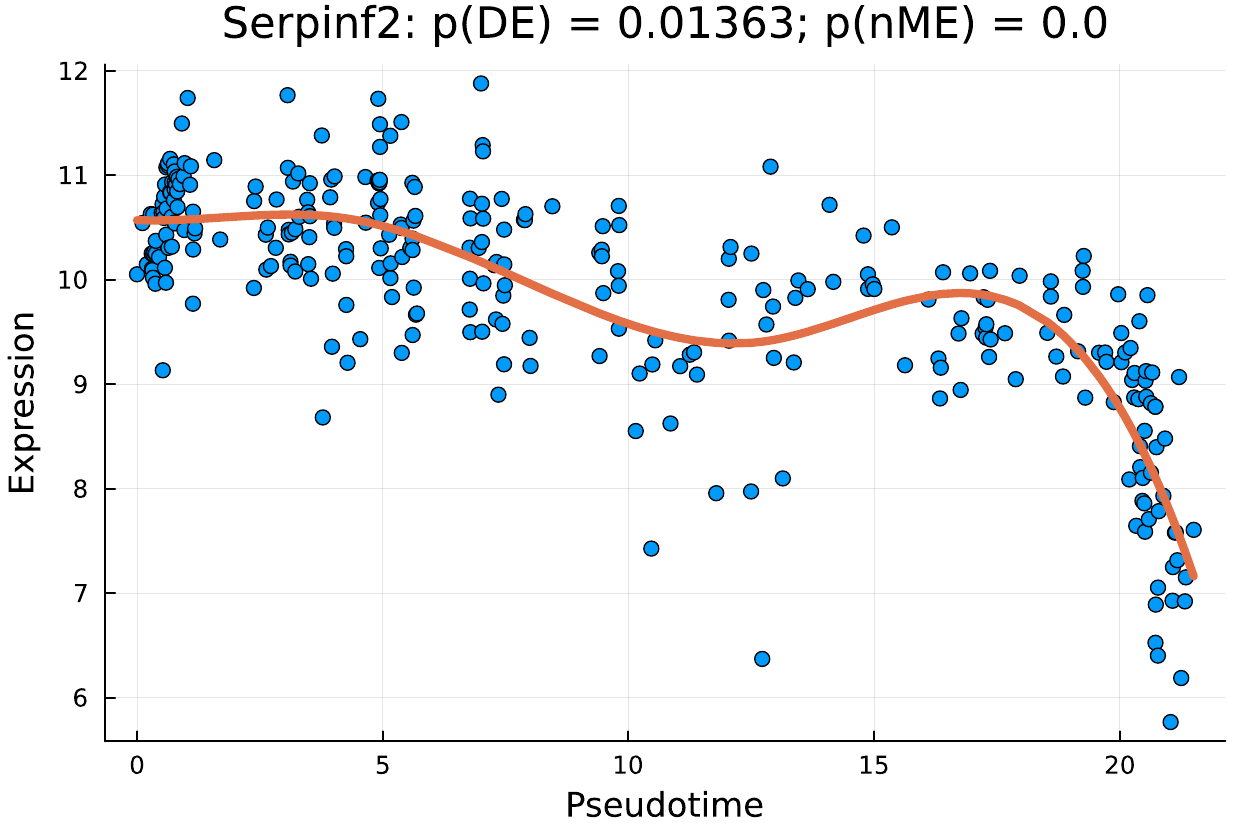}
        \caption{}
        \label{fig:liver2_4661}
    \end{subfigure}%
    \begin{subfigure}{0.25\textwidth}
        \includegraphics[width=\textwidth]{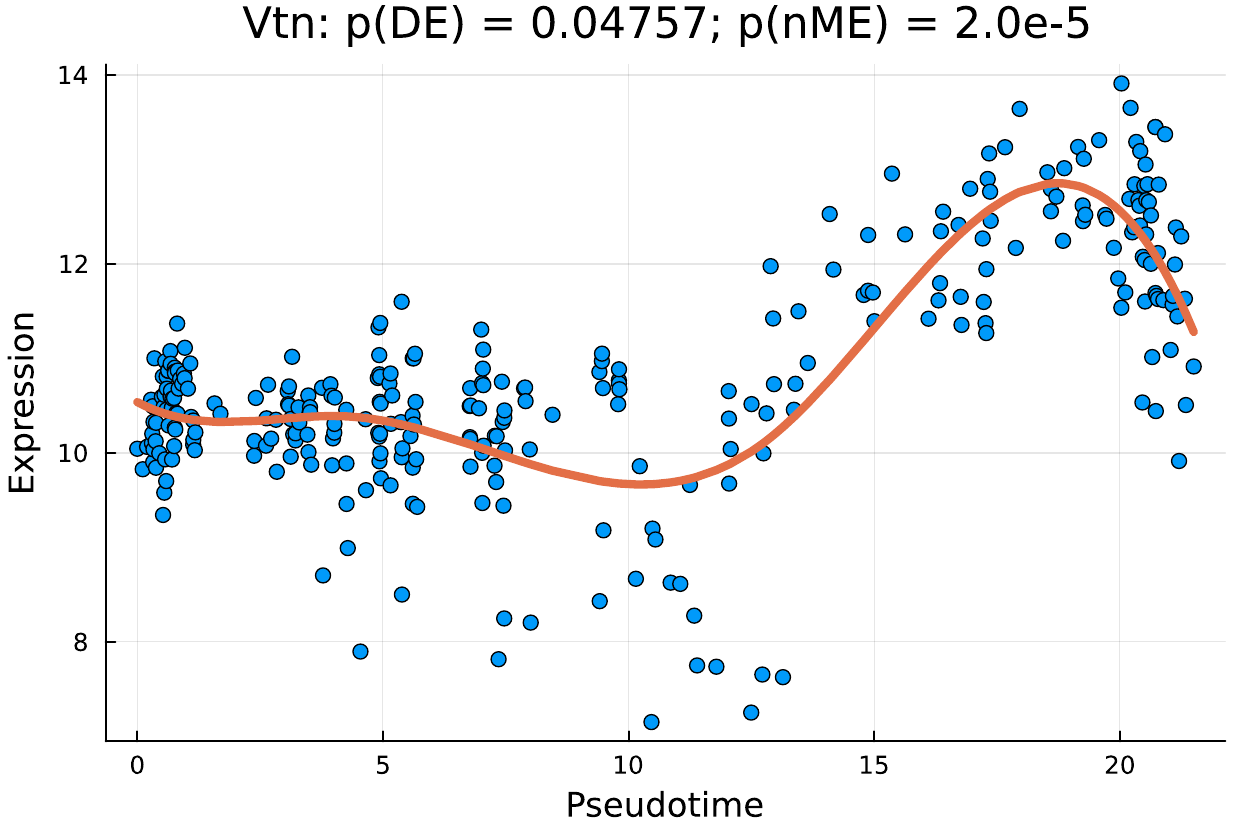}
        \caption{}
        \label{fig:liver2_5792}
    \end{subfigure}
    \caption{Hypothesis testing on the cholangiocyte lineage of the mouse liver data. \emph{(a)} The scatter plot of paired $p$-values $(p_{\text{nME}}, p_{\text{DE}})$, where the cutoffs selected by the BH procedure are 0.00053 and 0.00660, respectively. 
    \emph{(b)} Trajectory curves of all nME genes.
    \emph{(c)} and \emph{(d)} are two genes in the GO term ``regulation of blood coagulation'', which are enriched by the nME genes but not DE genes.}
    \label{fig:enter-label}
\end{figure}

We further check the new GO terms enriched by the nME genes, which are annotated with the star symbol in Figure~\ref{fig:liver_go}. These new GO terms might be contributed by new genes that are not identified as DE genes. For example, we take the GO term ``regulation of blood coagulation'' as an example, which contains 5 genes \{F11, Kng2, Serpinc1, Serpinf2, Vtn\} in the nME gene set, where the first three are also in the DE gene set, but the last two are only in the nME gene set. Figures~\ref{fig:liver2_4661} and \ref{fig:liver2_5792} display the fitted trajectory of the expression along pseudotime by our monotone decomposition fitting technique, and the $p$-values are also noted in the title of each figure. We observe that the $p$-value for the DE gene is not quite as significant as the one for the nME gene. In other words, these two $p$-values lie in the bottom right green block of Figure~\ref{fig:liver2_pval_grid}. Using the same data at the hepatoblast stage (an earlier stage than the cholangiocyte stage we considered), \textcite{ghazanfarInvestigatingHigherorderInteractions2020} identified 68 \emph{differentially variable} (DV) genes, which indicates that the \emph{variances} (instead of the \emph{mean} expression considered in DE genes and nME genes) of gene expressions change along the pseudotime. Accidentally, Serpinf2 and Vtn are the two and the only two which are both in the 68 DV gene set and the 69 nME gene set. The coexistence of DV and nME characteristics in these genes suggests intricate and dynamic expression patterns, which might indicate significant biological interest. The dual nature of being both DV and nME genes underscores the complexity of the regulatory mechanisms governing these specific genetic expressions.

Furthermore, we check all genes that are identified as nME but not DE. Figure~\ref{fig:liver2_curves} displays the trajectory curves of all nME genes, and the curves of nME but not DE genes are highlighted, while others are transparent. To facilitate comparative analysis, all curves are centered and different colors denote different clusters (see Section~\ref{sec:cluster}). Note that the variations of highlighted curves are relatively small compared to curves in transparent, so different methods might make different conclusions. As a result, some non-DE genes are treated as nME genes despite the initial expectation that nME genes should naturally encompass a subset of DE genes.

\subsection{nME genes can highlight non-monotonic patterns while DE genes blur them}\label{sec:cluster}

The clustering of genes based on their fitted expression patterns can reveal intriguing insights for biologists. However, a potential limitation of clustering based on DE genes is the tendency of clustering methods to amalgamate pure monotonic patterns with somewhat intricate non-monotonic patterns (e.g., Figure 5 of \textcite{vandenbergeTrajectorybasedDifferentialExpression2020}). To mitigate the amalgamation of non-monotonic patterns and maintain their clarity, one possible direction is to tailor clustering approaches, such as constructing more suitable similarity measurements. Another direct approach is to focus only on non-monotonic patterns from nME genes.

\begin{figure}[h]
    \centering
    \begin{subfigure}{0.25\textwidth}
        \includegraphics[width=\textwidth]{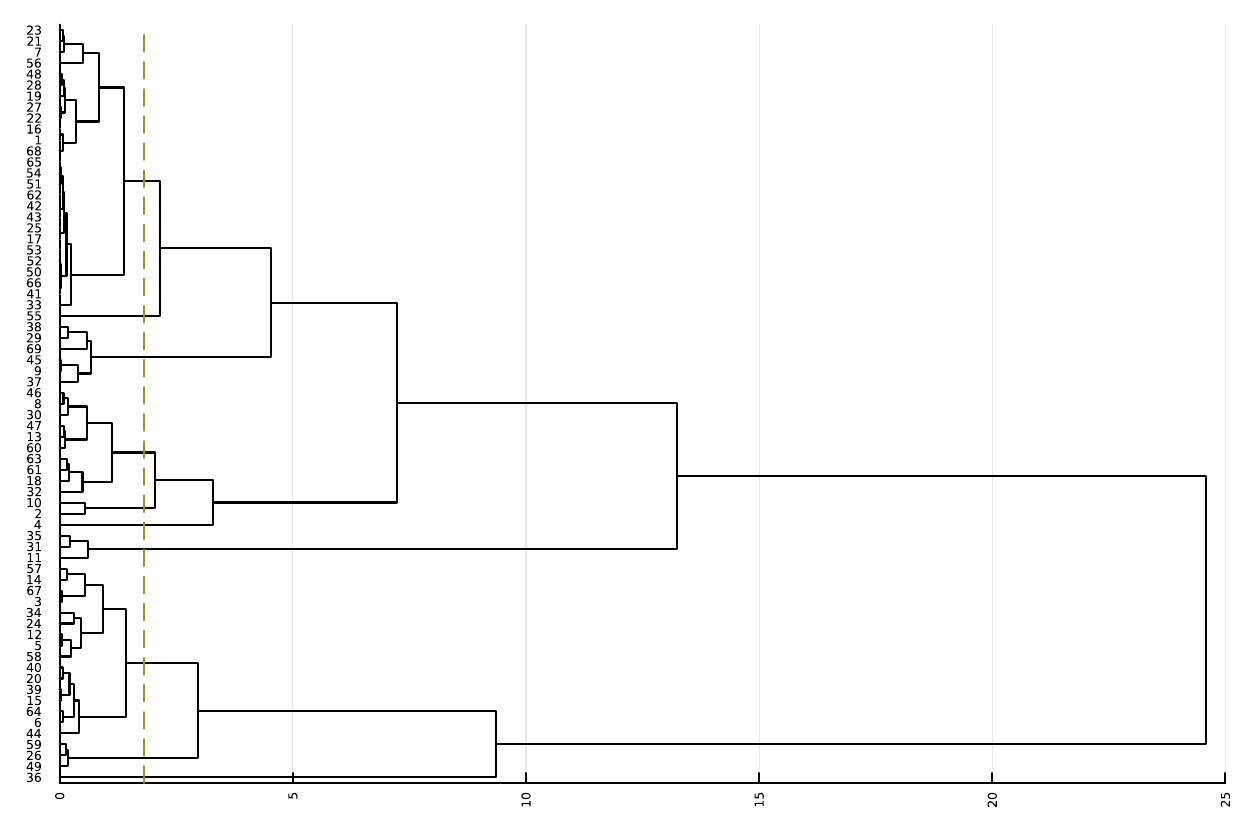}
        \caption{}
        \label{fig:nme-hc}
    \end{subfigure}%
    \begin{subfigure}{0.25\textwidth}
        \includegraphics[width=\textwidth, page=11]{res2/no1se-hclust-complete-h2.pdf}
        \caption{}
        \label{fig:cluster-all}
    \end{subfigure}%
    \begin{subfigure}{0.25\textwidth}
        \includegraphics[width=\textwidth, page=7]{res2/no1se-hclust-complete-h2.pdf}
        \caption{C9, Hp, Kng2}
        \label{fig:cluster7}
    \end{subfigure}%
    \begin{subfigure}{0.25\textwidth}
        \includegraphics[width=\textwidth, page=8]{res2/no1se-hclust-complete-h2.pdf}        
        \caption{Hgd, Proz, Sord}
        \label{fig:cluster8}
    \end{subfigure}
    \begin{subfigure}{0.25\textwidth}
        \includegraphics[width=\textwidth]{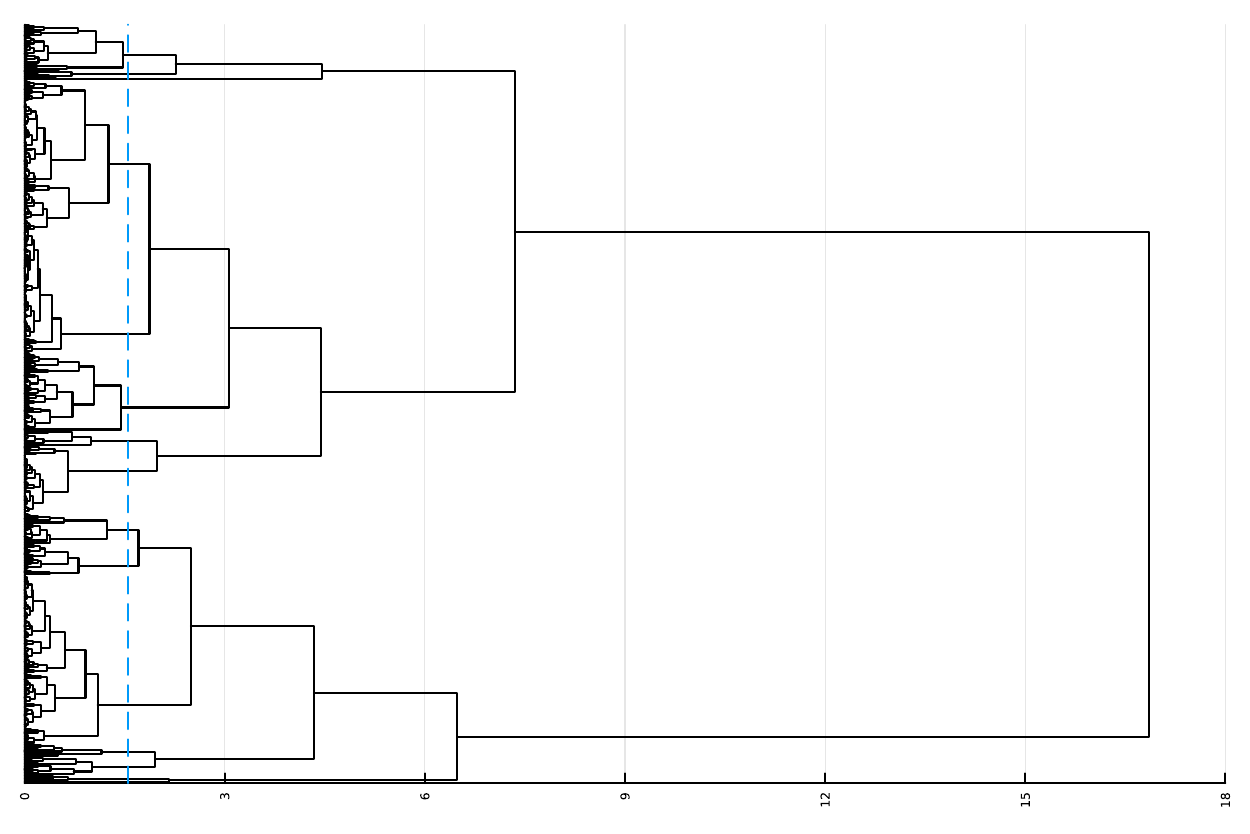}
        \caption{}
        \label{fig:de-hc}
    \end{subfigure}%
        \begin{subfigure}{0.25\textwidth}
        \includegraphics[width=\textwidth, page=13]{res2/de-hclust-complete-h1.5.pdf}        
        \caption{}
        \label{fig:cluster-de-all}
    \end{subfigure}%
    \begin{subfigure}{0.25\textwidth}
        \includegraphics[width=\textwidth, page=3]{res2/de-hclust-complete-h1.5.pdf}
        \caption{}
        \label{fig:cluster-de-3}
    \end{subfigure}%
    \begin{subfigure}{0.25\textwidth}
        \includegraphics[width=\textwidth, page=10]{res2/de-hclust-complete-h1.5.pdf}        
        \caption{}
        \label{fig:cluster-de-10}
    \end{subfigure}
    \caption{Clustering of trajectory curves of nME genes (first row) and DE genes (second row). \emph{(a):} Hierarchical dendrogram of 69 nME genes. The cutoff (2.0) at the dashed line results in 10 clusters. \emph{(b):} Trajectory curves of 69 nME genes. \emph{(c)-(d)} display the curves of 2 out of 10 clusters from \emph{(b)}, respectively. The gene symbols are noted for clusters with few genes. \emph{(e):} Hierarchical dendrogram of 801 DE genes. The cutoff (1.5) at the dashed line results in 12 clusters. \emph{(f):} Trajectory curves of 801 DE genes. \emph{(g)} and \emph{(h)} are two distinct clusters from 12 clusters in \emph{(f)}. Different colors denote different clusters. The same color in \emph{(b)} to \emph{(d)} denotes the same cluster, and the same color in \emph{(g)} to \emph{(h)} denotes the same cluster, but there is no direct correspondence between the colors in \emph{(b)} and \emph{(f)}.}
    \label{fig:cluster}
\end{figure}

Here, we consider the liver dataset in Section~\ref{sec:fine-locate}. Figure~\ref{fig:nme-hc} shows the dendrogram from hierarchical clustering with \emph{complete} linkage on 69 nME genes. 
We take the cutoff 2.0 to obtain 10 clusters with clear patterns. Figure~\ref{fig:cluster-all} presents the trajectory curves of those 69 nME genes, and different colors denote different clusters. Notably, we have highlighted three representative clusters, each depicted in its respective figure, ranging from Figure~\ref{fig:cluster7} to Figure~\ref{fig:cluster8}. Figure~\ref{fig:cluster7} displays a peak on the right side, while Figure~\ref{fig:cluster8} showcases a peak on the left side. 

On the other hand, we perform hierarchical clustering with a similar cutoff 1.5 and identify 12 clusters, as shown in Figure~\ref{fig:de-hc}. The trajectory curves of 801 DE genes with different colors representing different clusters are displayed in Figure~\ref{fig:cluster-de-all}. It is worth noting that the presence of monotonic patterns has somewhat concealed the underlying wiggly patterns, as evidenced in Figure~\ref{fig:cluster-de-3}, which combines non-monotonic patterns similar to those found in Figure~\ref{fig:cluster7} with numerous ascending curves. Similarly, Figure~\ref{fig:cluster-de-10} combines the non-monotonic pattern observed in Figure~\ref{fig:cluster8}, illustrating the challenges in distinguishing these patterns.

Pure non-monotonic patterns hold the potential to identify significant patterns. For example, all three genes in Figure~\ref{fig:cluster7} are significantly annotated to GO terms ``defense response to other organism'', ``response to external biotic
stimulus'', and ``response to other organism'', each accompanied by FDR adjusted $p$-value of 0.00278.

\section{Discussions}\label{sec:md_discuss}

We formulate the monotone decomposition with monotone splines. It can serve as a fitting method when we sum up two monotone components, and it can be used to conduct a test of monotonicity by checking whether one component is constant.

As a fitting method, the experiments have shown that the monotone decomposition with cubic splines improves the performance, especially in high noise cases. We can explain the better performance in monotone functions theoretically.
Similar phenomena have been observed for the monotone decomposition with smoothing splines. However, there are also some limitations:
\begin{itemize}
    \item The cross-validation procedure for simultaneously tuning two parameters is computationally extensive. The generative bootstrap sampler (GBS) proposed by \textcite{shinGenerativeMultipurposeSampler2023} might be an alternative to speed up the cross-validation step.
    \item Currently, the theoretical guarantees are only for monotone functions. It would be great if the theoretical results could be extended to general functions.
\end{itemize}

For the test of monotonicity by monotone decomposition, the proposed statistics based on monotone decomposition show competitive performance and are even much better than the existing approaches. 

We also apply the fitting and testing based on monotone decomposition to investigate the monotonic and non-monotonic trajectory patterns in several scRNA-seq datasets. In parallel with the conventional analysis of differentially expressed (DE) genes, we propose the concept of non-monotonically expressed (nME) genes, which might lead to new biological insights.



\section*{Acknowledgement}
Lijun Wang was supported by the Hong Kong Ph.D. Fellowship Scheme from the University Grant Committee. Hongyu Zhao was supported in part by NIH grant P50 CA196530. JSL was supported in part by the NSF DMS/NIGMS 2 Collaborative Research grant (R01 GM152814).

\section*{Supplementary Material}

The \supp{} contains additional simulation results and technical proofs of propositions.

\printbibliography

\appendix
\newpage

\section{More Simulation Results}\label{app:sim}

\subsection{Candidate Kernels}

Random functions with kernel $K$, including Squared Exponential (SE) kernel $K_{SE}$, Rational Quadratic (RQ) kernel $K_{RQ}$, Mat\'{e}rn (Mat) kernel $K_{Mat}$ and Periodic kernel $K_{Periodic}$\parencite{rasmussenGaussianProcessesMachine2006}. 
    \begin{align*}
        K_{SE}(x, x') &= \exp\left(-\frac{(x-x')^2}{2\ell^2}\right),\; K_{Mat}(x, x') = \frac{2^{1-\nu}}{\Gamma(\nu)}\left(\frac{\sqrt{2\nu}\vert x-x'\vert}{\ell}\right)^\nu S_\nu\left(\frac{\sqrt{2\nu}\vert x-x'\vert}{\ell}\right),\\
        K_{RQ}(x, x') &= \left(1+\frac{(x-x')^2}{2\alpha \ell^2}\right)^{-\alpha},\; K_{Periodic}(x, x') =\exp\left(-\frac{2\sin^2(\vert x-x'\vert/T)}{\ell^2}\right)\,,
    \end{align*}
    where $\ell, \nu, \alpha, T$ are the parameters, and $S_\nu$ is a modified Bessel function. In particular, ``Mat12'' refers to the Mat\'{e}rn kernel with $\nu=1/2$, and similarly, ``Mat32'' and ``Mat52'' indicate $\nu=3/2$ and $5/2$, respectively. Any additional parameters are appended to the kernel name; for example, ``SE-1'' represents the Squared Exponential kernel with $\ell=1$, ``Mat12-1'' denotes the Mat\'{e}rn kernel with $\nu=1/2,\ell=1$, and ``RQ-0.1-0.5'' is the Rational Quadratic kernel with parameter $\ell=0.1, \alpha=0.5$. 

\subsection{Random Function Generation}
A random function $f$ with kernel $K$ is generated as follows,
\begin{enumerate}
    \item Generate $n$ random points, $x_i\sim U[-1,1], i=1,\ldots,n$.
    \item Calculate the covariance matrix $\Sigma$ based on kernel $K$, $\Sigma_{ij} = K(x_i, x_j)$. Practically, add a small constant, say $10^{-7}$, on the diagonal of $\Sigma$ to prevent numerically ill-conditioned matrix \parencite{gramacySurrogatesGaussianProcess2020}. 
    \item Generate a random Gaussian vector with the above covariance matrix, $f\sim N(0, \Sigma)$.
\end{enumerate}

\begin{figure}[H]
    \centering
    \includegraphics[width=\textwidth]{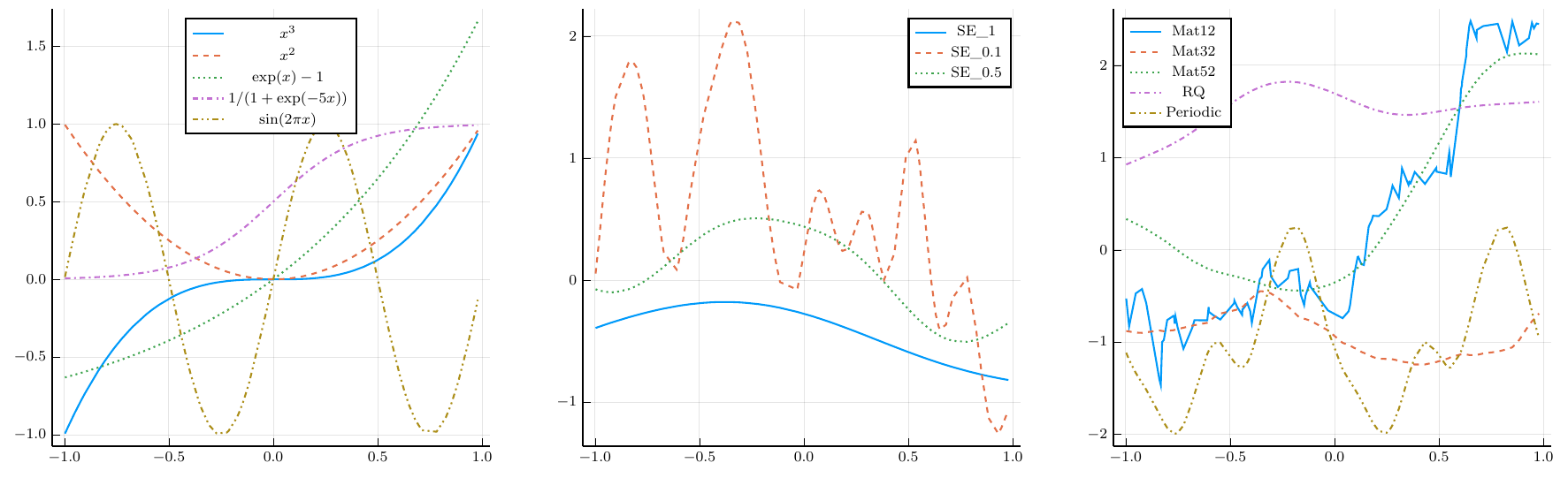}
    \caption[Candidate Functions for Monotone Decomposition.]{Candidate functions. The left panel shows several regular functions, the middle panel displays the random functions drawn from Gaussian processes with the square exponential kernel, and the right panel shows the random functions generated from Gaussian processes with other kernels.}
    \label{fig:functions}
\end{figure}

\subsection{Monotone Decomposition with Cubic Splines}

\begin{table}[H]
    \centering
    \caption[MSE for Monotone Decomposition with Cubic Splines using CV-tuned ($\mu, J$)]{A complete version with finer noise levels fro Table 1. Results for comparing the cubic splines and the fitting by monotone decomposition with CV-tuned $(\mu, J)$. The values are averages over 100 replications, with the standard error of the average in parentheses. The bold values highlight the smaller mean squared prediction error. }
    \label{tab:bspl_vs_mbss_mu_J}
    \resizebox{0.8\textwidth}{!}{%
        \begin{tabular}{ccccccll}
\toprule
\multirow{2}{*}{curve} & \multirow{2}{*}{$\sigma$}&\multicolumn{2}{c}{MSFE}&\multicolumn{2}{c}{MSPE}& \multirow{2}{*}{p-value}& \multirow{2}{*}{prop.}\tabularnewline
\cmidrule(lr){3-4}
\cmidrule(lr){5-6}
&&CubicSpline&MonoDecomp&CubicSpline&MonoDecomp\tabularnewline
\midrule
\multirow{4}{*}{$x^2$}&0.2& 1.95e+00 (1.4e-02)& 1.94e+00 (1.5e-02)& 1.46e+00 (6.1e-02)& \textbf{1.46e+00} (6.4e-02)& 4.44e-01& 0.57\tabularnewline
&0.4& 3.84e+00 (3.4e-02)& 3.83e+00 (3.0e-02)& 2.97e+00 (1.5e-01)& \textbf{2.90e+00} (1.3e-01)& 2.79e-01& 0.57\tabularnewline
&0.6& 5.80e+00 (4.3e-02)& 5.77e+00 (4.7e-02)& 4.45e+00 (2.0e-01)& \textbf{4.22e+00} (1.9e-01)& 7.97e-02 (.)& 0.63\tabularnewline
&1.0& 9.71e+00 (7.3e-02)& 9.73e+00 (7.3e-02)& 7.50e+00 (3.1e-01)& \textbf{6.93e+00} (2.6e-01)& 5.91e-03 (**)& 0.59\tabularnewline
\midrule
\multirow{4}{*}{$x^3$}&0.2& 1.89e+00 (1.6e-02)& 1.88e+00 (1.7e-02)& \textbf{1.53e+00} (6.9e-02)& 1.54e+00 (5.9e-02)& 3.81e-01& 0.47\tabularnewline
&0.4& 3.83e+00 (3.0e-02)& 3.80e+00 (3.0e-02)& \textbf{2.82e+00} (1.4e-01)& 2.89e+00 (1.2e-01)& 1.71e-01& 0.45\tabularnewline
&0.6& 5.74e+00 (4.6e-02)& 5.68e+00 (4.9e-02)& 4.78e+00 (2.1e-01)& \textbf{4.71e+00} (1.6e-01)& 2.93e-01& 0.49\tabularnewline
&1.0& 9.65e+00 (7.4e-02)& 9.60e+00 (7.2e-02)& 7.63e+00 (3.5e-01)& \textbf{7.17e+00} (2.7e-01)& 2.21e-02 (*)& 0.6\tabularnewline
\midrule
\multirow{4}{*}{$\exp(x)$}&0.2& 1.91e+00 (1.3e-02)& 1.88e+00 (1.5e-02)& \textbf{1.51e+00} (5.8e-02)& 1.61e+00 (6.6e-02)& 3.47e-02 (*)& 0.53\tabularnewline
&0.4& 3.83e+00 (2.9e-02)& 3.82e+00 (3.1e-02)& 2.79e+00 (1.3e-01)& \textbf{2.71e+00} (1.1e-01)& 1.97e-01& 0.57\tabularnewline
&0.6& 5.72e+00 (4.4e-02)& 5.68e+00 (4.4e-02)& 4.45e+00 (1.8e-01)& \textbf{4.27e+00} (1.7e-01)& 1.15e-01& 0.61\tabularnewline
&1.0& 9.57e+00 (5.9e-02)& 9.49e+00 (7.3e-02)& 7.56e+00 (2.8e-01)& \textbf{7.08e+00} (3.1e-01)& 3.59e-02 (*)& 0.57\tabularnewline
\midrule
\multirow{4}{*}{sigmoid}&0.2& 1.91e+00 (1.5e-02)& 1.91e+00 (1.6e-02)& 1.69e+00 (5.4e-02)& \textbf{1.59e+00} (5.0e-02)& 6.14e-03 (**)& 0.65\tabularnewline
&0.4& 3.89e+00 (3.2e-02)& 3.88e+00 (3.2e-02)& 2.99e+00 (1.2e-01)& \textbf{2.96e+00} (1.1e-01)& 3.53e-01& 0.47\tabularnewline
&0.6& 5.77e+00 (4.8e-02)& 5.76e+00 (4.8e-02)& 4.35e+00 (1.7e-01)& \textbf{4.14e+00} (1.4e-01)& 4.19e-02 (*)& 0.56\tabularnewline
&1.0& 9.51e+00 (8.5e-02)& 9.50e+00 (8.7e-02)& 7.33e+00 (3.2e-01)& \textbf{6.61e+00} (2.6e-01)& 1.45e-03 (**)& 0.56\tabularnewline
\midrule
\multirow{4}{*}{SE-1}&0.2& 1.93e+00 (1.4e-02)& 1.93e+00 (1.6e-02)& 1.62e+00 (7.3e-02)& \textbf{1.59e+00} (5.2e-02)& 3.21e-01& 0.52\tabularnewline
&0.4& 3.81e+00 (3.4e-02)& 3.78e+00 (3.6e-02)& 3.18e+00 (1.4e-01)& \textbf{3.16e+00} (1.2e-01)& 3.93e-01& 0.54\tabularnewline
&0.6& 5.77e+00 (4.4e-02)& 5.77e+00 (4.6e-02)& 4.67e+00 (2.0e-01)& \textbf{4.34e+00} (1.6e-01)& 2.88e-02 (*)& 0.57\tabularnewline
&1.0& 9.55e+00 (7.0e-02)& 9.51e+00 (7.6e-02)& 7.29e+00 (3.2e-01)& \textbf{6.62e+00} (2.7e-01)& 5.51e-03 (**)& 0.63\tabularnewline
\midrule
\multirow{4}{*}{SE-0.1}&0.2& 1.76e+00 (2.0e-02)& 1.78e+00 (2.5e-02)& \textbf{3.54e+00} (5.3e-02)& 3.54e+00 (6.7e-02)& 4.92e-01& 0.6\tabularnewline
&0.4& 3.54e+00 (3.6e-02)& 3.55e+00 (3.8e-02)& 6.59e+00 (1.1e-01)& \textbf{6.25e+00} (1.0e-01)& 1.36e-04 (***)& 0.66\tabularnewline
&0.6& 5.57e+00 (5.4e-02)& 5.59e+00 (6.0e-02)& 9.20e+00 (1.6e-01)& \textbf{9.13e+00} (1.6e-01)& 3.14e-01& 0.59\tabularnewline
&1.0& 9.29e+00 (8.5e-02)& 9.20e+00 (9.3e-02)& 1.44e+01 (2.6e-01)& \textbf{1.38e+01} (2.4e-01)& 5.93e-04 (***)& 0.7\tabularnewline
\midrule
\multirow{4}{*}{Mat12-1}&0.2& 2.12e+00 (3.2e-02)& 2.15e+00 (2.9e-02)& 5.43e+00 (5.2e-02)& \textbf{5.29e+00} (6.5e-02)& 7.85e-03 (**)& 0.72\tabularnewline
&0.4& 4.08e+00 (4.7e-02)& 4.02e+00 (4.8e-02)& 7.55e+00 (9.5e-02)& \textbf{7.20e+00} (8.9e-02)& 7.51e-09 (***)& 0.72\tabularnewline
&0.6& 5.82e+00 (6.5e-02)& 5.79e+00 (5.8e-02)& 9.34e+00 (1.4e-01)& \textbf{8.94e+00} (1.1e-01)& 1.35e-04 (***)& 0.7\tabularnewline
&1.0& 9.79e+00 (8.9e-02)& 9.73e+00 (9.1e-02)& 1.26e+01 (2.9e-01)& \textbf{1.17e+01} (2.4e-01)& 9.31e-07 (***)& 0.68\tabularnewline
\midrule
\multirow{4}{*}{Mat12-0.1}&0.2& 3.87e+00 (6.8e-02)& 3.88e+00 (7.3e-02)& 1.26e+01 (1.6e-01)& \textbf{1.25e+01} (1.8e-01)& 2.28e-01& 0.57\tabularnewline
&0.4& 5.15e+00 (8.3e-02)& 5.08e+00 (6.9e-02)& 1.47e+01 (1.5e-01)& \textbf{1.42e+01} (1.3e-01)& 1.25e-03 (**)& 0.66\tabularnewline
&0.6& 6.77e+00 (1.0e-01)& 6.64e+00 (8.8e-02)& 1.67e+01 (1.8e-01)& \textbf{1.61e+01} (1.7e-01)& 2.66e-04 (***)& 0.67\tabularnewline
&1.0& 1.04e+01 (1.3e-01)& 1.04e+01 (1.3e-01)& 2.07e+01 (2.4e-01)& \textbf{2.01e+01} (2.4e-01)& 1.85e-04 (***)& 0.73\tabularnewline
\midrule
\multirow{4}{*}{Mat32-1}&0.2& 1.87e+00 (1.6e-02)& 1.87e+00 (1.6e-02)& 2.40e+00 (4.5e-02)& \textbf{2.29e+00} (4.3e-02)& 1.30e-03 (**)& 0.66\tabularnewline
&0.4& 3.82e+00 (3.5e-02)& 3.79e+00 (3.6e-02)& 3.99e+00 (1.0e-01)& \textbf{3.92e+00} (9.6e-02)& 2.08e-01& 0.63\tabularnewline
&0.6& 5.77e+00 (4.4e-02)& 5.76e+00 (4.3e-02)& 5.60e+00 (1.6e-01)& \textbf{5.28e+00} (1.3e-01)& 4.84e-03 (**)& 0.68\tabularnewline
&1.0& 9.62e+00 (8.2e-02)& 9.61e+00 (8.3e-02)& 9.01e+00 (3.5e-01)& \textbf{8.00e+00} (2.6e-01)& 1.46e-04 (***)& 0.56\tabularnewline
\midrule
\multirow{4}{*}{Mat32-0.1}&0.2& 2.05e+00 (3.5e-02)& 2.11e+00 (4.4e-02)& \textbf{5.46e+00} (7.6e-02)& 5.54e+00 (1.1e-01)& 1.82e-01& 0.57\tabularnewline
&0.4& 3.84e+00 (5.4e-02)& 3.77e+00 (6.0e-02)& 8.83e+00 (1.1e-01)& \textbf{8.41e+00} (1.3e-01)& 8.21e-05 (***)& 0.74\tabularnewline
&0.6& 5.64e+00 (7.0e-02)& 5.58e+00 (7.6e-02)& 1.17e+01 (1.5e-01)& \textbf{1.13e+01} (1.4e-01)& 1.68e-05 (***)& 0.69\tabularnewline
&1.0& 9.62e+00 (1.1e-01)& 9.53e+00 (9.7e-02)& 1.68e+01 (2.5e-01)& \textbf{1.58e+01} (2.1e-01)& 4.67e-10 (***)& 0.72\tabularnewline
\midrule
\multirow{4}{*}{RQ-0.1-0.5}&0.2& 1.72e+00 (2.9e-02)& 1.74e+00 (3.1e-02)& 4.14e+00 (6.5e-02)& \textbf{4.12e+00} (7.4e-02)& 3.53e-01& 0.61\tabularnewline
&0.4& 3.77e+00 (4.6e-02)& 3.68e+00 (4.6e-02)& 7.33e+00 (1.1e-01)& \textbf{6.92e+00} (9.9e-02)& 1.44e-06 (***)& 0.77\tabularnewline
&0.6& 5.63e+00 (7.2e-02)& 5.54e+00 (6.4e-02)& 1.03e+01 (1.7e-01)& \textbf{9.61e+00} (1.6e-01)& 2.72e-07 (***)& 0.72\tabularnewline
&1.0& 9.55e+00 (1.1e-01)& 9.50e+00 (1.2e-01)& 1.50e+01 (2.4e-01)& \textbf{1.44e+01} (2.6e-01)& 5.01e-03 (**)& 0.68\tabularnewline
\midrule
\multirow{4}{*}{Periodic-0.1-4}&0.2& 1.68e+00 (2.8e-02)& 1.70e+00 (2.7e-02)& 4.27e+00 (7.0e-02)& \textbf{4.21e+00} (6.6e-02)& 2.00e-01& 0.6\tabularnewline
&0.4& 3.40e+00 (4.2e-02)& 3.48e+00 (5.2e-02)& 7.90e+00 (1.1e-01)& \textbf{7.84e+00} (1.3e-01)& 2.81e-01& 0.65\tabularnewline
&0.6& 5.42e+00 (7.3e-02)& 5.45e+00 (7.0e-02)& 1.12e+01 (1.6e-01)& \textbf{1.08e+01} (1.8e-01)& 2.52e-02 (*)& 0.65\tabularnewline
&1.0& 9.41e+00 (1.2e-01)& 9.31e+00 (1.1e-01)& 1.72e+01 (3.0e-01)& \textbf{1.60e+01} (2.5e-01)& 2.03e-10 (***)& 0.63\tabularnewline
\bottomrule
\end{tabular}

    }
\end{table}

\begin{table}[H]
    \centering
    \caption[MSE for Monotone Decomposition with Cubic Splines using CV-tuned $\mu$]{Results for comparing the cubic splines and the fitting by monotone decomposition with CV-tuned $\mu$ and fixed $J$. The values are averages over 100 replications, with the standard error of the average in parentheses. The bold values highlight the smaller mean squared prediction error. }
    \label{tab:bspl_vs_mbss_mu}
    \resizebox{0.8\textwidth}{!}{%
        \begin{tabular}{ccccccll}
\toprule
\multirow{2}{*}{curve} & \multirow{2}{*}{$\sigma$}&\multicolumn{2}{c}{MSFE}&\multicolumn{2}{c}{MSPE}& \multirow{2}{*}{p-value}& \multirow{2}{*}{prop.}\tabularnewline
\cmidrule(lr){3-4}
\cmidrule(lr){5-6}
&&CubicSpline&MonoDecomp&CubicSpline&MonoDecomp\tabularnewline
\midrule
\multirow{4}{*}{$x^2$}&0.1& 9.68e-01 (8.2e-03)& 9.71e-01 (8.2e-03)& 7.38e-01 (3.6e-02)& \textbf{7.07e-01} (3.3e-02)& 1.47e-03 (**)& 0.65\tabularnewline
&0.2& 1.92e+00 (1.6e-02)& 1.93e+00 (1.6e-02)& 1.53e+00 (7.7e-02)& \textbf{1.45e+00} (7.2e-02)& 1.03e-03 (**)& 0.74\tabularnewline
&0.5& 4.88e+00 (3.6e-02)& 4.90e+00 (3.6e-02)& 3.67e+00 (1.5e-01)& \textbf{3.37e+00} (1.3e-01)& 2.20e-06 (***)& 0.77\tabularnewline
&1.0& 9.63e+00 (8.2e-02)& 9.73e+00 (8.0e-02)& 7.12e+00 (3.2e-01)& \textbf{5.92e+00} (2.5e-01)& 1.73e-10 (***)& 0.77\tabularnewline
\midrule
\multirow{4}{*}{$x^3$}&0.1& 9.56e-01 (7.6e-03)& 9.58e-01 (7.1e-03)& 7.32e-01 (3.5e-02)& \textbf{6.98e-01} (2.9e-02)& 1.01e-03 (**)& 0.61\tabularnewline
&0.2& 1.93e+00 (1.6e-02)& 1.94e+00 (1.5e-02)& 1.47e+00 (7.1e-02)& \textbf{1.38e+00} (5.8e-02)& 2.88e-05 (***)& 0.69\tabularnewline
&0.5& 4.83e+00 (4.3e-02)& 4.86e+00 (4.2e-02)& 3.66e+00 (1.5e-01)& \textbf{3.41e+00} (1.4e-01)& 7.49e-03 (**)& 0.64\tabularnewline
&1.0& 9.55e+00 (8.1e-02)& 9.68e+00 (7.8e-02)& 8.03e+00 (3.7e-01)& \textbf{7.02e+00} (2.7e-01)& 2.01e-06 (***)& 0.69\tabularnewline
\midrule
\multirow{4}{*}{$\exp(x)$}&0.1& 9.56e-01 (7.4e-03)& 9.57e-01 (7.3e-03)& 7.51e-01 (3.3e-02)& \textbf{7.26e-01} (3.0e-02)& 3.12e-04 (***)& 0.61\tabularnewline
&0.2& 1.93e+00 (1.5e-02)& 1.94e+00 (1.5e-02)& 1.40e+00 (6.6e-02)& \textbf{1.32e+00} (5.9e-02)& 4.32e-07 (***)& 0.67\tabularnewline
&0.5& 4.82e+00 (3.9e-02)& 4.84e+00 (3.7e-02)& 3.52e+00 (1.7e-01)& \textbf{3.02e+00} (1.4e-01)& 9.36e-11 (***)& 0.8\tabularnewline
&1.0& 9.74e+00 (7.6e-02)& 9.82e+00 (7.5e-02)& 7.47e+00 (3.3e-01)& \textbf{6.09e+00} (2.5e-01)& 6.35e-11 (***)& 0.8\tabularnewline
\midrule
\multirow{4}{*}{sigmoid}&0.1& 9.53e-01 (7.3e-03)& 9.60e-01 (7.5e-03)& 8.84e-01 (2.4e-02)& \textbf{8.55e-01} (2.5e-02)& 3.16e-03 (**)& 0.68\tabularnewline
&0.2& 1.87e+00 (1.4e-02)& 1.89e+00 (1.4e-02)& 1.81e+00 (6.2e-02)& \textbf{1.67e+00} (5.2e-02)& 5.12e-05 (***)& 0.71\tabularnewline
&0.5& 4.78e+00 (3.8e-02)& 4.82e+00 (3.6e-02)& 3.83e+00 (1.7e-01)& \textbf{3.51e+00} (1.3e-01)& 2.42e-04 (***)& 0.66\tabularnewline
&1.0& 9.50e+00 (7.7e-02)& 9.62e+00 (7.7e-02)& 7.62e+00 (3.4e-01)& \textbf{6.36e+00} (2.4e-01)& 5.51e-08 (***)& 0.61\tabularnewline
\midrule
\multirow{4}{*}{SE-1}&0.1& 9.69e-01 (6.9e-03)& 9.74e-01 (6.7e-03)& 8.68e-01 (3.5e-02)& \textbf{8.18e-01} (3.0e-02)& 1.44e-04 (***)& 0.73\tabularnewline
&0.2& 1.92e+00 (1.6e-02)& 1.92e+00 (1.6e-02)& 1.62e+00 (6.1e-02)& \textbf{1.55e+00} (5.6e-02)& 3.30e-04 (***)& 0.66\tabularnewline
&0.5& 4.84e+00 (3.6e-02)& 4.88e+00 (3.5e-02)& 3.77e+00 (1.7e-01)& \textbf{3.52e+00} (1.3e-01)& 2.43e-03 (**)& 0.61\tabularnewline
&1.0& 9.69e+00 (6.9e-02)& 9.77e+00 (6.7e-02)& 7.31e+00 (3.1e-01)& \textbf{6.15e+00} (2.5e-01)& 2.24e-09 (***)& 0.63\tabularnewline
\midrule
\multirow{4}{*}{SE-0.1}&0.1& 8.51e-01 (1.1e-02)& 9.02e-01 (1.9e-02)& \textbf{1.88e+00} (3.1e-02)& 1.98e+00 (6.3e-02)& 3.83e-02 (*)& 0.57\tabularnewline
&0.2& 1.73e+00 (1.6e-02)& 1.79e+00 (3.1e-02)& \textbf{3.43e+00} (4.8e-02)& 3.48e+00 (1.1e-01)& 2.94e-01& 0.73\tabularnewline
&0.5& 4.47e+00 (5.0e-02)& 4.58e+00 (6.0e-02)& 7.77e+00 (1.4e-01)& \textbf{7.73e+00} (1.7e-01)& 3.74e-01& 0.67\tabularnewline
&1.0& 9.31e+00 (8.4e-02)& 9.52e+00 (8.4e-02)& 1.45e+01 (2.6e-01)& \textbf{1.41e+01} (2.5e-01)& 3.91e-04 (***)& 0.76\tabularnewline
\midrule
\multirow{4}{*}{Mat12-1}&0.1& 1.45e+00 (2.5e-02)& 1.56e+00 (2.4e-02)& \textbf{4.42e+00} (5.5e-02)& 4.61e+00 (5.8e-02)& 2.37e-07 (***)& 0.37\tabularnewline
&0.2& 2.19e+00 (3.5e-02)& 2.30e+00 (3.3e-02)& \textbf{5.46e+00} (5.7e-02)& 5.50e+00 (6.5e-02)& 1.78e-01& 0.64\tabularnewline
&0.5& 4.95e+00 (5.5e-02)& 5.05e+00 (5.2e-02)& 8.18e+00 (9.8e-02)& \textbf{8.04e+00} (1.1e-01)& 1.52e-02 (*)& 0.73\tabularnewline
&1.0& 9.80e+00 (9.1e-02)& 9.94e+00 (9.0e-02)& 1.21e+01 (2.6e-01)& \textbf{1.16e+01} (2.2e-01)& 6.76e-04 (***)& 0.74\tabularnewline
\midrule
\multirow{4}{*}{Mat12-0.1}&0.1& 3.41e+00 (6.6e-02)& 3.83e+00 (8.4e-02)& \textbf{1.16e+01} (1.4e-01)& 1.25e+01 (2.3e-01)& 1.85e-06 (***)& 0.23\tabularnewline
&0.2& 3.75e+00 (7.9e-02)& 4.17e+00 (9.4e-02)& \textbf{1.24e+01} (1.7e-01)& 1.32e+01 (2.3e-01)& 2.05e-07 (***)& 0.27\tabularnewline
&0.5& 5.91e+00 (9.0e-02)& 6.36e+00 (1.0e-01)& \textbf{1.56e+01} (1.6e-01)& 1.62e+01 (2.4e-01)& 1.01e-03 (**)& 0.35\tabularnewline
&1.0& 1.04e+01 (1.4e-01)& 1.08e+01 (1.2e-01)& 2.11e+01 (2.2e-01)& \textbf{2.11e+01} (2.3e-01)& 4.48e-01& 0.48\tabularnewline
\midrule
\multirow{4}{*}{Mat32-1}&0.1& 9.39e-01 (9.2e-03)& 9.51e-01 (8.3e-03)& 1.33e+00 (2.3e-02)& \textbf{1.29e+00} (2.0e-02)& 5.22e-04 (***)& 0.7\tabularnewline
&0.2& 1.91e+00 (1.7e-02)& 1.92e+00 (1.6e-02)& 2.27e+00 (4.7e-02)& \textbf{2.18e+00} (4.2e-02)& 1.76e-04 (***)& 0.71\tabularnewline
&0.5& 4.79e+00 (4.0e-02)& 4.84e+00 (3.8e-02)& 5.01e+00 (1.4e-01)& \textbf{4.53e+00} (1.1e-01)& 3.31e-07 (***)& 0.66\tabularnewline
&1.0& 9.59e+00 (8.2e-02)& 9.67e+00 (8.1e-02)& 8.49e+00 (2.9e-01)& \textbf{7.58e+00} (2.6e-01)& 3.94e-08 (***)& 0.67\tabularnewline
\midrule
\multirow{4}{*}{Mat32-0.1}&0.1& 1.25e+00 (3.3e-02)& 1.42e+00 (4.2e-02)& \textbf{3.86e+00} (8.2e-02)& 4.27e+00 (1.3e-01)& 1.67e-05 (***)& 0.42\tabularnewline
&0.2& 1.97e+00 (3.8e-02)& 2.23e+00 (6.7e-02)& \textbf{5.41e+00} (7.8e-02)& 5.95e+00 (1.9e-01)& 1.59e-03 (**)& 0.53\tabularnewline
&0.5& 4.66e+00 (5.9e-02)& 4.97e+00 (6.9e-02)& \textbf{1.03e+01} (1.2e-01)& 1.06e+01 (1.8e-01)& 2.05e-02 (*)& 0.55\tabularnewline
&1.0& 9.55e+00 (1.2e-01)& 9.82e+00 (1.2e-01)& 1.69e+01 (2.3e-01)& \textbf{1.67e+01} (2.7e-01)& 4.00e-02 (*)& 0.68\tabularnewline
\midrule
\multirow{4}{*}{RQ-0.1-0.5}&0.1& 8.92e-01 (1.9e-02)& 9.95e-01 (3.3e-02)& \textbf{2.37e+00} (4.1e-02)& 2.58e+00 (1.0e-01)& 1.91e-02 (*)& 0.51\tabularnewline
&0.2& 1.82e+00 (2.7e-02)& 1.95e+00 (3.7e-02)& \textbf{4.15e+00} (7.0e-02)& 4.36e+00 (1.1e-01)& 6.07e-03 (**)& 0.57\tabularnewline
&0.5& 4.60e+00 (6.2e-02)& 4.84e+00 (7.4e-02)& 9.08e+00 (1.5e-01)& \textbf{8.99e+00} (2.0e-01)& 2.66e-01& 0.61\tabularnewline
&1.0& 9.30e+00 (9.9e-02)& 9.57e+00 (9.8e-02)& 1.48e+01 (2.3e-01)& \textbf{1.43e+01} (2.2e-01)& 6.64e-03 (**)& 0.74\tabularnewline
\midrule
\multirow{4}{*}{Periodic-0.1-4}&0.1& 8.82e-01 (3.6e-02)& 9.46e-01 (3.8e-02)& \textbf{2.44e+00} (1.1e-01)& 2.58e+00 (1.2e-01)& 2.75e-03 (**)& 0.54\tabularnewline
&0.2& 1.65e+00 (2.6e-02)& 1.80e+00 (4.4e-02)& \textbf{4.26e+00} (6.2e-02)& 4.47e+00 (1.3e-01)& 3.92e-02 (*)& 0.67\tabularnewline
&0.5& 4.34e+00 (5.1e-02)& 4.54e+00 (6.0e-02)& 9.44e+00 (1.3e-01)& \textbf{9.41e+00} (1.9e-01)& 4.14e-01& 0.59\tabularnewline
&1.0& 9.25e+00 (1.3e-01)& 9.64e+00 (1.3e-01)& 1.76e+01 (2.6e-01)& \textbf{1.72e+01} (2.9e-01)& 3.24e-02 (*)& 0.66\tabularnewline
\bottomrule
\end{tabular}

    }
\end{table}

\subsection{Monotone Decomposition Fitting with Smoothing Splines}

\subsubsection{A complete version for Table 2}

\begin{table}[H]
    \centering
    \caption[MSE for Monotone Decomposition with Smoothing Splines using CV-tuned ($\mu,\lambda$)]{Results for comparing the smoothing splines with CV-tuned $\lambda$ and the fitting by monotone decomposition with CV-tuned $(\lambda, \mu)$. The values are averages over 100 replications, with the standard error of the average in parentheses. The bold values highlight the smaller mean squared prediction error.}
    \label{tab:ss_vs_md_lambda_mu}
    \resizebox{0.75\textwidth}{!}{%
        \begin{tabular}{ccccccll}
\toprule
\multirow{2}{*}{curve} & \multirow{2}{*}{$\sigma$}&\multicolumn{2}{c}{MSFE}&\multicolumn{2}{c}{MSPE}& \multirow{2}{*}{p-value}& \multirow{2}{*}{prop.}\tabularnewline
\cmidrule(lr){3-4}
\cmidrule(lr){5-6}
&&SmoothSpline&MonoDecomp&SmoothSpline&MonoDecomp\tabularnewline
\midrule
\multirow{5}{*}{$x^2$}&0.1& 9.57e-01 (8.3e-03)& 9.73e-01 (8.2e-03)& \textbf{7.38e-01} (2.4e-02)& 8.67e-01 (2.4e-02)& 1.74e-14 (***)& 0.25\tabularnewline
&0.5& 4.79e+00 (4.1e-02)& 4.80e+00 (4.1e-02)& 3.41e+00 (1.2e-01)& \textbf{3.39e+00} (1.1e-01)& 3.61e-01& 0.46\tabularnewline
&1.0& 9.65e+00 (8.9e-02)& 9.69e+00 (8.5e-02)& 6.44e+00 (2.9e-01)& \textbf{6.35e+00} (2.5e-01)& 1.68e-01& 0.49\tabularnewline
&1.5& 1.44e+01 (1.1e-01)& 1.45e+01 (1.1e-01)& 9.31e+00 (3.9e-01)& \textbf{9.14e+00} (3.4e-01)& 4.68e-02 (*)& 0.6\tabularnewline
&2.0& 1.92e+01 (1.6e-01)& 1.92e+01 (1.5e-01)& 1.23e+01 (4.9e-01)& \textbf{1.15e+01} (4.3e-01)& 4.45e-06 (***)& 0.81\tabularnewline
\midrule
\multirow{5}{*}{$x^3$}&0.1& 9.46e-01 (8.0e-03)& 9.89e-01 (8.0e-03)& \textbf{8.65e-01} (2.4e-02)& 1.11e+00 (2.5e-02)& 0.00e+00 (***)& 0.2\tabularnewline
&0.5& 4.86e+00 (4.4e-02)& 4.88e+00 (4.3e-02)& 3.65e+00 (1.2e-01)& \textbf{3.55e+00} (1.2e-01)& 9.06e-03 (**)& 0.58\tabularnewline
&1.0& 9.75e+00 (8.2e-02)& 9.77e+00 (8.2e-02)& 6.47e+00 (1.8e-01)& \textbf{6.21e+00} (1.7e-01)& 1.18e-04 (***)& 0.65\tabularnewline
&1.5& 1.45e+01 (1.1e-01)& 1.46e+01 (1.1e-01)& 9.16e+00 (3.1e-01)& \textbf{8.67e+00} (2.8e-01)& 1.67e-06 (***)& 0.81\tabularnewline
&2.0& 1.92e+01 (1.8e-01)& 1.92e+01 (1.6e-01)& 1.16e+01 (5.9e-01)& \textbf{1.09e+01} (4.9e-01)& 1.15e-04 (***)& 0.78\tabularnewline
\midrule
\multirow{5}{*}{$\exp(x)$}&0.1& 9.53e-01 (7.6e-03)& 9.64e-01 (7.7e-03)& \textbf{7.96e-01} (2.6e-02)& 9.13e-01 (2.6e-02)& 5.02e-12 (***)& 0.26\tabularnewline
&0.5& 4.80e+00 (4.3e-02)& 4.82e+00 (3.9e-02)& 3.33e+00 (1.5e-01)& \textbf{3.30e+00} (1.3e-01)& 2.60e-01& 0.51\tabularnewline
&1.0& 9.74e+00 (8.4e-02)& 9.75e+00 (8.4e-02)& 5.94e+00 (2.3e-01)& \textbf{5.82e+00} (2.1e-01)& 3.12e-02 (*)& 0.58\tabularnewline
&1.5& 1.45e+01 (1.3e-01)& 1.46e+01 (1.3e-01)& 9.10e+00 (4.4e-01)& \textbf{8.83e+00} (4.1e-01)& 7.93e-03 (**)& 0.69\tabularnewline
&2.0& 1.95e+01 (1.5e-01)& 1.95e+01 (1.5e-01)& 1.08e+01 (4.9e-01)& \textbf{1.06e+01} (4.4e-01)& 3.80e-02 (*)& 0.57\tabularnewline
\midrule
\multirow{5}{*}{sigmoid}&0.1& 9.58e-01 (8.0e-03)& 9.58e-01 (7.9e-03)& 7.75e-01 (2.7e-02)& \textbf{7.69e-01} (2.5e-02)& 1.81e-01& 0.56\tabularnewline
&0.5& 4.82e+00 (4.2e-02)& 4.82e+00 (4.2e-02)& 3.55e+00 (1.3e-01)& \textbf{3.49e+00} (1.2e-01)& 2.45e-02 (*)& 0.61\tabularnewline
&1.0& 9.60e+00 (7.8e-02)& 9.64e+00 (7.5e-02)& 5.99e+00 (2.9e-01)& \textbf{5.68e+00} (2.4e-01)& 4.47e-04 (***)& 0.67\tabularnewline
&1.5& 1.43e+01 (1.6e-01)& 1.44e+01 (1.5e-01)& 8.94e+00 (5.4e-01)& \textbf{8.36e+00} (4.9e-01)& 5.86e-07 (***)& 0.65\tabularnewline
&2.0& 1.93e+01 (1.6e-01)& 1.94e+01 (1.5e-01)& 1.15e+01 (6.4e-01)& \textbf{1.08e+01} (5.4e-01)& 2.06e-05 (***)& 0.7\tabularnewline
\midrule
\multirow{5}{*}{SE-1}&0.1& 9.66e-01 (7.5e-03)& 9.71e-01 (7.4e-03)& \textbf{7.54e-01} (2.4e-02)& 8.19e-01 (2.6e-02)& 7.24e-08 (***)& 0.26\tabularnewline
&0.5& 4.88e+00 (3.9e-02)& 4.89e+00 (3.8e-02)& 3.15e+00 (1.2e-01)& \textbf{3.10e+00} (1.1e-01)& 5.92e-02 (.)& 0.55\tabularnewline
&1.0& 9.67e+00 (8.3e-02)& 9.70e+00 (8.1e-02)& 6.32e+00 (2.8e-01)& \textbf{6.11e+00} (2.5e-01)& 3.86e-03 (**)& 0.6\tabularnewline
&1.5& 1.47e+01 (1.1e-01)& 1.47e+01 (1.1e-01)& 8.65e+00 (3.8e-01)& \textbf{8.39e+00} (3.4e-01)& 2.48e-02 (*)& 0.55\tabularnewline
&2.0& 1.93e+01 (1.5e-01)& 1.94e+01 (1.5e-01)& 1.15e+01 (4.3e-01)& \textbf{1.10e+01} (3.8e-01)& 6.24e-04 (***)& 0.7\tabularnewline
\midrule
\multirow{5}{*}{SE-0.1}&0.1& 7.68e-01 (9.1e-03)& 7.94e-01 (9.4e-03)& \textbf{1.71e+00} (2.7e-02)& 1.79e+00 (3.0e-02)& 3.41e-09 (***)& 0.24\tabularnewline
&0.5& 4.36e+00 (4.4e-02)& 4.40e+00 (4.5e-02)& \textbf{6.65e+00} (1.0e-01)& 6.72e+00 (1.0e-01)& 2.24e-03 (**)& 0.42\tabularnewline
&1.0& 8.92e+00 (9.2e-02)& 8.99e+00 (9.0e-02)& 1.23e+01 (2.1e-01)& \textbf{1.23e+01} (2.1e-01)& 4.32e-01& 0.57\tabularnewline
&1.5& 1.41e+01 (1.3e-01)& 1.42e+01 (1.3e-01)& 1.76e+01 (3.1e-01)& \textbf{1.72e+01} (3.1e-01)& 5.96e-05 (***)& 0.64\tabularnewline
&2.0& 1.89e+01 (1.7e-01)& 1.91e+01 (1.7e-01)& 2.16e+01 (4.3e-01)& \textbf{2.11e+01} (4.3e-01)& 2.13e-04 (***)& 0.68\tabularnewline
\midrule
\multirow{5}{*}{Mat12-1}&0.1& 1.16e+00 (2.0e-02)& 1.23e+00 (1.9e-02)& \textbf{3.71e+00} (2.7e-02)& 3.80e+00 (3.1e-02)& 5.72e-14 (***)& 0.15\tabularnewline
&0.5& 4.80e+00 (4.9e-02)& 4.87e+00 (4.5e-02)& \textbf{7.50e+00} (8.3e-02)& 7.51e+00 (8.3e-02)& 3.83e-01& 0.4\tabularnewline
&1.0& 9.65e+00 (9.5e-02)& 9.69e+00 (9.2e-02)& 1.15e+01 (2.3e-01)& \textbf{1.13e+01} (2.1e-01)& 3.40e-04 (***)& 0.56\tabularnewline
&1.5& 1.43e+01 (1.1e-01)& 1.43e+01 (1.1e-01)& 1.40e+01 (3.0e-01)& \textbf{1.38e+01} (2.8e-01)& 3.13e-03 (**)& 0.64\tabularnewline
&2.0& 1.96e+01 (1.6e-01)& 1.97e+01 (1.6e-01)& 1.69e+01 (4.4e-01)& \textbf{1.64e+01} (3.7e-01)& 1.97e-04 (***)& 0.69\tabularnewline
\midrule
\multirow{5}{*}{Mat12-0.1}&0.1& 2.60e+00 (3.9e-02)& 2.88e+00 (3.8e-02)& \textbf{9.86e+00} (7.0e-02)& 1.03e+01 (8.1e-02)& 0.00e+00 (***)& 0.06\tabularnewline
&0.5& 4.81e+00 (6.4e-02)& 5.11e+00 (6.1e-02)& \textbf{1.35e+01} (8.5e-02)& 1.37e+01 (9.4e-02)& 6.90e-06 (***)& 0.3\tabularnewline
&1.0& 9.76e+00 (1.2e-01)& 9.92e+00 (1.1e-01)& 1.90e+01 (2.2e-01)& \textbf{1.90e+01} (2.2e-01)& 2.06e-01& 0.58\tabularnewline
&1.5& 1.46e+01 (1.7e-01)& 1.48e+01 (1.6e-01)& 2.29e+01 (2.8e-01)& \textbf{2.25e+01} (2.7e-01)& 1.21e-05 (***)& 0.69\tabularnewline
&2.0& 1.93e+01 (1.8e-01)& 1.95e+01 (1.7e-01)& 2.70e+01 (4.1e-01)& \textbf{2.65e+01} (3.6e-01)& 3.05e-04 (***)& 0.66\tabularnewline
\midrule
\multirow{5}{*}{Mat32-1}&0.1& 9.15e-01 (8.0e-03)& 9.20e-01 (7.7e-03)& \textbf{1.18e+00} (2.1e-02)& 1.19e+00 (2.2e-02)& 5.89e-02 (.)& 0.45\tabularnewline
&0.5& 4.86e+00 (4.2e-02)& 4.88e+00 (4.0e-02)& 4.35e+00 (1.3e-01)& \textbf{4.32e+00} (1.2e-01)& 2.45e-01& 0.52\tabularnewline
&1.0& 9.59e+00 (8.5e-02)& 9.64e+00 (7.5e-02)& 7.20e+00 (2.4e-01)& \textbf{7.04e+00} (2.0e-01)& 3.65e-02 (*)& 0.49\tabularnewline
&1.5& 1.45e+01 (1.3e-01)& 1.46e+01 (1.3e-01)& 1.03e+01 (3.7e-01)& \textbf{1.00e+01} (3.3e-01)& 7.41e-03 (**)& 0.65\tabularnewline
&2.0& 1.95e+01 (1.7e-01)& 1.96e+01 (1.7e-01)& 1.26e+01 (5.1e-01)& \textbf{1.19e+01} (4.4e-01)& 6.29e-05 (***)& 0.68\tabularnewline
\midrule
\multirow{5}{*}{Mat32-0.1}&0.1& 8.83e-01 (1.4e-02)& 9.47e-01 (1.4e-02)& \textbf{2.92e+00} (2.4e-02)& 3.02e+00 (2.6e-02)& 3.54e-10 (***)& 0.21\tabularnewline
&0.5& 4.22e+00 (5.3e-02)& 4.37e+00 (5.0e-02)& \textbf{8.99e+00} (9.9e-02)& 9.03e+00 (9.9e-02)& 1.82e-01& 0.45\tabularnewline
&1.0& 8.96e+00 (1.1e-01)& 9.14e+00 (1.0e-01)& 1.48e+01 (2.2e-01)& \textbf{1.47e+01} (2.1e-01)& 1.44e-01& 0.54\tabularnewline
&1.5& 1.41e+01 (1.6e-01)& 1.43e+01 (1.6e-01)& 1.99e+01 (3.5e-01)& \textbf{1.97e+01} (3.4e-01)& 8.29e-02 (.)& 0.59\tabularnewline
&2.0& 1.92e+01 (1.9e-01)& 1.94e+01 (1.8e-01)& 2.36e+01 (4.4e-01)& \textbf{2.29e+01} (3.9e-01)& 6.38e-08 (***)& 0.74\tabularnewline
\midrule
\multirow{5}{*}{RQ-0.1-0.5}&0.1& 7.21e-01 (9.5e-03)& 7.52e-01 (1.0e-02)& \textbf{2.04e+00} (2.1e-02)& 2.07e+00 (2.2e-02)& 1.35e-04 (***)& 0.38\tabularnewline
&0.5& 4.31e+00 (4.8e-02)& 4.41e+00 (4.7e-02)& \textbf{7.50e+00} (9.8e-02)& 7.55e+00 (9.5e-02)& 1.36e-01& 0.46\tabularnewline
&1.0& 9.10e+00 (1.1e-01)& 9.22e+00 (1.1e-01)& 1.27e+01 (2.4e-01)& \textbf{1.25e+01} (2.2e-01)& 1.21e-03 (**)& 0.6\tabularnewline
&1.5& 1.43e+01 (1.5e-01)& 1.44e+01 (1.5e-01)& 1.75e+01 (3.3e-01)& \textbf{1.74e+01} (3.1e-01)& 1.29e-01& 0.5\tabularnewline
&2.0& 1.91e+01 (1.9e-01)& 1.92e+01 (1.8e-01)& 2.10e+01 (5.4e-01)& \textbf{2.07e+01} (5.0e-01)& 1.08e-02 (*)& 0.63\tabularnewline
\midrule
\multirow{5}{*}{Periodic-0.1-4}&0.1& 7.11e-01 (8.8e-03)& 7.44e-01 (1.0e-02)& \textbf{2.07e+00} (2.7e-02)& 2.11e+00 (2.6e-02)& 3.37e-06 (***)& 0.27\tabularnewline
&0.5& 4.07e+00 (4.6e-02)& 4.20e+00 (4.6e-02)& \textbf{8.20e+00} (1.2e-01)& 8.35e+00 (1.1e-01)& 1.66e-03 (**)& 0.27\tabularnewline
&1.0& 8.69e+00 (1.0e-01)& 8.89e+00 (9.1e-02)& 1.44e+01 (2.1e-01)& \textbf{1.43e+01} (1.9e-01)& 3.03e-01& 0.49\tabularnewline
&1.5& 1.39e+01 (1.7e-01)& 1.42e+01 (1.6e-01)& 2.06e+01 (2.7e-01)& \textbf{2.02e+01} (2.7e-01)& 1.96e-05 (***)& 0.67\tabularnewline
&2.0& 1.90e+01 (1.8e-01)& 1.92e+01 (1.8e-01)& 2.47e+01 (3.9e-01)& \textbf{2.45e+01} (3.8e-01)& 4.46e-02 (*)& 0.63\tabularnewline
\bottomrule
\end{tabular}

    }
\end{table}

\subsubsection{Tune $\mu$ with fixed $\lambda$}

\begin{figure}[H]
    \centering
    \begin{subfigure}{0.5\textwidth}
    \includegraphics[width=\textwidth]{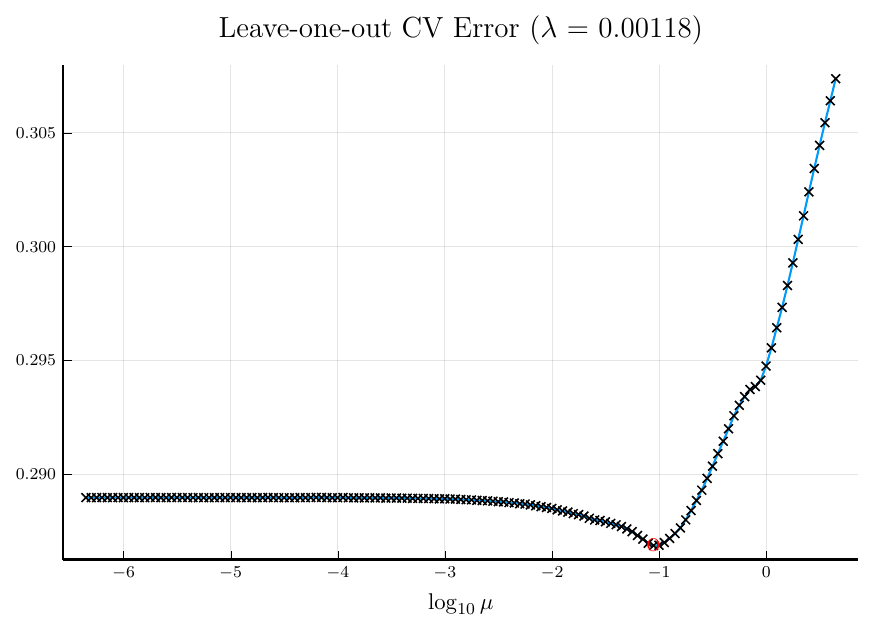}
    \end{subfigure}%
    \begin{subfigure}{0.5\textwidth}
    \includegraphics[width=\textwidth]{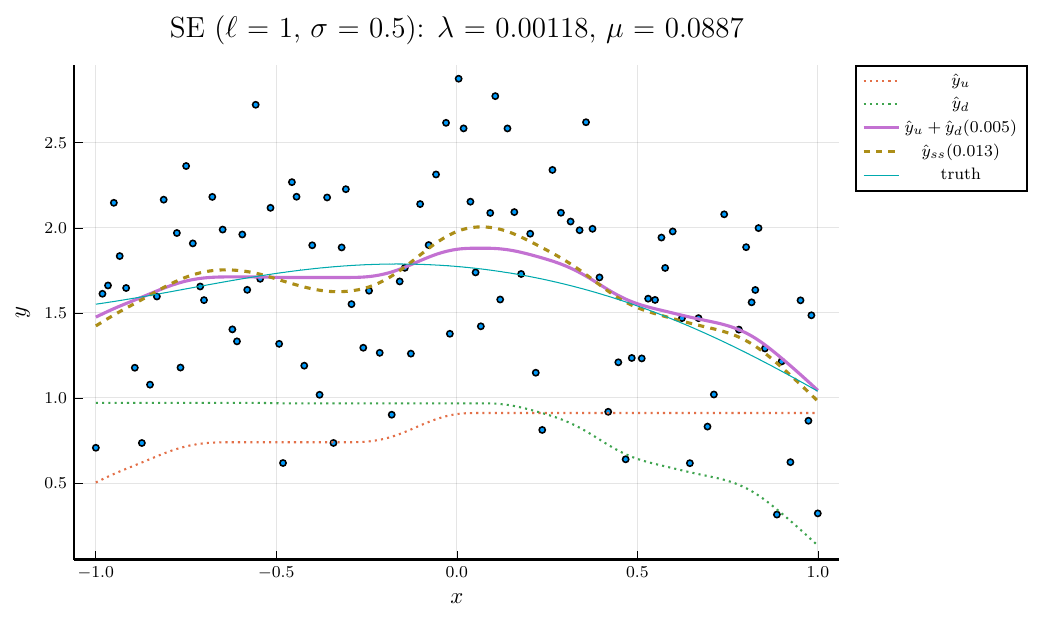}
    \end{subfigure}
    \caption[Monotone Decomposition with Smoothing Splines using CV-tuned $\mu$]{Demo for monotone decomposition with smoothing splines. The left panel shows the leave-one-out cross-validation error curve for each candidate $\mu$ when $\lambda$ is fixed to the one of smoothing splines. The right panel shows the corresponding fitting curves, together with the truth and the noised observations.}
    \label{fig:demo_ss}
\end{figure}

First, we fix the parameter $\lambda$ for the roughness penalty as the CV-tuned one for the smoothing splines. Then, we choose the parameter $\mu$ to minimize the CV error. Figure~\ref{fig:demo_ss} demonstrates the procedure. The left panel shows the CV-error curve for each candidate parameter $\mu$, and the right panel compares the monotone decomposition fitting given the parameter which minimized the curve in the left panel to the smoothing spline fitting. The monotone decomposition achieves a better MSPE, and it is obvious that the better performance is mainly due to the shrinkage on the local modes based on the shapes of fitting curves.

The average results based on 100 repetitions are summarized in Table \ref{tab:ss_vs_md_mu}. The table shows that we can obtain better performance in the high noise cases, and comparable results in the smaller noise scenarios.

\begin{table}[H]
    \centering
    \caption[MSE for Monotone Decomposition with Smoothing Splines using CV-tuned $\mu$]{Results for comparing the smoothing splines with CV-tuned $\lambda$ and the fitting by monotone decomposition with CV-tuned $\mu$ and the same $\lambda$. The values are averages over 100 replications, with the standard error of the average in parentheses. The bold values highlight the smaller mean square prediction error. }
    \label{tab:ss_vs_md_mu}
    \resizebox{0.85\textwidth}{!}{%
        \begin{tabular}{cccccccll}
\toprule
\multirow{2}{*}{curve} & \multirow{2}{*}{$\sigma$}& \multirow{2}{*}{SNR}&\multicolumn{2}{c}{MSFE}&\multicolumn{2}{c}{MSPE}& \multirow{2}{*}{p-value}& \multirow{2}{*}{prop.}\tabularnewline
\cmidrule(lr){4-5}
\cmidrule(lr){6-7}
&&&SmoothSpline&MonoDecomp&SmoothSpline&MonoDecomp\tabularnewline
\midrule
\multirow{6}{*}{$x^2$}&0.1& 1.03e+01 (1.9e-01)& 9.30e-03 (1.4e-04)& 9.55e-03 (1.4e-04)& \textbf{6.78e-04} (4.8e-05)& 9.03e-04 (5.4e-05)& 7.17e-12 (***)& 0.24\tabularnewline
&0.2& 2.68e+00 (7.4e-02)& 3.72e-02 (6.4e-04)& 3.76e-02 (6.3e-04)& \textbf{2.15e-03} (1.8e-04)& 2.29e-03 (1.7e-04)& 1.03e-02 (*)& 0.33\tabularnewline
&0.5& 5.04e-01 (2.1e-02)& 2.35e-01 (3.8e-03)& 2.37e-01 (3.6e-03)& 1.36e-02 (1.4e-03)& \textbf{1.23e-02} (9.5e-04)& 3.49e-02 (*)& 0.44\tabularnewline
&1.0& 1.75e-01 (1.0e-02)& 9.48e-01 (1.5e-02)& 9.59e-01 (1.5e-02)& 5.17e-02 (3.0e-03)& \textbf{4.81e-02} (2.7e-03)& 7.22e-03 (**)& 0.48\tabularnewline
&1.5& 1.23e-01 (1.2e-02)& 2.07e+00 (3.5e-02)& 2.09e+00 (3.6e-02)& 1.03e-01 (1.0e-02)& \textbf{9.57e-02} (8.8e-03)& 1.14e-02 (*)& 0.57\tabularnewline
&2.0& 1.13e-01 (2.3e-02)& 3.82e+00 (5.9e-02)& 3.87e+00 (5.6e-02)& 1.94e-01 (2.4e-02)& \textbf{1.55e-01} (1.6e-02)& 1.98e-04 (***)& 0.63\tabularnewline
\midrule
\multirow{6}{*}{$x^3$}&0.1& 1.76e+01 (3.3e-01)& 8.80e-03 (1.3e-04)& 9.92e-03 (1.4e-04)& \textbf{8.51e-04} (4.8e-05)& 1.70e-03 (7.3e-05)& 0.00e+00 (***)& 0.15\tabularnewline
&0.2& 4.51e+00 (1.1e-01)& 3.62e-02 (6.0e-04)& 3.74e-02 (6.1e-04)& \textbf{2.89e-03} (1.5e-04)& 3.11e-03 (1.6e-04)& 4.74e-02 (*)& 0.51\tabularnewline
&0.5& 7.70e-01 (2.3e-02)& 2.33e-01 (3.5e-03)& 2.36e-01 (3.5e-03)& 1.51e-02 (9.3e-04)& \textbf{1.36e-02} (8.0e-04)& 3.25e-04 (***)& 0.52\tabularnewline
&1.0& 2.43e-01 (1.6e-02)& 9.44e-01 (1.6e-02)& 9.54e-01 (1.6e-02)& 5.32e-02 (3.9e-03)& \textbf{4.63e-02} (2.9e-03)& 1.77e-04 (***)& 0.69\tabularnewline
&1.5& 1.58e-01 (1.6e-02)& 2.09e+00 (3.7e-02)& 2.12e+00 (3.5e-02)& 1.11e-01 (1.1e-02)& \textbf{9.53e-02} (1.1e-02)& 1.90e-05 (***)& 0.69\tabularnewline
&2.0& 1.36e-01 (2.1e-02)& 3.75e+00 (7.0e-02)& 3.80e+00 (6.6e-02)& 1.74e-01 (2.1e-02)& \textbf{1.37e-01} (1.6e-02)& 1.55e-04 (***)& 0.74\tabularnewline
\midrule
\multirow{6}{*}{$\exp(x)$}&0.1& 5.10e+01 (1.1e+00)& 9.06e-03 (1.7e-04)& 9.42e-03 (1.6e-04)& \textbf{6.66e-04} (5.0e-05)& 9.11e-04 (5.6e-05)& 3.13e-06 (***)& 0.36\tabularnewline
&0.2& 1.26e+01 (3.1e-01)& 3.65e-02 (6.4e-04)& 3.69e-02 (6.0e-04)& 2.35e-03 (2.6e-04)& \textbf{2.34e-03} (1.8e-04)& 4.72e-01& 0.37\tabularnewline
&0.5& 1.98e+00 (4.5e-02)& 2.40e-01 (3.9e-03)& 2.41e-01 (3.8e-03)& 1.16e-02 (9.4e-04)& \textbf{1.10e-02} (7.9e-04)& 4.03e-02 (*)& 0.52\tabularnewline
&1.0& 5.89e-01 (2.2e-02)& 9.07e-01 (1.4e-02)& 9.14e-01 (1.3e-02)& 5.39e-02 (5.7e-03)& \textbf{4.79e-02} (4.5e-03)& 3.84e-04 (***)& 0.55\tabularnewline
&1.5& 3.02e-01 (1.9e-02)& 2.11e+00 (3.6e-02)& 2.13e+00 (3.5e-02)& 1.02e-01 (1.0e-02)& \textbf{9.10e-02} (8.7e-03)& 5.87e-04 (***)& 0.62\tabularnewline
&2.0& 2.00e-01 (2.9e-02)& 3.79e+00 (6.3e-02)& 3.81e+00 (6.2e-02)& 1.71e-01 (2.7e-02)& \textbf{1.53e-01} (2.4e-02)& 1.71e-04 (***)& 0.72\tabularnewline
\midrule
\multirow{6}{*}{sigmoid}&0.1& 1.70e+01 (3.2e-01)& 9.31e-03 (1.4e-04)& 9.36e-03 (1.4e-04)& 6.33e-04 (4.3e-05)& \textbf{6.03e-04} (3.3e-05)& 5.12e-02 (.)& 0.55\tabularnewline
&0.2& 4.38e+00 (6.9e-02)& 3.66e-02 (4.6e-04)& 3.68e-02 (4.6e-04)& 2.44e-03 (1.7e-04)& \textbf{2.39e-03} (1.6e-04)& 7.53e-02 (.)& 0.61\tabularnewline
&0.5& 7.87e-01 (2.7e-02)& 2.32e-01 (4.0e-03)& 2.33e-01 (3.9e-03)& 1.50e-02 (1.0e-03)& \textbf{1.34e-02} (7.9e-04)& 1.59e-05 (***)& 0.63\tabularnewline
&1.0& 2.31e-01 (1.3e-02)& 9.44e-01 (1.6e-02)& 9.51e-01 (1.6e-02)& 3.96e-02 (3.9e-03)& \textbf{3.43e-02} (3.1e-03)& 8.84e-06 (***)& 0.67\tabularnewline
&1.5& 1.70e-01 (1.5e-02)& 2.09e+00 (3.3e-02)& 2.12e+00 (3.1e-02)& 1.12e-01 (1.0e-02)& \textbf{8.82e-02} (7.8e-03)& 4.95e-05 (***)& 0.64\tabularnewline
&2.0& 1.26e-01 (1.9e-02)& 3.69e+00 (6.5e-02)& 3.73e+00 (6.3e-02)& 1.67e-01 (2.4e-02)& \textbf{1.32e-01} (1.7e-02)& 1.58e-04 (***)& 0.75\tabularnewline
\midrule
\multirow{6}{*}{SE-1}&0.1& 3.43e+01 (4.3e+00)& 9.02e-03 (1.6e-04)& 9.16e-03 (1.5e-04)& 7.53e-04 (7.0e-05)& \textbf{7.32e-04} (5.5e-05)& 2.40e-01& 0.43\tabularnewline
&0.2& 5.41e+00 (6.9e-01)& 3.77e-02 (6.6e-04)& 3.78e-02 (6.5e-04)& \textbf{2.31e-03} (2.0e-04)& 2.36e-03 (1.9e-04)& 2.04e-01& 0.4\tabularnewline
&0.5& 1.20e+00 (1.4e-01)& 2.38e-01 (3.9e-03)& 2.39e-01 (3.8e-03)& 1.24e-02 (1.2e-03)& \textbf{1.12e-02} (9.5e-04)& 7.34e-03 (**)& 0.51\tabularnewline
&1.0& 3.59e-01 (3.9e-02)& 9.38e-01 (1.6e-02)& 9.50e-01 (1.6e-02)& 4.74e-02 (3.6e-03)& \textbf{4.12e-02} (2.8e-03)& 4.67e-04 (***)& 0.63\tabularnewline
&1.5& 1.68e-01 (1.9e-02)& 2.13e+00 (3.2e-02)& 2.15e+00 (3.2e-02)& 9.58e-02 (8.1e-03)& \textbf{8.60e-02} (6.7e-03)& 3.72e-04 (***)& 0.6\tabularnewline
&2.0& 1.18e-01 (1.2e-02)& 3.68e+00 (5.9e-02)& 3.71e+00 (5.7e-02)& 1.69e-01 (1.6e-02)& \textbf{1.48e-01} (1.2e-02)& 2.36e-04 (***)& 0.75\tabularnewline
\midrule
\multirow{6}{*}{SE-0.1}&0.1& 1.50e+02 (6.4e+00)& 6.36e-03 (1.3e-04)& 7.06e-03 (1.8e-04)& \textbf{3.00e-03} (7.7e-05)& 3.51e-03 (1.6e-04)& 3.81e-05 (***)& 0.25\tabularnewline
&0.2& 3.55e+01 (1.9e+00)& 2.79e-02 (5.5e-04)& 2.90e-02 (5.7e-04)& \textbf{1.00e-02} (3.1e-04)& 1.05e-02 (3.3e-04)& 6.47e-05 (***)& 0.39\tabularnewline
&0.5& 4.98e+00 (2.3e-01)& 1.97e-01 (3.9e-03)& 2.02e-01 (3.7e-03)& \textbf{4.70e-02} (1.7e-03)& 4.74e-02 (1.5e-03)& 2.98e-01& 0.38\tabularnewline
&1.0& 1.30e+00 (6.9e-02)& 8.23e-01 (1.6e-02)& 8.54e-01 (1.5e-02)& \textbf{1.54e-01} (5.3e-03)& 1.55e-01 (4.6e-03)& 3.62e-01& 0.35\tabularnewline
&1.5& 6.35e-01 (3.9e-02)& 1.96e+00 (4.0e-02)& 2.02e+00 (4.0e-02)& \textbf{3.20e-01} (1.2e-02)& 3.23e-01 (1.3e-02)& 2.79e-01& 0.51\tabularnewline
&2.0& 3.69e-01 (2.3e-02)& 3.71e+00 (6.7e-02)& 3.79e+00 (6.8e-02)& 4.69e-01 (1.8e-02)& \textbf{4.62e-01} (1.7e-02)& 1.57e-01& 0.54\tabularnewline
\midrule
\multirow{6}{*}{Mat12-1}&0.1& 3.51e+01 (2.7e+00)& 1.41e-02 (4.1e-04)& 1.58e-02 (4.3e-04)& \textbf{1.40e-02} (2.3e-04)& 1.48e-02 (2.6e-04)& 3.30e-10 (***)& 0.2\tabularnewline
&0.2& 1.29e+01 (9.7e-01)& 3.80e-02 (1.0e-03)& 4.17e-02 (9.8e-04)& \textbf{2.39e-02} (4.6e-04)& 2.44e-02 (5.0e-04)& 1.39e-02 (*)& 0.43\tabularnewline
&0.5& 1.97e+00 (1.5e-01)& 2.27e-01 (4.6e-03)& 2.38e-01 (4.4e-03)& 5.88e-02 (1.7e-03)& \textbf{5.78e-02} (1.5e-03)& 1.09e-01& 0.54\tabularnewline
&1.0& 6.51e-01 (5.1e-02)& 9.19e-01 (1.6e-02)& 9.40e-01 (1.5e-02)& 1.26e-01 (4.8e-03)& \textbf{1.20e-01} (4.1e-03)& 1.12e-03 (**)& 0.57\tabularnewline
&1.5& 3.52e-01 (4.4e-02)& 2.10e+00 (3.4e-02)& 2.14e+00 (3.1e-02)& 2.13e-01 (1.2e-02)& \textbf{1.97e-01} (9.0e-03)& 2.02e-02 (*)& 0.62\tabularnewline
&2.0& 1.71e-01 (1.9e-02)& 3.76e+00 (6.0e-02)& 3.81e+00 (5.9e-02)& 2.94e-01 (2.1e-02)& \textbf{2.75e-01} (1.8e-02)& 1.00e-03 (**)& 0.61\tabularnewline
\midrule
\multirow{6}{*}{Mat12-0.1}&0.1& 1.51e+01 (8.5e-01)& 6.84e-02 (2.5e-03)& 8.71e-02 (3.1e-03)& \textbf{9.73e-02} (1.8e-03)& 1.10e-01 (2.3e-03)& 7.39e-13 (***)& 0.1\tabularnewline
&0.2& 9.80e+00 (5.0e-01)& 9.35e-02 (2.9e-03)& 1.12e-01 (3.5e-03)& \textbf{1.14e-01} (1.7e-03)& 1.26e-01 (2.5e-03)& 3.16e-11 (***)& 0.1\tabularnewline
&0.5& 3.90e+00 (2.3e-01)& 2.55e-01 (7.8e-03)& 2.90e-01 (7.9e-03)& \textbf{1.92e-01} (3.0e-03)& 2.02e-01 (3.5e-03)& 8.52e-08 (***)& 0.2\tabularnewline
&1.0& 1.24e+00 (7.2e-02)& 9.43e-01 (2.5e-02)& 1.01e+00 (2.3e-02)& \textbf{3.62e-01} (7.5e-03)& 3.65e-01 (7.5e-03)& 2.24e-01& 0.25\tabularnewline
&1.5& 5.39e-01 (4.7e-02)& 2.15e+00 (5.0e-02)& 2.23e+00 (4.6e-02)& 5.26e-01 (1.7e-02)& \textbf{5.14e-01} (1.3e-02)& 8.29e-02 (.)& 0.51\tabularnewline
&2.0& 3.22e-01 (3.3e-02)& 3.85e+00 (8.0e-02)& 3.94e+00 (7.4e-02)& 7.69e-01 (2.3e-02)& \textbf{7.36e-01} (2.2e-02)& 7.18e-04 (***)& 0.5\tabularnewline
\midrule
\multirow{6}{*}{Mat32-1}&0.1& 3.78e+01 (3.8e+00)& 8.48e-03 (1.6e-04)& 8.70e-03 (1.6e-04)& \textbf{1.50e-03} (5.1e-05)& 1.57e-03 (5.1e-05)& 1.71e-02 (*)& 0.4\tabularnewline
&0.2& 1.11e+01 (1.0e+00)& 3.50e-02 (6.3e-04)& 3.57e-02 (6.0e-04)& \textbf{4.32e-03} (2.0e-04)& 4.37e-03 (1.9e-04)& 2.68e-01& 0.47\tabularnewline
&0.5& 1.70e+00 (1.6e-01)& 2.32e-01 (3.6e-03)& 2.35e-01 (3.6e-03)& 1.95e-02 (1.1e-03)& \textbf{1.87e-02} (9.8e-04)& 1.28e-02 (*)& 0.48\tabularnewline
&1.0& 4.05e-01 (3.9e-02)& 9.54e-01 (1.4e-02)& 9.65e-01 (1.4e-02)& 6.44e-02 (4.2e-03)& \textbf{6.09e-02} (3.9e-03)& 5.44e-03 (**)& 0.57\tabularnewline
&1.5& 2.24e-01 (2.4e-02)& 2.17e+00 (3.9e-02)& 2.18e+00 (3.8e-02)& 1.13e-01 (1.1e-02)& \textbf{1.01e-01} (8.2e-03)& 1.09e-03 (**)& 0.6\tabularnewline
&2.0& 1.55e-01 (1.4e-02)& 3.79e+00 (6.3e-02)& 3.84e+00 (6.2e-02)& 1.94e-01 (1.6e-02)& \textbf{1.62e-01} (1.2e-02)& 4.61e-05 (***)& 0.74\tabularnewline
\midrule
\multirow{6}{*}{Mat32-0.1}&0.1& 1.30e+02 (6.8e+00)& 7.72e-03 (2.2e-04)& 9.77e-03 (3.8e-04)& \textbf{8.67e-03} (2.0e-04)& 9.84e-03 (3.0e-04)& 2.79e-05 (***)& 0.27\tabularnewline
&0.2& 4.25e+01 (2.6e+00)& 2.47e-02 (7.1e-04)& 2.88e-02 (8.5e-04)& \textbf{2.04e-02} (3.6e-04)& 2.15e-02 (5.4e-04)& 3.83e-03 (**)& 0.32\tabularnewline
&0.5& 5.49e+00 (2.8e-01)& 1.81e-01 (4.9e-03)& 2.01e-01 (5.2e-03)& \textbf{8.16e-02} (2.2e-03)& 8.51e-02 (2.4e-03)& 1.74e-03 (**)& 0.4\tabularnewline
&1.0& 1.34e+00 (6.7e-02)& 8.44e-01 (2.0e-02)& 9.01e-01 (1.9e-02)& 2.44e-01 (7.8e-03)& \textbf{2.40e-01} (7.6e-03)& 6.21e-02 (.)& 0.41\tabularnewline
&1.5& 6.73e-01 (4.4e-02)& 1.99e+00 (4.1e-02)& 2.07e+00 (3.9e-02)& 3.98e-01 (1.6e-02)& \textbf{3.82e-01} (1.3e-02)& 4.94e-03 (**)& 0.5\tabularnewline
&2.0& 3.67e-01 (3.0e-02)& 3.76e+00 (6.6e-02)& 3.87e+00 (6.2e-02)& 5.92e-01 (2.6e-02)& \textbf{5.64e-01} (2.0e-02)& 8.86e-03 (**)& 0.6\tabularnewline
\midrule
\multirow{6}{*}{RQ-0.1-0.5}&0.1& 1.23e+02 (6.6e+00)& 5.70e-03 (1.5e-04)& 6.56e-03 (1.9e-04)& \textbf{4.33e-03} (8.3e-05)& 4.71e-03 (1.0e-04)& 2.57e-07 (***)& 0.28\tabularnewline
&0.2& 3.44e+01 (2.0e+00)& 2.47e-02 (6.1e-04)& 2.66e-02 (6.3e-04)& \textbf{1.30e-02} (3.2e-04)& 1.34e-02 (3.3e-04)& 2.92e-03 (**)& 0.34\tabularnewline
&0.5& 4.27e+00 (2.4e-01)& 1.94e-01 (4.5e-03)& 2.10e-01 (5.0e-03)& \textbf{5.97e-02} (1.6e-03)& 6.17e-02 (1.7e-03)& 6.21e-02 (.)& 0.41\tabularnewline
&1.0& 1.07e+00 (6.2e-02)& 8.53e-01 (1.8e-02)& 8.92e-01 (1.7e-02)& 1.72e-01 (6.8e-03)& \textbf{1.70e-01} (6.1e-03)& 3.52e-01& 0.43\tabularnewline
&1.5& 4.43e-01 (3.3e-02)& 2.11e+00 (3.9e-02)& 2.15e+00 (3.8e-02)& 3.05e-01 (1.0e-02)& \textbf{3.02e-01} (1.0e-02)& 3.03e-01& 0.49\tabularnewline
&2.0& 3.07e-01 (2.4e-02)& 3.83e+00 (6.6e-02)& 3.92e+00 (6.6e-02)& 4.56e-01 (1.8e-02)& \textbf{4.43e-01} (1.6e-02)& 3.16e-02 (*)& 0.54\tabularnewline
\midrule
\multirow{6}{*}{Periodic-0.1-4}&0.1& 1.90e+02 (7.8e+00)& 5.22e-03 (1.3e-04)& 6.09e-03 (2.5e-04)& \textbf{4.37e-03} (9.4e-05)& 4.99e-03 (2.3e-04)& 1.19e-03 (**)& 0.34\tabularnewline
&0.2& 4.32e+01 (2.3e+00)& 2.35e-02 (4.9e-04)& 2.60e-02 (6.4e-04)& \textbf{1.45e-02} (3.6e-04)& 1.54e-02 (4.9e-04)& 2.14e-03 (**)& 0.3\tabularnewline
&0.5& 6.84e+00 (3.2e-01)& 1.66e-01 (4.2e-03)& 1.76e-01 (4.6e-03)& \textbf{6.68e-02} (1.8e-03)& 6.84e-02 (1.9e-03)& 3.44e-02 (*)& 0.51\tabularnewline
&1.0& 1.59e+00 (8.1e-02)& 7.92e-01 (1.9e-02)& 8.49e-01 (1.9e-02)& \textbf{2.27e-01} (7.4e-03)& 2.34e-01 (7.8e-03)& 5.08e-02 (.)& 0.48\tabularnewline
&1.5& 6.94e-01 (5.0e-02)& 1.96e+00 (4.3e-02)& 2.06e+00 (4.4e-02)& 4.40e-01 (1.4e-02)& \textbf{4.35e-01} (1.4e-02)& 2.60e-01& 0.55\tabularnewline
&2.0& 4.22e-01 (3.9e-02)& 3.62e+00 (8.5e-02)& 3.72e+00 (8.4e-02)& 6.51e-01 (2.3e-02)& \textbf{6.30e-01} (2.0e-02)& 1.19e-02 (*)& 0.62\tabularnewline
\bottomrule
\end{tabular}

    }
\end{table}



\subsubsection{Tune the shrinkage factor $k$}

In light of Proposition 6
, we tune the shrinkage factor $k$ instead of $(\lambda, \mu)$. It shows that the monotone decomposition with the shrinkage factor can achieve better performance in high noise cases for monotone functions such as $x^3, \exp(x)$ and sigmoid function. Promisingly, it also works for $x^2$ and SE\_1, although Proposition 6 
is intended for pure monotone functions.

\begin{figure}[H]
    \centering
    \includegraphics[width=\textwidth]{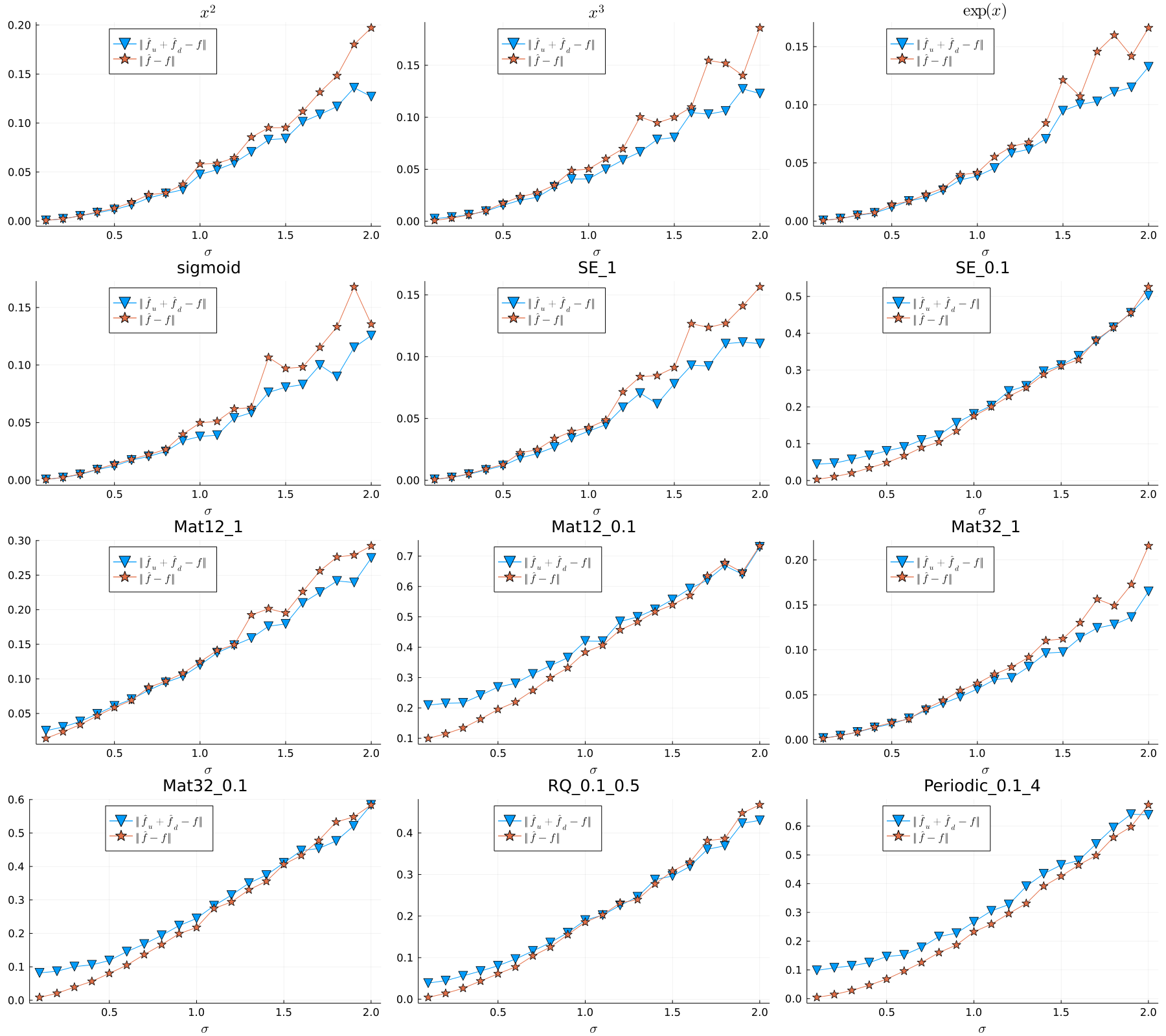}
    \caption[MSE for Monotone Decomposition with Smoothing Splines via $k$]{Results for comparing the smoothing splines and the fitting by monotone decomposition where the parameters are tuning by varying the shrinkage factor $k$. The values are averages over 100 replications. }
\end{figure}

\subsection{QQ plots of p-values}

\begin{figure}[H]
\centering
\includegraphics[width=0.8\textwidth]{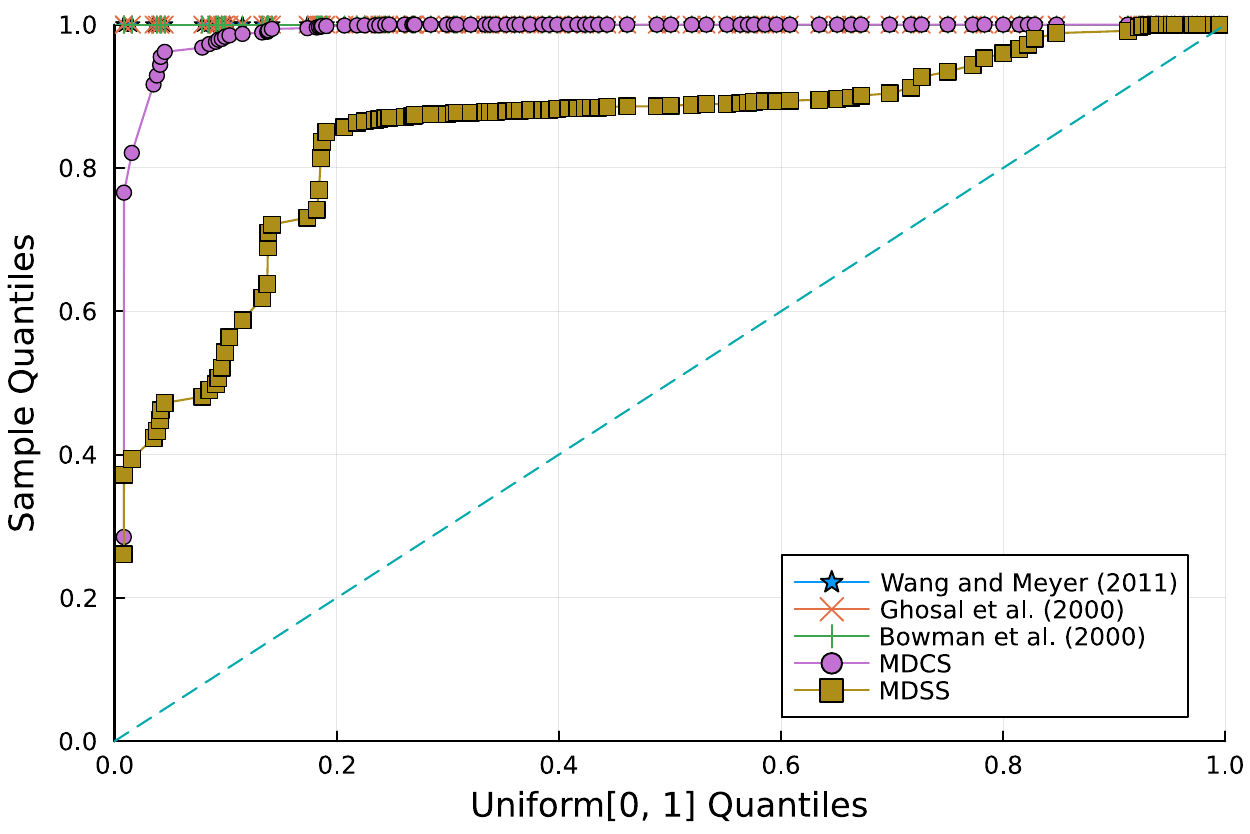}       
\caption{Uniform QQ plot of $p$-values for five approaches on curve $x$ with $n=200, \sigma=0.01$.}
\end{figure}

\begin{figure}[H]
\centering
\includegraphics[width=0.8\textwidth]{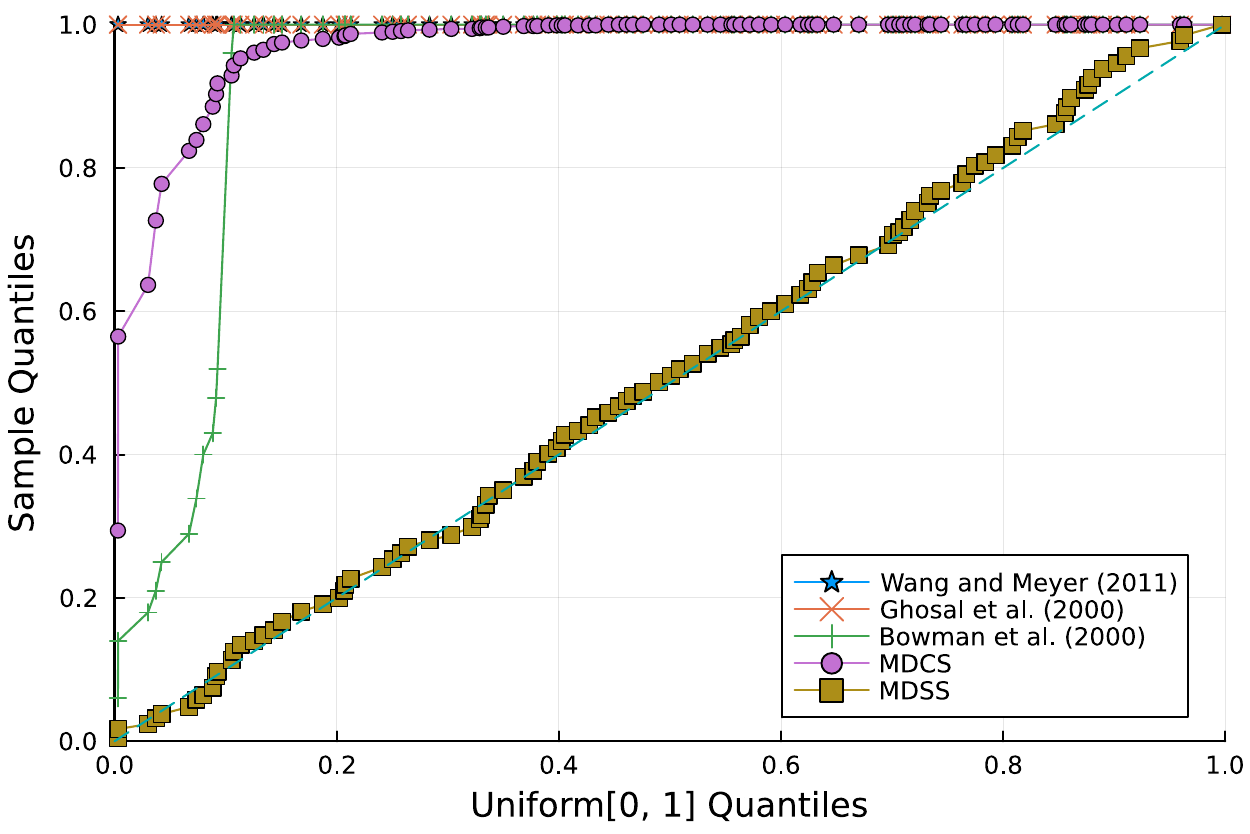}       
\caption{Uniform QQ plot of $p$-values for five approaches on curve $x^{1/3}$ with $n=200,\sigma=0.01$.}
\end{figure}

\begin{figure}[H]
\centering
\includegraphics[width=0.8\textwidth]{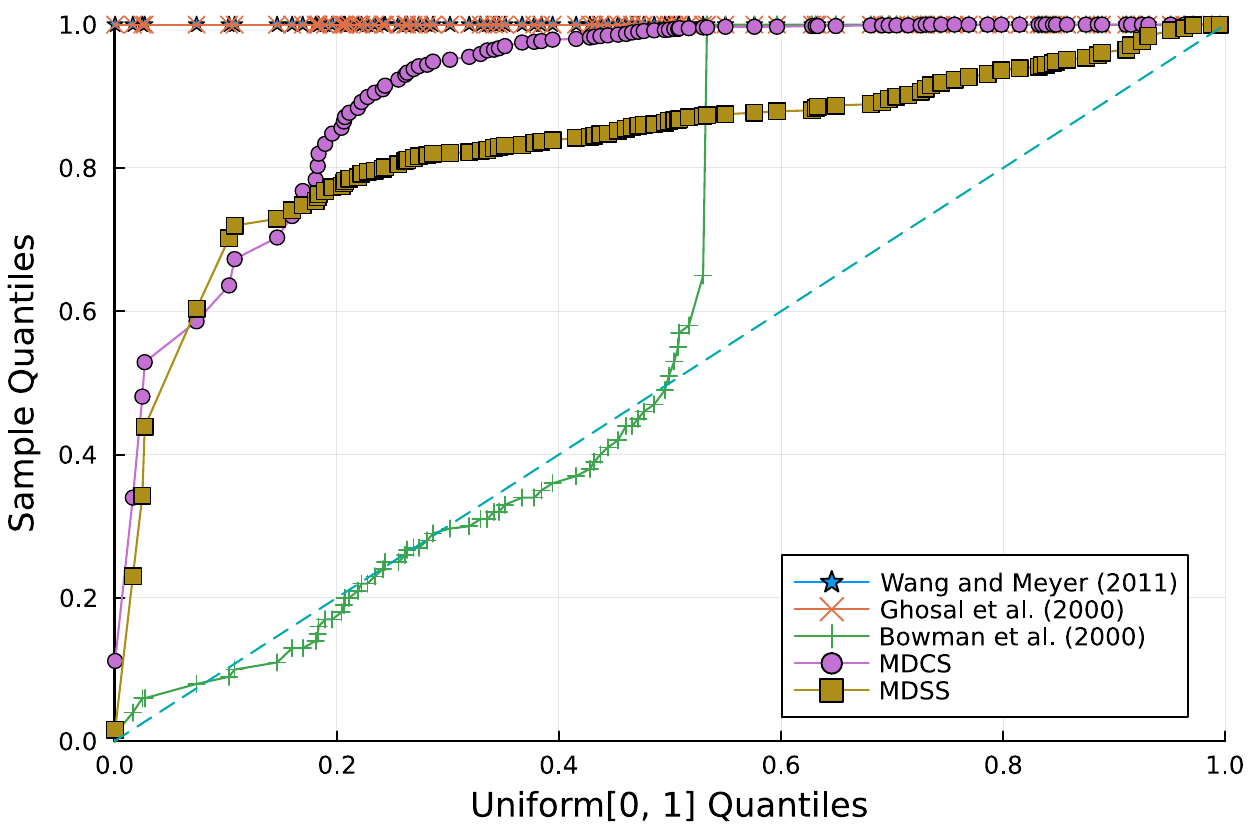}       
\caption{Uniform QQ plot of $p$-values for five approaches on curve $1/(1+e^{-x})$ with $n=200,\sigma=0.01$.}
\end{figure}

\begin{figure}[H]
\centering
\includegraphics[width=0.8\textwidth]{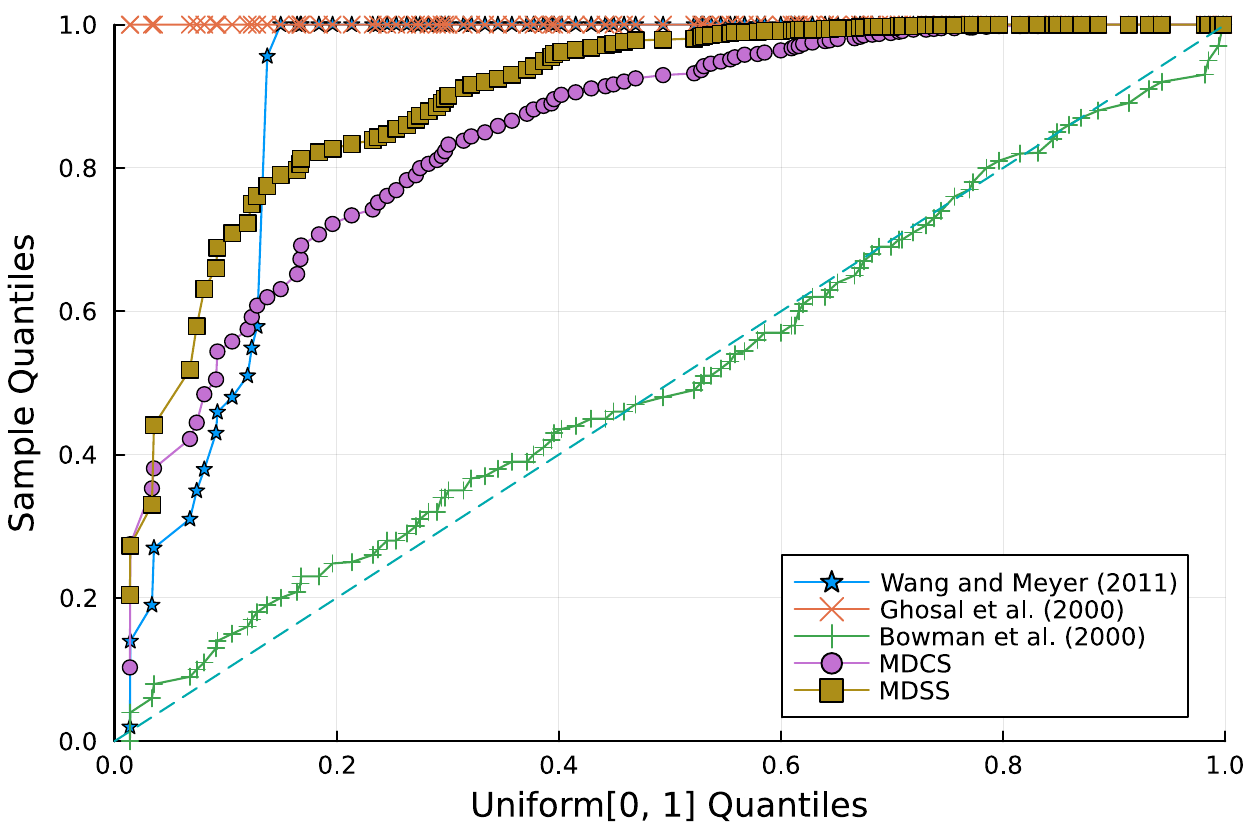}       
\caption{Uniform QQ plot of $p$-values for five approaches on curve $x$ with $n=200, \sigma=0.1$.}
\end{figure}

\begin{figure}[H]
\centering
\includegraphics[width=0.8\textwidth]{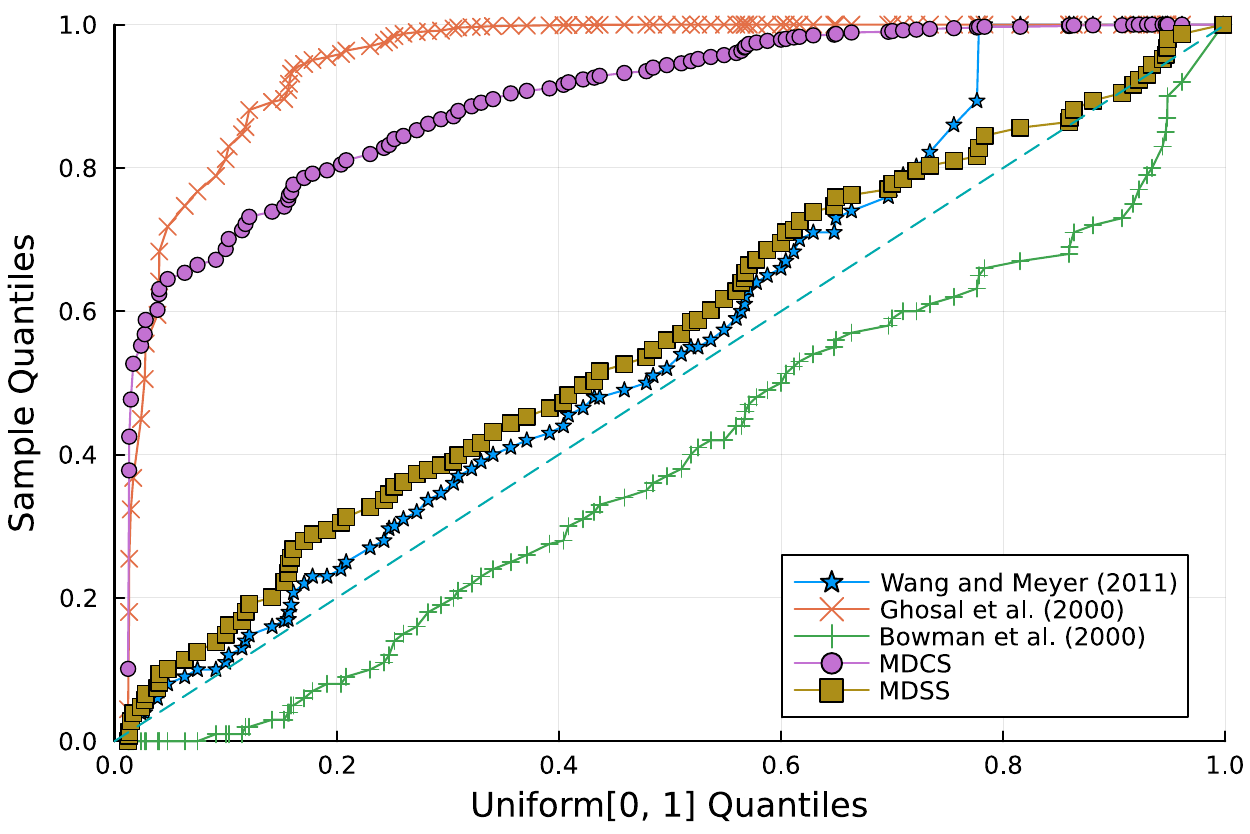}
\caption{Uniform QQ plot of $p$-values for five approaches on curve $x^3$ with $n=200, \sigma=0.1$.}
\end{figure}

\begin{figure}[H]
\centering
\includegraphics[width=0.8\textwidth]{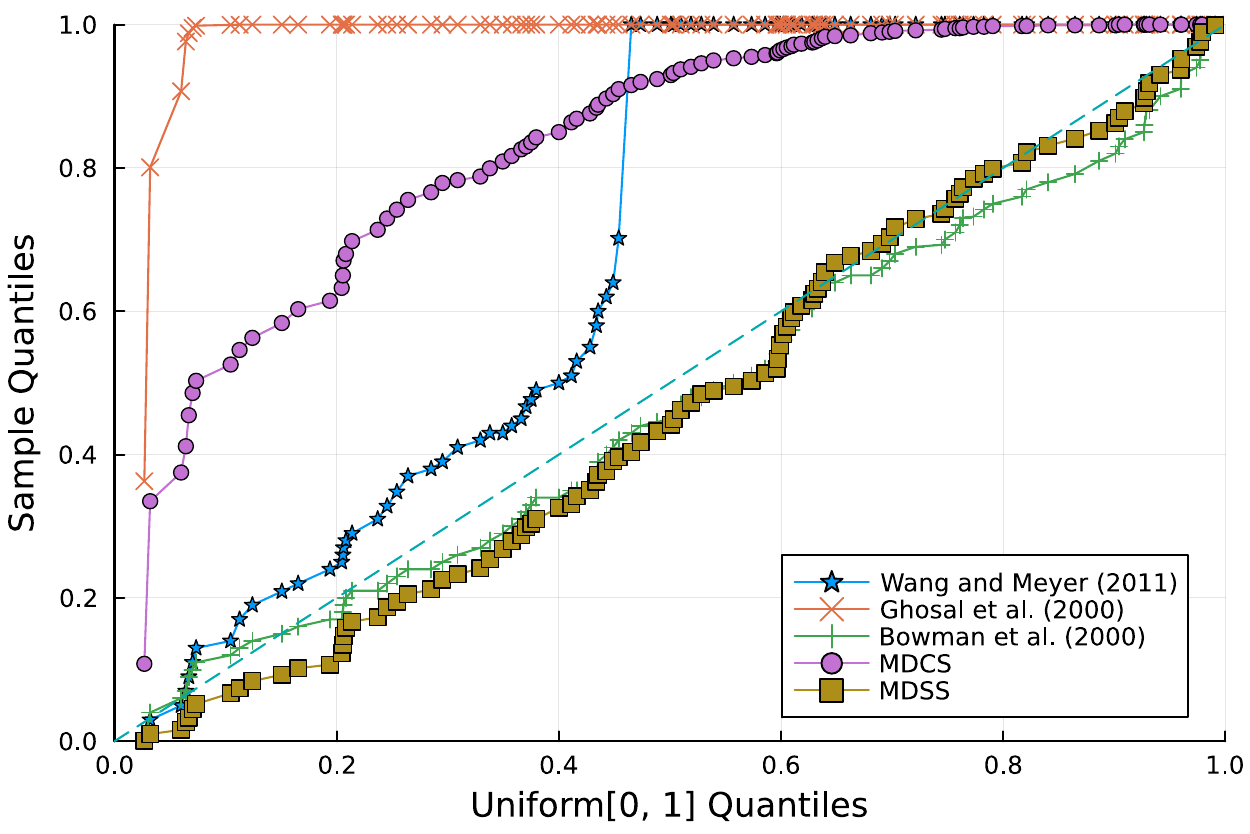}       
\caption{Uniform QQ plot of $p$-values for five approaches on curve $x^{1/3}$ with $n=200, \sigma=0.1$.}
\end{figure}

\begin{figure}[H]
\centering
\includegraphics[width=0.8\textwidth]{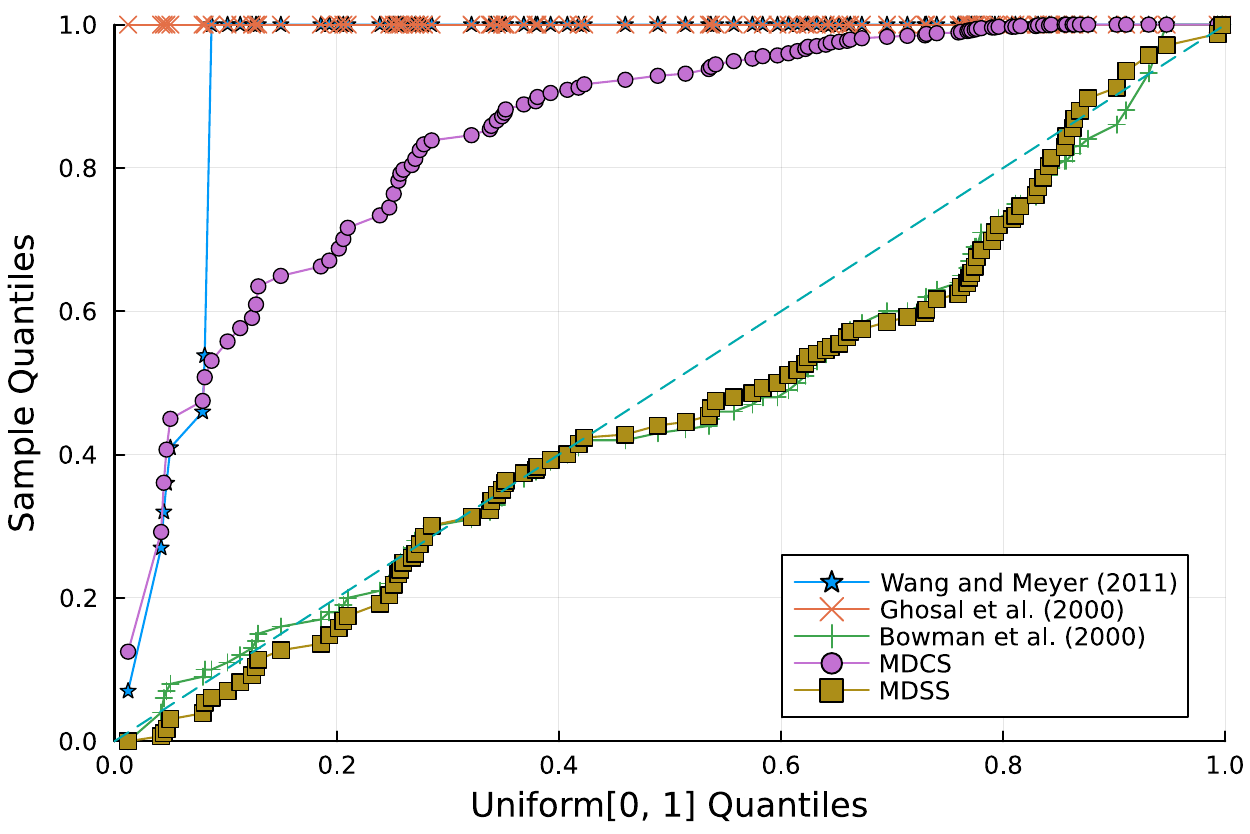}       
\caption{Uniform QQ plot of $p$-values for five approaches on curve $e^x$ with $n=200, \sigma=0.1$.}
\end{figure}

\begin{figure}[H]
\centering
\includegraphics[width=0.8\textwidth]{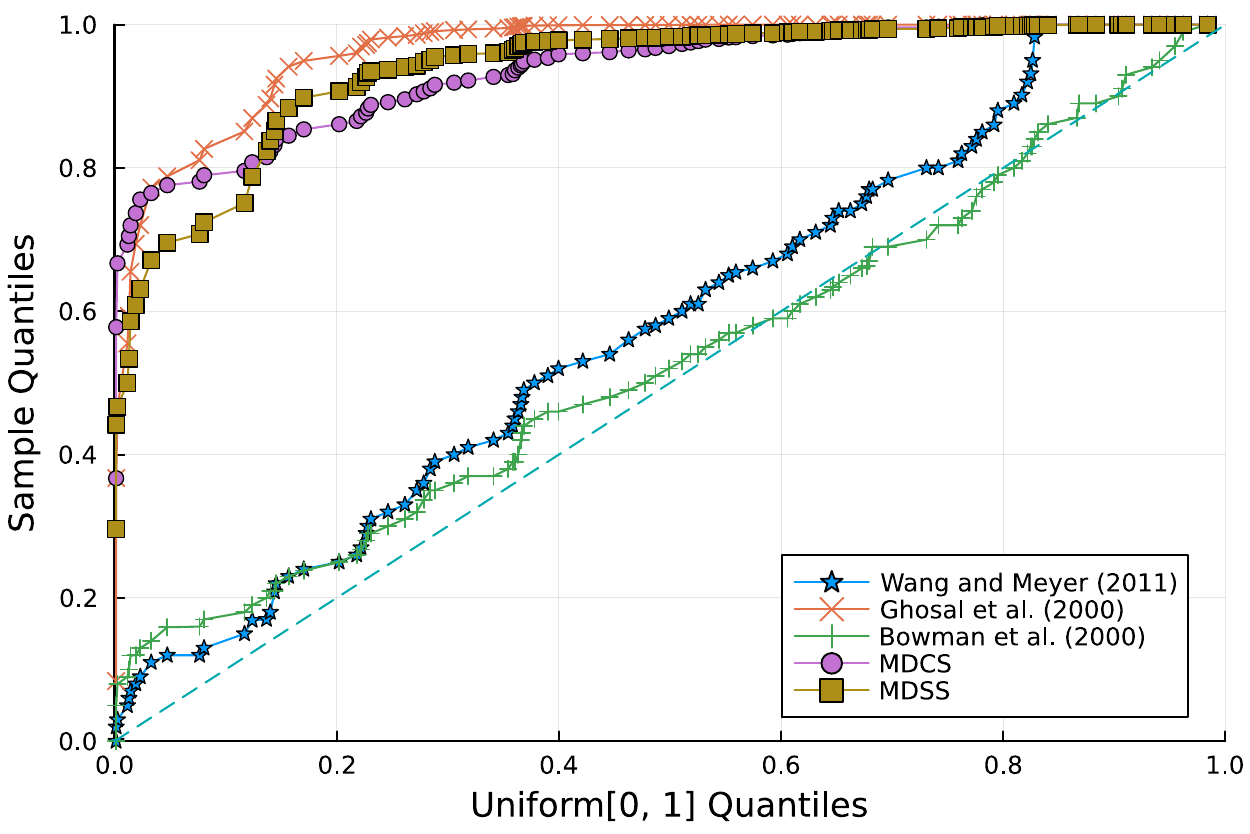}       
\caption{Uniform QQ plot of $p$-values for five approaches on curve $1/(1+e^{-x})$ with $n=200,\sigma=0.1$.}
\end{figure}

\subsection{Different Curves for Testing the Monotonicity}

\begin{figure}[H]
    \centering
    \begin{subfigure}{0.5\textwidth}
    \includegraphics[width=\textwidth]{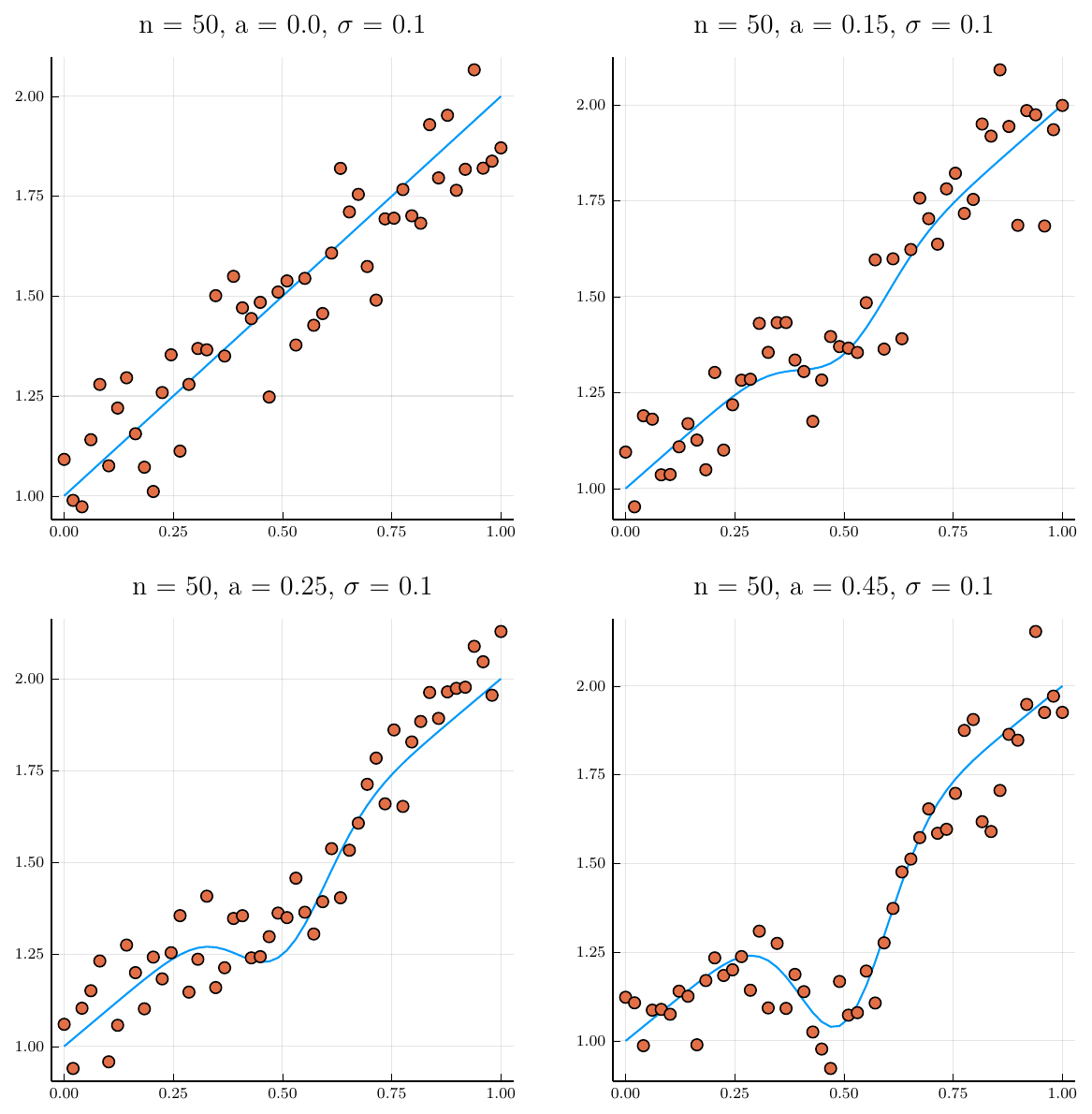}
    \caption{\textcite{bowmanTestingMonotonicityRegression1998}}
    \end{subfigure}%
    \begin{subfigure}{0.5\textwidth}
    \includegraphics[width=\textwidth]{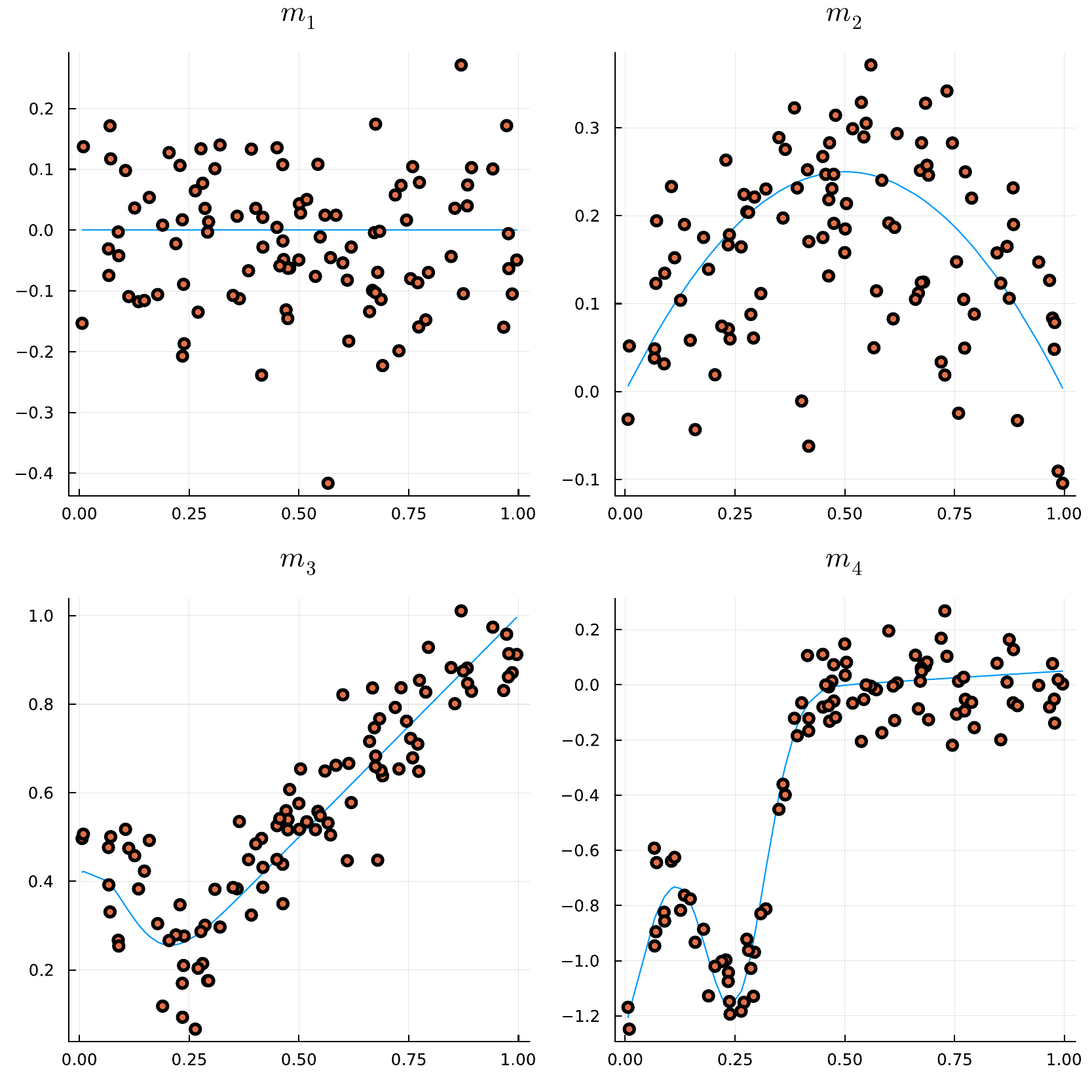}
    \caption{\textcite{ghosalTestingMonotonicityRegression2000}}
    \end{subfigure}
    \caption[Different Curves for Testing the Monotonicity]{Different curves for testing the monotonicity.}
    \label{fig:sim-monotest}
\end{figure}

\section{More GO Analysis}

\begin{figure}[H]
    \centering
    \begin{subfigure}{0.5\textwidth}
        \includegraphics[width=\textwidth]{res2/ego-paul-100000-2023-10-17T22_36_11+08_00-nfold5-J6-1se_0.05-L1_1.pdf}
        \caption{}
        \label{fig:paul1_1}
    \end{subfigure}%
    \begin{subfigure}{0.5\textwidth}
        \includegraphics[width=\textwidth]{res2/ego-paul-100000-2023-10-17T22_36_11+08_00-nfold5-J6-1se_0.05-L1_2.pdf}
        \caption{}
        \label{fig:paul1_2}
    \end{subfigure}
    \begin{subfigure}{0.5\textwidth}
        \includegraphics[width=\textwidth]{res2/ego-paul-100000-2023-10-17T22_36_11+08_00-nfold5-J6-1se_0.05-L1_9.pdf}
        \caption{}
        \label{fig:paul1_9}
    \end{subfigure}%
    \begin{subfigure}{0.5\textwidth}
        \includegraphics[width=\textwidth]{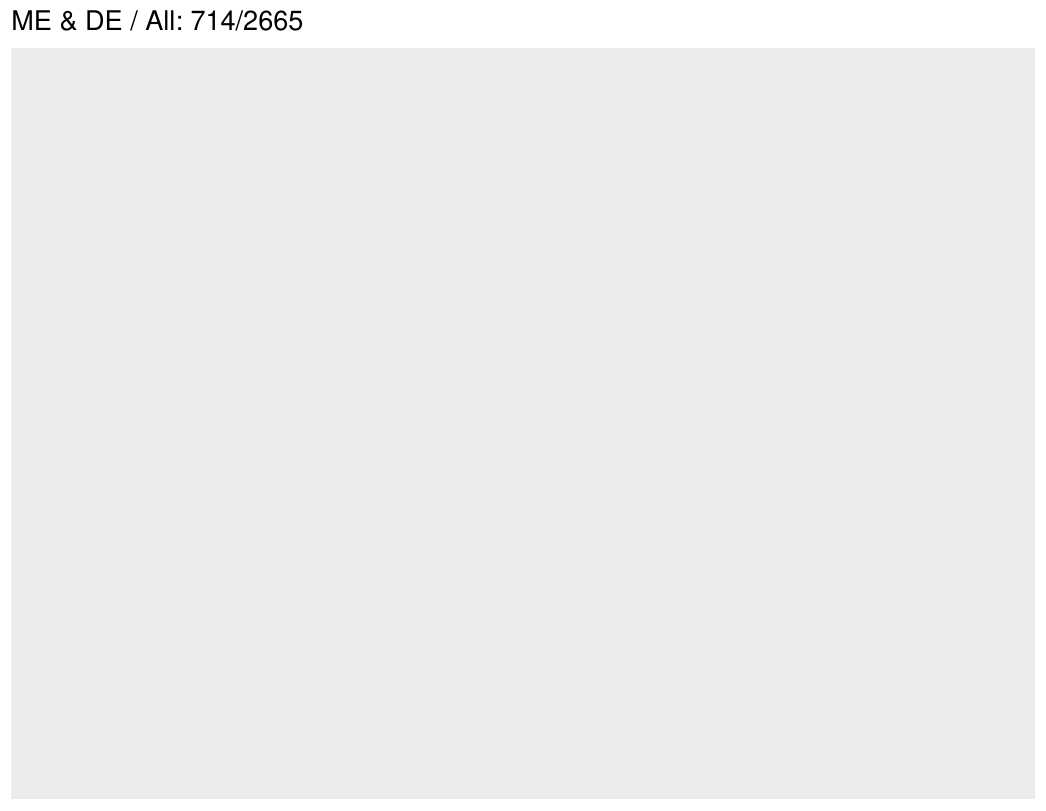}
        \caption{}
        \label{fig:paul1_10}
    \end{subfigure}

    \caption{GO enrichment analysis on the \textcite{paulTranscriptionalHeterogeneityLineage2015} dataset of four gene sets (\emph{(a)}: nME gene set; \emph{(b)}: DE gene set; \emph{(c)}: the intersection of nME and DE genes; \emph{(d)}: the intersection of ME and DE genes) with BH FDR cutoff $\alpha = 0.05$. The numbers of genes are noted in the title of each subfigure. The enriched GO terms are sorted by the adjusted $p$-value. \emph{(b)} and \emph{(d)} are left empty deliberately since no significant GO terms are selected.}
    \label{fig:paul1}
\end{figure}

Figure~\ref{fig:paul1} is the same as Figure 8
except that panel (d) displays the GO terms identified by the interaction of DE genes and ME genes (the complementary of nME genes). Similar to DE genes, the interaction of DE genes and ME genes failed to identify significant GO terms.
 
\begin{figure}[H]
    \centering
    \begin{subfigure}{0.5\textwidth}
        \includegraphics[width=\textwidth]{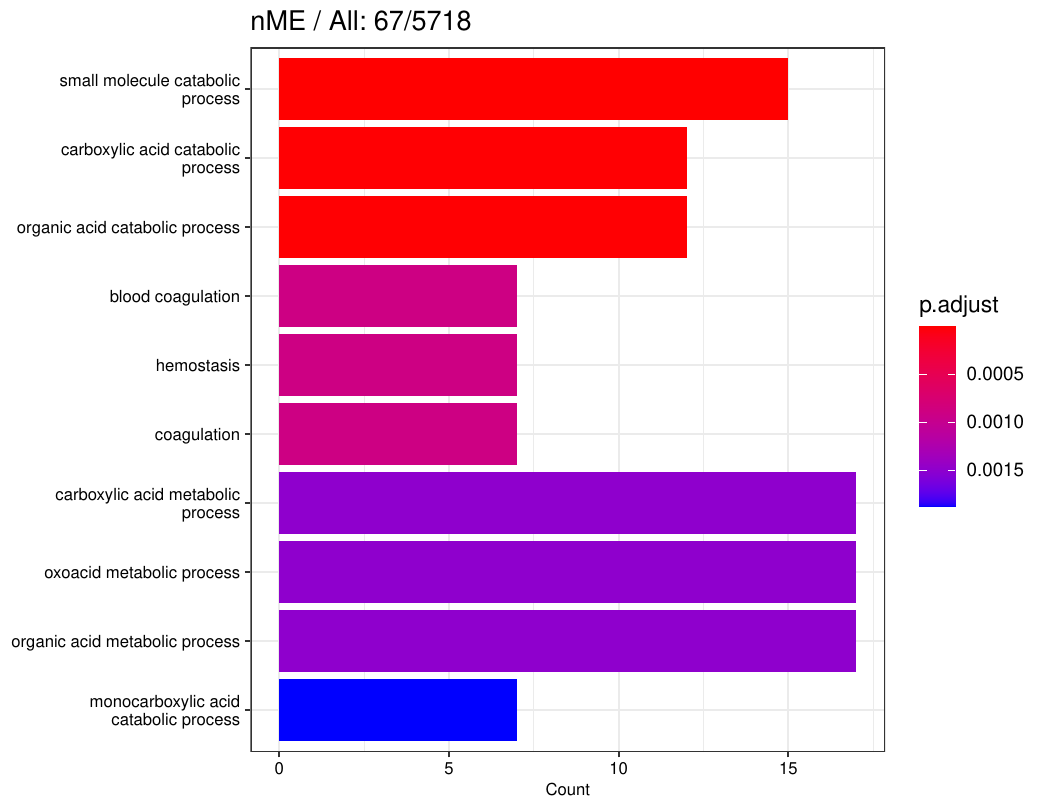}
        \caption{}
        \label{fig:liver2_1}
    \end{subfigure}%
    \begin{subfigure}{0.5\textwidth}
        \includegraphics[width=\textwidth]{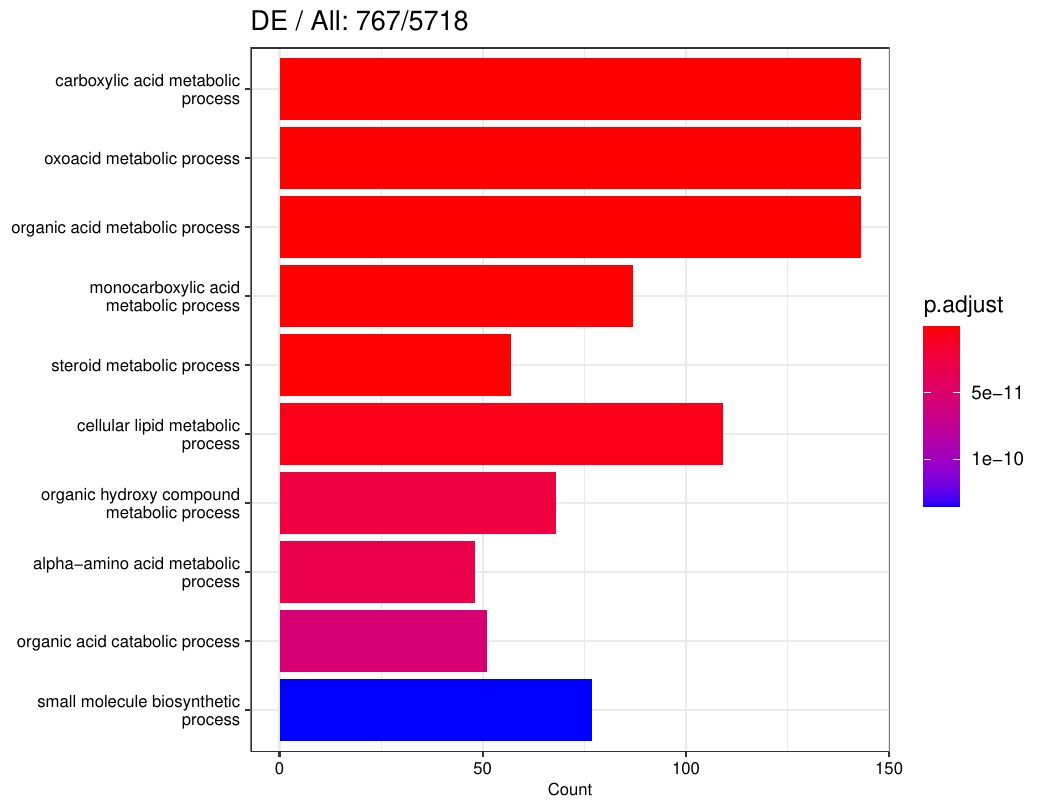}
        \caption{}
        \label{fig:liver2_2}
    \end{subfigure}
    \begin{subfigure}{0.5\textwidth}
        \includegraphics[width=\textwidth]{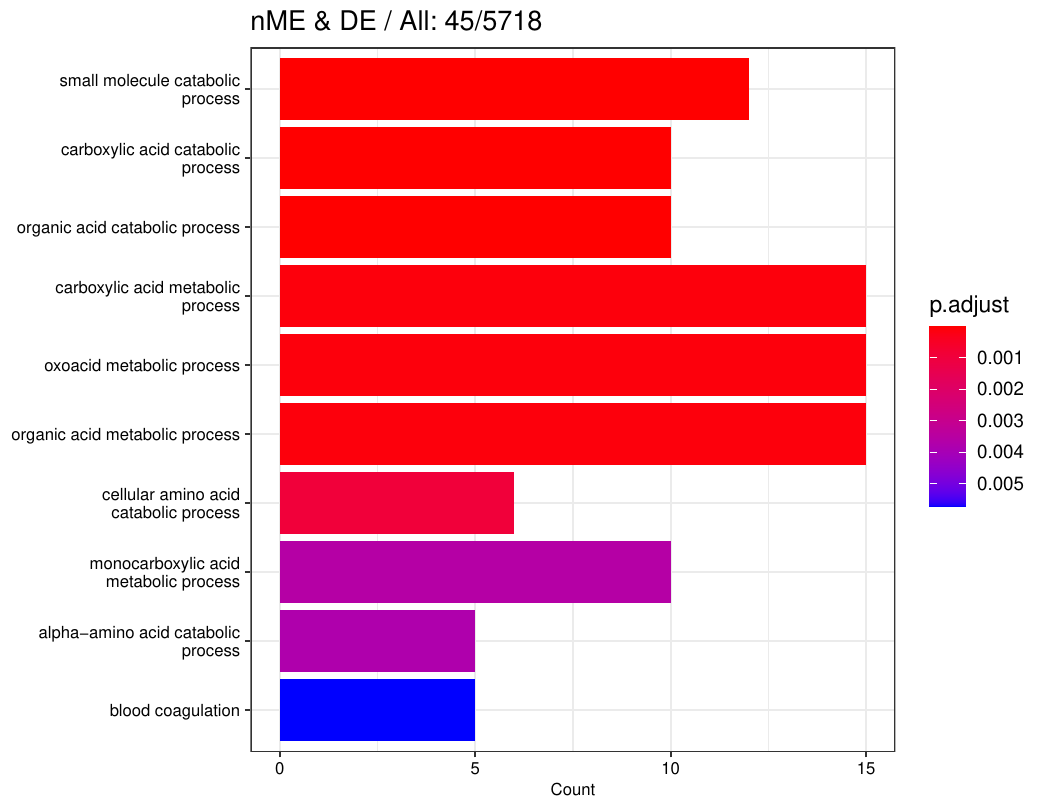}
        \caption{}
        \label{fig:liver2_9}
    \end{subfigure}%
    \begin{subfigure}{0.5\textwidth}
        \includegraphics[width=\textwidth]{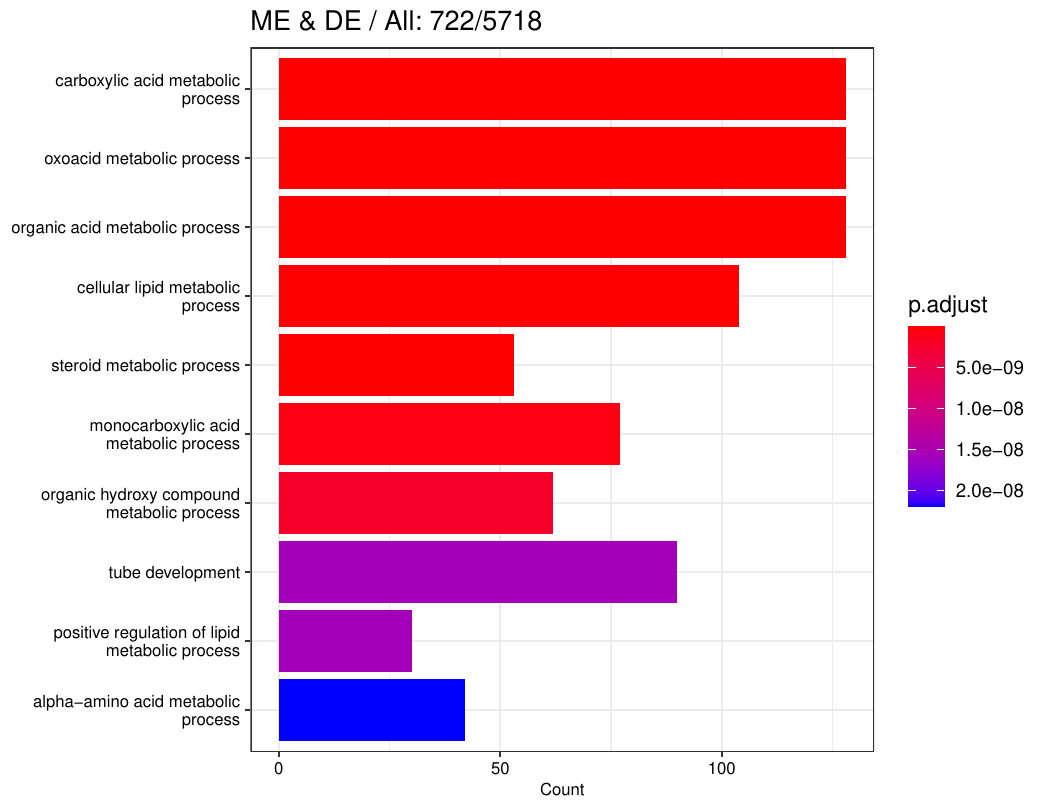}
        \caption{}
        \label{fig:liver2_10}
    \end{subfigure}

    \caption{GO enrichment analysis on the liver dataset of four gene sets (\emph{(a)}: nME gene set; \emph{(b)}: DE gene set; \emph{(c)}: the intersection of nME and DE genes; \emph{(d)}: the intersection of ME and DE genes) with BH FDR cutoff $\alpha = 0.05$. The numbers of genes are noted in the title of each subfigure. The enriched GO terms are sorted by the adjusted $p$-value. \emph{(b)} is left empty deliberately since no significant GO terms are selected.}
    \label{fig:liver2}
\end{figure}

\begin{figure}[H]
    \centering
    \includegraphics[width=0.5\textwidth]{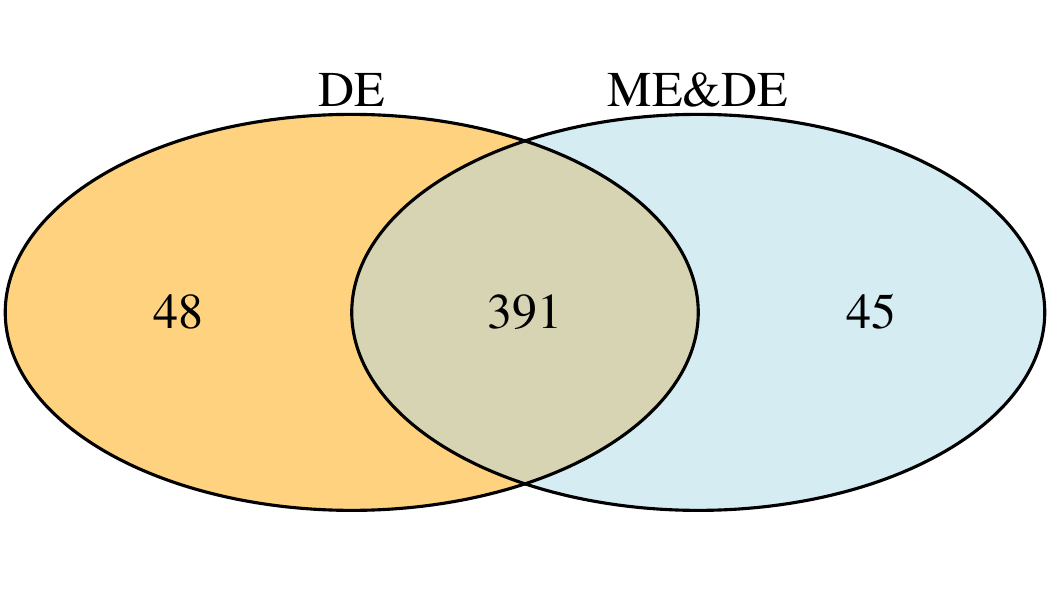}
    \caption{Venn diagram of GO terms identified by DE genes and DE\&ME genes.}
    \label{fig:venn-me}
\end{figure}

For the liver dataset, the GO terms identified by different gene sets are shown in Figure~\ref{fig:liver2}. Most GO terms selected by the DE genes (Figure~\ref{fig:liver2_2}) and by the ME\&DE genes (Figure~\ref{fig:liver2_10}) are the same. The numbers of different GO terms are illustrated in the Venn diagram of Figure~\ref{fig:venn-me}.

\begin{figure}[H]
    \centering
    \includegraphics[width=0.8\textwidth]{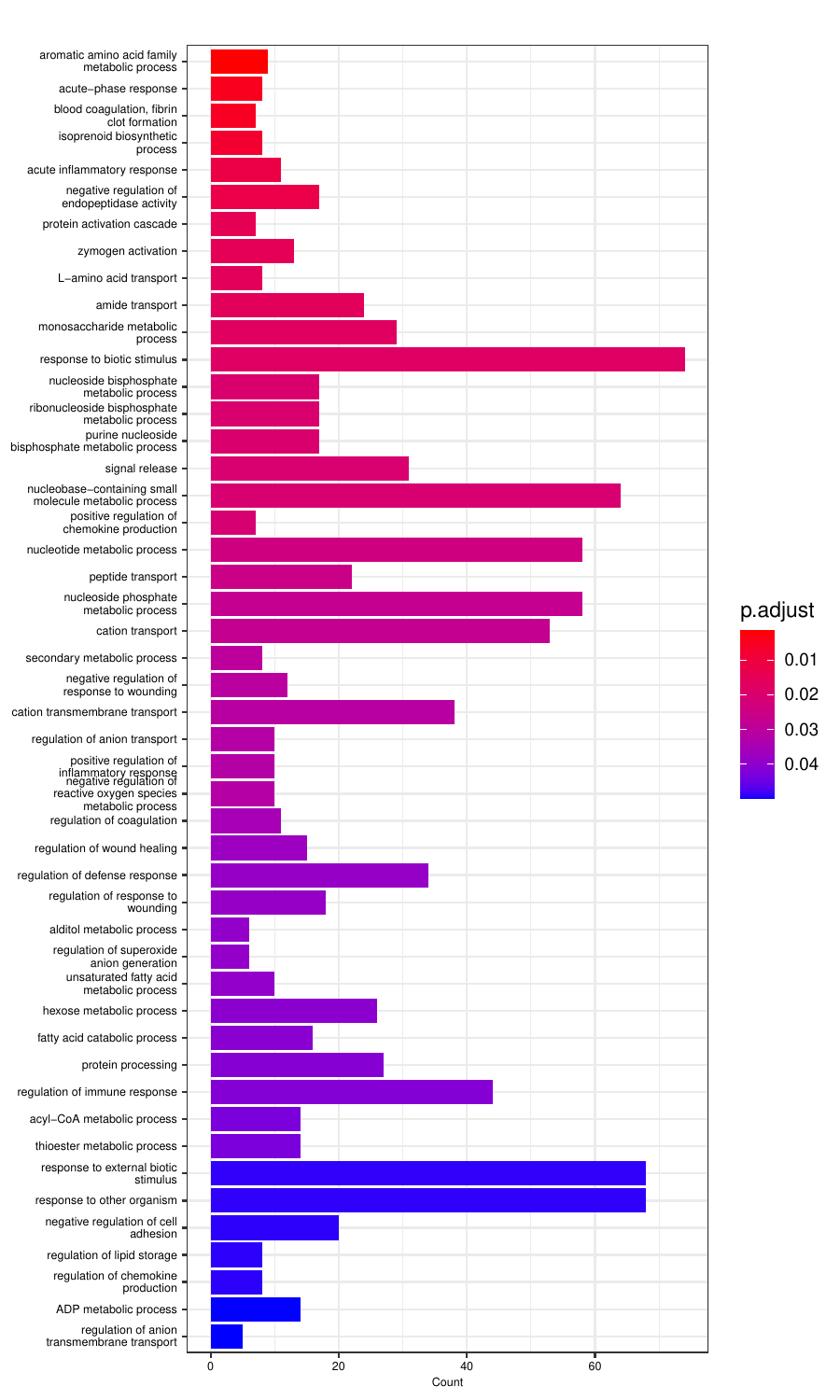}
    \caption{GO terms identified by DE genes but not ME genes.}
    \label{fig:go2vs10}
\end{figure}

\begin{figure}[H]
    \centering
    \includegraphics[width=0.8\textwidth]{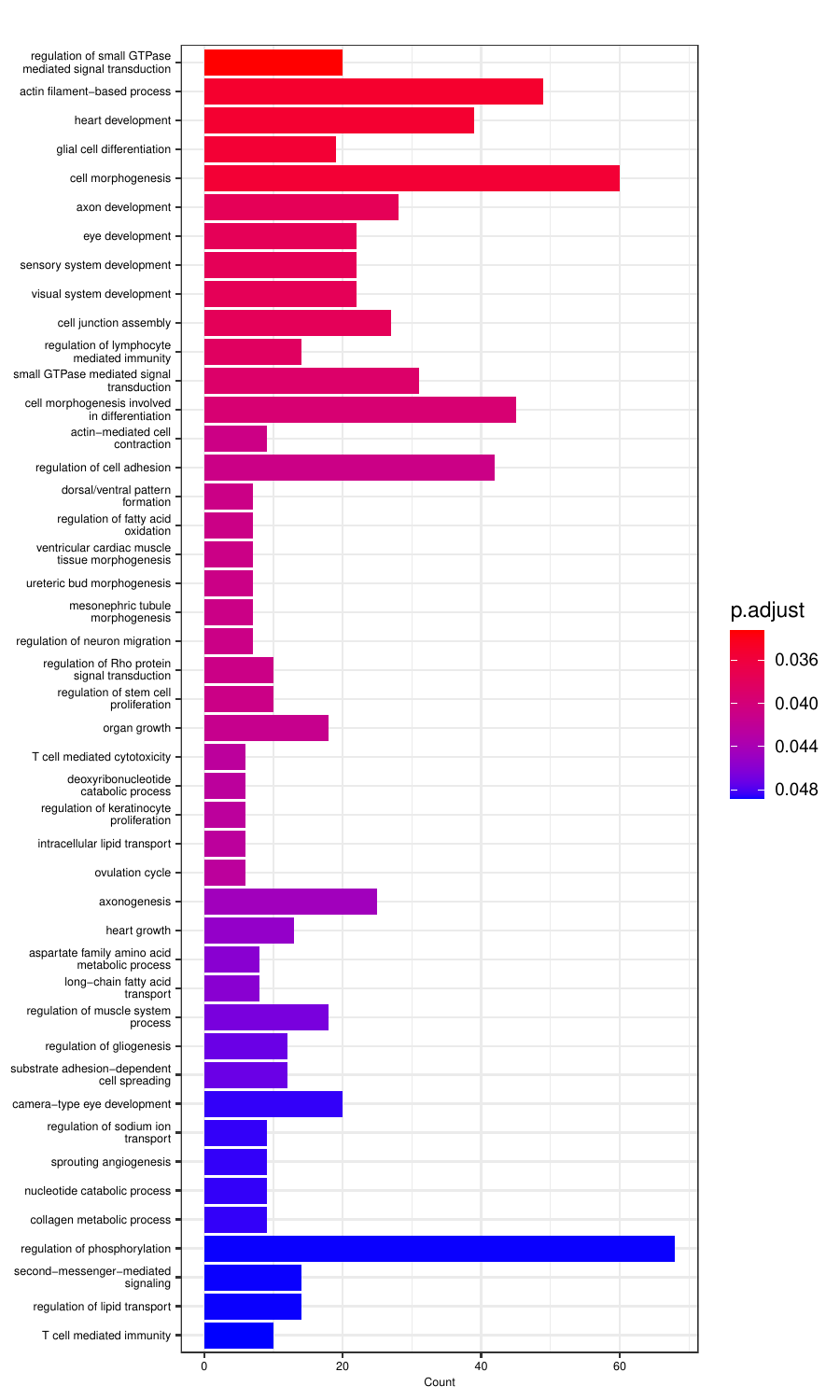}
    \caption{GO terms identified by ME genes but not DE genes.}
    \label{fig:go10vs2}
\end{figure}

Figures~\ref{fig:go2vs10} and \ref{fig:go10vs2} display the distinct GO terms selected by DE genes and ME genes, respectively. Note that the counts of GO terms and associated $p$-values are relatively small compared to their common GO terms.

\section{Theoretical proofs of some basic propositions}
\subsection{A proposition}
\begin{proposition}
    Suppose $f(x)$ is a univariate continuous function in an interval $I$, there exists a decomposition
\begin{equation}
f(x) = f_\up(x) + f_\down(x).
\label{eq:decomp}
\end{equation}
    where $f_\up(x)$ and $f_\down(x)$ are continuous increasing and decreasing functions, respectively.
\end{proposition}

\begin{proof}
    Decompose the interval $I$ as
    $$
    I = \cup_{k=1}^n [a_{2k-1}, a_{2k}]\,.
    $$
    Without loss of generality, suppose $f(x)$ is increasing in $[a_{2k-1}, a_{2k}]$ and decreasing in $[a_{2k}, a_{2k+1}]$. Then one feasible decomposition is
    $$
    f_\up(x) = \begin{cases}
    f(x) & x\in [a_1, a_2]\\
    f(a_2) & x\in [a_2, a_3]\\
    f(x) - f(a_3) + f(a_2) & x\in [a_3, a_4]\\
    f(a_4)-f(a_3) + f(a_2) & x\in [a_4, a_5]\\
    \cdots & \cdots\\
    f(x) + \sum_{i=2}^k (f(a_{2i-2}) - f(a_{2i-1}))& x\in [a_{2k-1}, a_{2k}], k\ge 2\\
    f(a_{2k}) + \sum_{i=2}^k (f(a_{2i-2}) - f(a_{2i-1}))& x\in [a_{2k}, a_{2k+1}], k\ge 2
    \end{cases}\,,
    $$
    and $f_\down(x)=f(x) - f_\up(x)$.
\end{proof}

\subsection{Proposition 1}

\begin{proof}
Consider the problem
\begin{equation}
    \begin{split}
\min_{\gamma^u,\gamma^d}&\;\Vert y-\bB(\gamma^u + \gamma^d)\Vert_2\\
\text{s.t. }&\gamma_1^u\le \gamma_2^u\le\cdots\le\gamma_J^u;\; \gamma_1^d\ge \gamma_2^d\ge\cdots\ge\gamma_J^d\\
&\gamma^u+\gamma^d =\hat\gamma\,,
    \end{split}
\end{equation}
where $\hat\gamma$ is the unconstrained solution. Since we can always decompose a vector into the sum of an increasing vector and a decreasing vector. For $2\le i\le J$, let
\begin{align}
\hat\gamma^u_i &= \hat\gamma_{i-1}^u + (\hat\gamma_i-\hat\gamma_{i-1})_+\\
\hat\gamma^d_i &= \hat\gamma_{i-1}^d + (\hat\gamma_i-\hat\gamma_{i-1})_-\,,
\end{align}
where $t_+=\max(0, t)$ and $t_-=\min(0, t)$. And 
$$
\hat\gamma^u_1+\hat\gamma^d_1 = \hat\gamma_1\,,
$$
for example, take $\hat\gamma_1^u=\hat\gamma_1$ and $\hat\gamma_1^d=0$. Thus the constraints are feasible, i.e., non-empty, and then any solution would satisfy
$$
\hat\gamma^u+\hat\gamma^d = \hat\gamma\,.
$$
\end{proof}

\subsection{Proposition 2}\label{thm:proof_exist_ties}

\begin{proof}
    
Here we only prove the second property since the first property has been incorporated into the proof for the main propositions in Sections~\ref{sec:proof_md_cs_increase2} and \ref{sec:proof_md_ss_increase2}.

We can also put the constraint $\gamma\in\Upsilon$ into the Lagrangian form.
Define
$$
\bA \triangleq \begin{bmatrix}
    \bI_{J-1} & 0
\end{bmatrix} - \begin{bmatrix}
    0 &\bI_{J-1}
\end{bmatrix} =
\begin{bmatrix}
1 & -1 & 0 & \cdots & 0 & 0\\
0 & 1 & -1 &\cdots & 0 & 0\\
0 & 0 & 1 & \cdots & 0 & 0\\
\vdots & \vdots & \vdots & \ddots & \vdots & \vdots\\
0 & 0 & 0 & \cdots & 1 & -1
\end{bmatrix}_{(J-1)\times J}\,,\quad
\bH \triangleq \begin{bmatrix}
\bA & \zero\\
\zero & -\bA
\end{bmatrix}\,,
$$
and the Lagrange form is
\begin{equation}
L_\ls(\gamma, \nu, \mu) = \Vert \bfy-\bZ\gamma\Vert_2^2 + 2\nu^T\bH\gamma + \mu\gamma^T\bW\gamma\,,
\end{equation}
where $2(J-1)$-vector $\nu$ and scalar $\mu$ are the Lagrange multipliers, and factor $2$ is just for convenient formulas after taking the first derivatives.
Take the derivatives on the Lagrange form, 
$$
-2\bZ^T(y-\bZ\gamma) + 2\bH^T\nu + 2\mu\bW\gamma + 2\lambda \begin{bmatrix}
\bOmega & \bOmega\\
\bOmega & \bOmega
\end{bmatrix}\gamma = 0\,,
$$
where if $\lambda=0$, it reduces to the cubic spline. Then
\begin{align}
-\bB^T_i(\bfy-\bZ\gamma) + \bH_{i}^T\nu + \mu \bB_i^T\bB(\gamma^u-\gamma^d) +\lambda\bOmega_i(\gamma^u+\gamma^d) = 0\\
-\bB^T_i(\bfy-\bZ\gamma) + \bH_{i+J}^T\nu + \mu \bB_i^T\bB(-\gamma^u+\gamma^d)+\lambda\bOmega_i(\gamma^u+\gamma^d) = 0
\end{align}
Note that
$$
\sum_{i=1}^J\bH_{i}^T\nu = 0\,,
$$
and
$$
\sum_{i=1}^J\bH_{i+J}^T\nu = 0\,,
$$
then we have
$$
\mu\one^T\bB^T\bB(\gamma^u-\gamma^d)=\mu\one^T\bB^T\bB(-\gamma^u+\gamma^d)\,,
$$
which implies
$$
\mu \one^T\bB(\gamma^u - \gamma^d) = \mu \one^T\bB(-\gamma^u + \gamma^d)\,,
$$
so
$$
2\mu\one^T\bB\gamma^u = 2\mu\one^T\bB\gamma^d\,.
$$
If $\mu = 0$, by KKT condition, then $\bB\gamma^u -\bB\gamma^d = 0$. If $\mu \neq 0$, then $\one^T\bB\gamma^u = \one^T\bB\gamma^d$.

Thus, $\one^T\bB\gamma^u = \one^T\bB\gamma^d$ always holds.
\end{proof}

\section{Proof of Proposition 3}
\label{sec:proof_md_cs_sol}

\begin{lemma}[Chebyshev Sum Inequality]\label{fact:chebyshev_sum_ieq}
If $a_1\ge a_2\ge\cdots\ge a_n$, and $b_1\ge b_2\ge\cdots\ge b_n$, then
$$
n\sum_{k=1}^na_kb_k \ge \left(\sum_{k=1}^na_k\right)\left(\sum_{k=1}^nb_k\right)
$$
\end{lemma}

\begin{proof}
Since these two sequences are decreasing, then the sum
$$
S = \sum_{j=1}^n\sum_{k=1}^n(a_j-a_k)(b_j-b_k)\ge 0\,.
$$
Note that
\begin{align*}
    S &= \sum_{j=1}^n\left(na_jb_j-b_j\sum_{k=1}^na_k-a_j\sum_{k=1}^nb_k+\sum_{k=1}^na_kb_k\right)\\
    &=n\sum_{j=1}^na_jb_j +n\sum_{k=1}^na_kb_k - \sum_{j=1}^nb_j\sum_{k=1}^na_k - \sum_{j=1}^na_j\sum_{k=1}^nb_k\\
    &=2n\sum_{k=1}^na_kb_k - 2\sum_{k=1}^nb_k\sum_{k=1}^na_k\,,
\end{align*}
hence 
$$
n\sum_{k=1}^na_kb_k \ge \left(\sum_{k=1}^na_k\right)\left(\sum_{k=1}^nb_k\right)\,.
$$
\end{proof}

\subsection{(i)} \label{sec:md_cs_constant}

\begin{proof}
Suppose $\gamma = \gamma^u+\gamma^d$ is increasing, then we have a decomposition in which $\gamma^d$ is a vector with identical elements. Write the decomposition as $\gamma=\gamma^u+c\one$, where $c$ is a constant. If there exists a non-zero non-decreasing vector $\delta$ such that the non-increasing part is not a constant, i.e.,
$$
\gamma = (\gamma^u + \delta) + (c\one-\delta) \triangleq \tilde\gamma^u +\tilde\gamma^d\,.
$$
Then
$$
\Vert y-\bB(\gamma^u+c\one)\Vert = \Vert y-\bB(\tilde\gamma^u+\tilde\gamma^d)\Vert\,,
$$
and if we impose the roughness penalty, we have
$$
(\gamma^u+\gamma^d)^T\bOmega(\gamma^u+\gamma^d) = (\tilde\gamma^u+\tilde\gamma^d)^T\bOmega(\tilde\gamma^u+\tilde\gamma^d)\,.
$$
So the first two terms in the Lagrange objective function (13) are the same, and we only need to compare the difference between these two components,
\begin{align*}
    &\Vert \bB(\tilde\gamma^u - \tilde\gamma^d)\Vert_2^2 - \Vert \bB(\gamma^u - \gamma^d)\Vert_2^2 \\
    =&(\tilde\gamma^u-\tilde\gamma^d- (\gamma^u-\gamma^d))^T\bB^T\bB(\tilde\gamma^u-\tilde\gamma^d+(\gamma^u-\gamma^d))\\
    =&2\delta^T\bB^T\bB(2\gamma^u+2\delta-2c\one)\\
    =&4\left\{\delta^T\bB^T\bB\delta + \delta^T\bB^T\bB(\gamma^u-c\one) \right\}\,.
\end{align*}
By Proposition 2
, we have
$$
\one^T\bB\gamma^u=\one^T\bB \gamma^d\,\qquad\one^T\bB\tilde\gamma^u = \one^T\bB\tilde\gamma^d\,,
$$
then
$$
\one^T\bB(\gamma^u-c\one) = \one^T\bB\delta=0\,.
$$
Let $\bfa = \bB\delta,\bfb = \bB(\gamma^u-c\one)$, then by Fact \ref{fact:chebyshev_sum_ieq}, 
$$
\bfa^T\bfb \ge \frac 1n(\one^T\bfa)(\one^T\bfb) = 0\,.
$$
And $\delta^T\bB^T\bB\delta > 0$ for non-zero $\delta$, thus,
$$
\Vert \bB(\tilde\gamma^u - \tilde\gamma^d)\Vert_2^2 > \Vert \bB(\gamma^u - \gamma^d)\Vert_2^2\,.
$$
So if $\gamma^u+\gamma^d$ is increasing, the best decomposition is $\gamma^d=c\one$.

Note that 
$$
\one^T\bB\gamma^u = \one^T\bB c\one = c\one^T\one\,,
$$
the constant $c$ is
$$
c = \frac{\one^T\bB\gamma^u}{n}\,.
$$

\end{proof}

\subsection{(ii)}\label{sec:proof_md_cs_increase2}

\begin{proof}
Take the derivatives on the Lagrange form, 
\begin{align}
-2\bZ^T(y-\bZ\gamma) + 2\bH^T\nu + 2\mu\bW\gamma = 0
\label{eq:deriv_lag}
\end{align}
then
\begin{align*}
\hat\gamma &= (\bZ^T\bZ + \mu\bW)^{-1}(\bZ^Ty-\bH^T\nu) \\
&\triangleq(\bZ^T\bZ + \mu\bW)^{-1}\bZ^Ty - (\bZ^T\bZ + \mu\bW)^{-1}\xi\,,
\end{align*}
where $\xi\triangleq \bH^T\nu$.
Note that
$$
(\bZ^T\bZ+\mu \bW)^{-1} = \begin{bmatrix}
1 + \mu & 1 - \mu \\
1 - \mu & 1 + \mu
\end{bmatrix}^{-1} \otimes (\bB^T\bB)^{-1} = \frac{1}{4\mu}\begin{bmatrix}
\mu + 1 & \mu -1 \\
\mu - 1 & \mu +1
\end{bmatrix} \otimes (\bB^T\bB)^{-1}\,,
$$
then
$$
(\bZ^T\bZ+\mu \bW)^{-1}\bZ^Ty = \left\{
\frac{1}{4\mu}\begin{bmatrix}
\mu + 1 & \mu -1 \\
\mu - 1 & \mu +1
\end{bmatrix}
\begin{bmatrix}
1\\
1
\end{bmatrix}
\right\} \otimes (\bB^T\bB)^{-1}\bB^Ty = \frac 12\begin{bmatrix}
1\\
1
\end{bmatrix}\otimes (\bB^T\bB)^{-1}\bB^Ty\,.
$$

Consider the $i$-th element of \eqref{eq:deriv_lag},
\begin{equation}
-\bZ_i^T(y-\bZ\gamma) + \bH_i^T\nu + \mu \bW_i^T\gamma = 0\,,
\label{eq:deriv_lag_i}
\end{equation}
where $\bZ_i,\bH_i,\bW_i$ are the $i$-th column of $\bZ, \bH,\bW$, respectively. Note that for $1\le i\le J$,
$$
\bW_i =
\begin{bmatrix}
\bB^T\bB_i\\
-\bB^T\bB_i
\end{bmatrix}\,,\qquad
\bW_{i+J} = \begin{bmatrix}
-\bB^T\bB_i\\
\bB^T\bB_i
\end{bmatrix}\,,
$$
and
$$
\bZ_i = \bB_i\,,\qquad \bZ_{i+J} = \bB_i\,,
$$
where $\bB_i$ is the $i$-th column of $\bB$.
Then \eqref{eq:deriv_lag_i} turns out to be
\begin{align}
    -\bB^T_i(y-\bZ\gamma) + \bH_{i}^T\nu + \mu \bB_i^T\bB(\gamma^u-\gamma^d) &= 0\,,
\label{eq:first_deriv_J}\\
        -\bB^T_i(y-\bZ\gamma) + \bH_{i+J}^T\nu + \mu \bB^T_i\bB(-\gamma^u+\gamma^d) &= 0\,.
\label{eq:first_deriv_2J}
\end{align}
\eqref{eq:first_deriv_J} + \eqref{eq:first_deriv_2J} yields
\begin{equation}
(\bH_{i}^T - \bH_{i+J}^T)\nu = -2\mu \bB_i^T\bB(\gamma^u-\gamma^d)\,,
\label{eq:first_deriv_sum}
\end{equation}
and \eqref{eq:first_deriv_J} - \eqref{eq:first_deriv_2J} yields
$$
(\bH_{i}^T + \bH_{i+J}^T)\nu = 2\bB^T_i(y-\bZ\gamma)\,.
$$
If there is no ties in the solution, i.e.,
$$
\gamma_1^u < \gamma_2^u < \cdots < \gamma_J^u\,,\qquad \gamma_1^d > \gamma_2^d > \cdots > \gamma_J^d\,,
$$
by the KKT condition, $\nu = 0$, and then \eqref{eq:first_deriv_sum} implies $\mu = 0$. Then it reduces to unconstrained B-spline fitting. This argument proves the first point of Proposition 2.

If 
$$
\gamma^u_1 < \gamma^u_2 < \ldots < \gamma^u_{J}\,,
$$
then from \eqref{eq:first_deriv_sum}, we have
$$
\xi_{i+J} = 2\mu \bB_i^T\bB(\gamma^u - \gamma^d)\,,
$$
then
$$
\xi = \begin{bmatrix}
0 \\
1
\end{bmatrix}\otimes
2\mu\bB^T\bB(\gamma^u-\gamma^d)\,,
$$
it follows that 
$$
(\bZ^T\bZ + \mu\bW)^{-1}\xi = \frac{1}{2}\begin{bmatrix}
\mu - 1\\
\mu + 1
\end{bmatrix} \otimes (\gamma^u - \gamma^d)\,.
$$
So
$$
\begin{bmatrix}
\gamma^u\\
\gamma^d
\end{bmatrix}
= \frac 12\begin{bmatrix}
(\bB^T\bB)^{-1}\bB^Ty - (\mu-1)(\gamma^u-\gamma^d)\\
(\bB^T\bB)^{-1}\bB^Ty - (\mu+1)(\gamma^u-\gamma^d)
\end{bmatrix}\,,
$$
both yield
$$
(\mu+1)\gamma^u - (\mu - 1)\gamma^d = (\bB^T\bB)^{-1}\bB^Ty\,.
$$
On the other hand, from \eqref{eq:first_deriv_J}, we have
$$
-\bB^T(y-\bZ\gamma) + \mu\bB^T\bB(\gamma^u-\gamma^d) = 0\,,
$$
that is,
$$
(\gamma^u+\gamma^d) + \mu(\gamma^u-\gamma^d) = (\bB^T\bB)^{-1}\bB^Ty\,,
$$
that is,
$$
(\mu + 1)\gamma^u - (\mu - 1)\gamma^d = (\bB^T\bB)^{-1}\bB^Ty \triangleq \hat\gamma^\ls\,.
$$
Incorporate with the result in Section \ref{sec:md_cs_constant}, we have
$$
\gamma^d = c\one\,, \gamma^u = \frac{1}{\mu+1}\hat\gamma^\ls +\frac{\mu-1}{\mu+1}c\one\,.
$$
\end{proof}

\subsection{(iii)}

\begin{proof}
Note that
\begin{equation}
    \begin{cases}
    \xi_1 &=\nu_1\\
    \xi_2 &=\nu_2-\nu_1\\
    \vdots & \\
    \xi_{J-1}&=\nu_{J-1}-\nu_{J-2}\\
    \xi_J &=-\nu_{J-1}
    \end{cases}
    \label{eq:xi_nu}
\end{equation}
If $\hat\gamma^u_1 < \cdots < \hat\gamma^u_{k_1}=\cdots=\hat\gamma^u_{k_2}<\cdots <\hat\gamma^u_{k_{2m-1}} =\cdots = \hat\gamma^u_{k_{2m}}< \hat\gamma_J$, then
\begin{equation*}
    \begin{cases}
\nu_1 &= \cdots = \nu_{k_1-1} =0    \\
\vdots & \\
\nu_{k_{2m-2}} & = \cdots = \nu_{k_{2m-1}} = 0\\
\nu_{k_{2m}} & = \cdots = \nu_{J-1}=0
    \end{cases}
    \qquad\Longrightarrow\qquad
    \begin{cases}
    \xi_1 & = \cdots = \xi_{k_1-1} = 0\\
    \vdots &\\
    \displaystyle\sum_{i=k_1}^{k_2-1}\xi_i & =0\\
    \vdots &\\
    \displaystyle\sum_{i=k_{2m-1}}^{k_{2m}-1}\xi_{i} &=0\\
    \xi_{k_{2m}} &=\cdots = \xi_{J} = 0\\
    \end{cases}
\end{equation*}
Thus,
\begin{align*}
    \bB_1^Ty &= (\mu+1)\bB^T_1\bB\gamma^u - (\mu-1)\bB^T_1\bB\gamma^d\\
    \vdots & \\
    \bB_{k_1-1}^Ty &= (\mu+1)\bB^T_{k_1-1}\bB\gamma^u - (\mu-1)\bB^T_{k_1-1}\bB\gamma^d\\
    \sum_{i=k_1}^{k_2-1}\bB_i^Ty &=(\mu+1)\sum_{i=k_1}^{k_2-1}\bB^T_i\bB\gamma^u - (\mu-1)\sum_{i=k_1}^{k_2-1}\bB^T_i\bB\gamma^d\\
    \vdots & \\
    \sum_{i=k_{2m-2}}^{k_{2m}-1}\bB_i^Ty &=(\mu+1)\sum_{i=k_{2m-2}}^{k_{2m}-1}\bB^T_i\bB\gamma^u - (\mu-1)\sum_{i=k_{2m-2}}^{k_{2m}-1}\bB^T_i\bB\gamma^d\\
    \bB_{k_{2m}}^Ty &= (\mu+1)\bB^T_{k_{2m}}\bB\gamma^u - (\mu-1)\bB^T_{k_{2m}}\bB\gamma^d\\
    \vdots &\\
    \bB_J^Ty & = (\mu+1)\bB^T_J\bB\gamma^u - (\mu-1)\bB_J^T\bB\gamma^d\,.
\end{align*}
Then
$$
\bG\bB^Ty = (\mu+1)\bG\bB^T\bB\gamma^u - (\mu-1)\bG\bB^T\bB\gamma^d\,,
$$
where
$$
\bG = \begin{bmatrix}
\bI_{k_1-1} & & &&&\\
& \one^T_{k_2-k_1+1} & &&&\\
& & \ddots &&&\\
 &  &  & \bI_{k_{2m-1}-k_{2m-2}-1} &&\\
 &  & &  & \one^T_{k_{2m}-k_{2m-1}+1} &\\
 &  & &  & & \bI_{J-k_{2m}}
\end{bmatrix}\,.
$$

Let $\gamma^u = \bG^T\beta$, where $\beta$ is the sub-vector of $\gamma^u$ constructed by the unique elements. Since $\gamma^d = c\one$, and $\bG^T\one=\one$, then
$$
\bG\bB^Ty = (\mu+1)\bG\bB^T\bB\bG^T\beta - (\mu-1)\bG\bB^T\bB\bG^T\gamma^dc\one\,,
$$
then
$$
\beta = \frac{1}{\mu+1}(\bG\bB^T\bB\bG^T)^{-1}\bG\bB^Ty + \frac{\mu-1}{\mu+1}c\one\,.
$$
thus
$$
\gamma^u = \bG^T\beta = \frac{1}{\mu+1}\bG(\bG\bB^T\bB\bG^T)^{-1}\bG\bB^Ty + \frac{\mu-1}{\mu+1}c\one\,,\qquad \gamma^d = c\one\,.
$$
\end{proof}

\subsection{Corollary \ref{coro:md_cs_decre}}\label{sec:proof_coro_cs}
The monotone decomposition for decreasing functions can be straightforward to derive, as summarized in Corollary \ref{coro:md_cs_decre}.

\begin{corollary}\label{coro:md_cs_decre}
Let $\hat\gamma = (\hat\gamma^u, \hat\gamma^d)$ be the monotone decomposition to problem (6).
Suppose $\hat\gamma^u+\hat\gamma^d$ is decreasing, then
\begin{enumerate}[label=(\roman*)]
    \item $\hat\gamma^u$ is a vector with identical elements; 
    \item if $\hat\gamma^d_1 > \cdots > \hat\gamma^d_{k_1}=\cdots=\hat\gamma^d_{k_2} > \cdots > \hat\gamma^d_{k_{2m-1}} =\cdots = \hat\gamma^d_{k_{2m}}> \cdots > \hat\gamma_J$, where $1\le k_1 \le k_2\le \cdots\le k_{2m-1}\le k_{2m}\le J$, then 
    $$
\hat\gamma^d = \frac{1}{\mu+1}\bG^T(\bG\bB^T\bB\bG^T)^{-1}\bG\bB^Ty +\frac{\mu-1}{\mu+1}c\one\,,\qquad \hat\gamma^u = \frac{\one^T\bB\hat\gamma^d}{n}\one\,.
$$
\end{enumerate}
\end{corollary}
\begin{proof}
With $(X, y)$, the solution is $\hat\gamma^u+\hat\gamma^d$, then the solution for $(X, -y)$ is $-\hat\gamma^u-\hat\gamma^d$.

If $\gamma^u+\gamma^d$ is decreasing, then $-\gamma^u-\gamma^d$ is increasing. Note that $-\gamma^u$ is non-increasing, and $-\gamma^d$ is non-decreasing, then by Proposition 1,
firstly, $-\gamma^u$ is a constant, and hence $\gamma^u$ is a constant.

If $\hat\gamma^d_1 > \cdots > \hat\gamma^d_{k_1}=\cdots=\hat\gamma^d_{k_2} > \cdots > \hat\gamma^d_{k_{2m-1}} =\cdots = \hat\gamma^d_{k_{2m}}> \hat\gamma_J$, then
$$
-\hat\gamma^d_1 < \cdots < -\hat\gamma^d_{k_1}=\cdots=-\hat\gamma^d_{k_2} < \cdots < -\hat\gamma^d_{k_{2m-1}} =\cdots = -\hat\gamma^d_{k_{2m}}< -\hat\gamma_J\,.
$$

Thus,
$$
-\hat\gamma^d = \frac{1}{\mu+1}\bG(\bG\bB^T\bB\bG^T)^{-1}\bG\bB^T(-y) + \frac{\mu-1}{\mu+1}\tilde c\one\,,\qquad \tilde c =\frac{\one^T\bB(-\gamma^d)}{n}\,,
$$
it follows that
$$
\hat\gamma^d = \frac{1}{\mu+1}\bG(\bG\bB^T\bB\bG^T)^{-1}\bG\bB^Ty + \frac{\mu-1}{\mu+1}c\one\,,\qquad c = \frac{\one^T\bB\gamma^d}{n}\,,\qquad \hat\gamma^u=c\one\,.
$$
\end{proof}


\section{A Side Product: Unimodal Functions}

We explore the solution for more general unimodal functions, as summarized in Proposition~\ref{thm:unimodal}. Specifically, the second point of Proposition 3 can be viewed as a special instance with $k=\ell=1$ of Proposition~\ref{thm:unimodal}.
\begin{proposition}\label{thm:unimodal}
Let $\hat\gamma = (\hat\gamma^u, \hat\gamma^d)$ be the monotone decomposition to problem (8). If 
\begin{align*}
    \hat\gamma_1^u &= \hat\gamma_2^u = \cdots = \hat\gamma_k^u < \hat\gamma_{k+1}^u < \cdots < \hat\gamma_J^u\\
    \hat\gamma_1^d &> \hat\gamma_2^d > \cdots > \hat\gamma_\ell^d = \hat\gamma_{\ell+1}^d = \cdots = \hat\gamma_J^d\,,
\end{align*}
suppose $\ell \le k$, then
$$
\begin{bmatrix}
\bI_{\ell-1} & &\\
& \one_{k-\ell+1} &\\
& & \bI_{J-k}
\end{bmatrix}\bB^T\bfy = \bA\bB^T\bB\hat\gamma^u + (2\bI-\bA)\bB^T\bB\hat\gamma^d\,,
$$
where
$$
\bA = \begin{bmatrix}
(1-\mu)\bI_{\ell-1} & & \\
2\mu\one_{\ell-1}^T & 1+\mu &\\
& & (1+\mu)\bI_{J-k}
\end{bmatrix}\,.
$$
\end{proposition}

\begin{proof}

If 
$$
\gamma_1^u = \gamma_2^u = \cdots = \gamma_k^u < \gamma_{k+1}^u < \cdots < \gamma_J^u
$$
and
$$
\gamma_1^d > \gamma_2^d > \cdots > \gamma_\ell^d = \gamma_{\ell+1}^d = \cdots = \gamma_J^d\,.
$$
Suppose $\ell \le k$, then
$$
\nu_k = \nu_{k+1} = \cdots = \nu_{J-1} = 0\,,
$$
and
$$
\nu_{J-1+i} = 0, i=1,\ldots, \ell - 1\,.
$$
Then from \eqref{eq:xi_nu}, we have $\xi_i = 0, k+1\le i\le J$, and then
\begin{align}
    \sum_{i=1}^k\bB_i^Ty &=(\mu+1)\sum_{i=1}^k\bB^T_i\bB\gamma^u - (\mu-1)\sum_{i=1}^k\bB^T_i\bB\gamma^d\label{eq:unimodal_k_1}\,,\\
    \bB_i^Ty &= (\mu+1)\bB^T_i\bB\gamma^u - (\mu-1)\bB^T_i\bB\gamma^d \quad k+1\le i\le J\,.\label{eq:unimodal_k_2}
\end{align}
Similarly, $\xi_{i+J} = 0, 1\le i\le \ell-1$, and hence
\begin{align}
    \bB_i^Ty &= (\mu+1)\bB^T_i\bB\gamma^d - (\mu - 1)\bB^T_i\bB\gamma^u\quad 1\le i\le \ell-1\label{eq:unimodal_l_1}\\
    \sum_{i=\ell}^J \bB_i^Ty &= (\mu+1)\sum_{i=\ell}^J\bB^T_i\bB\gamma^d - (\mu-1)\sum_{i=\ell}^J\bB^T_i\bB\gamma^u\label{eq:unimodal_l_2}\,.
\end{align}

Note that
\begin{align}
\sum_{i=1}^J\bB_i^Ty &= (\mu+1)\sum_{i=1}^J\bB_i^T\bB\gamma^u - (\mu-1)\sum_{i=1}^J\bB_i^T\bB\gamma^d\notag\\
&=(\mu+1)\sum_{i=1}^J\bB_i^T\bB\gamma^d - (\mu+1)\sum_{i=1}^J\bB_i^T\bB\gamma^u\label{eq:unimodal_sum}\,,
\end{align}
then \eqref{eq:unimodal_k_1} + \eqref{eq:unimodal_l_2} - \eqref{eq:unimodal_sum} yields
\begin{align*}
    \sum_{\ell}^k\bB_i^Ty &= \sum_{i=1}^k\bB_i^Ty + \sum_{i=\ell}^J\bB_i^Ty - \sum_{i=1}^J\bB_i^Ty\\
    &=(\mu+1)\sum_{i=1}^k\bB^T_i\bB\gamma^u - (\mu-1)\sum_{i=1}^k\bB^T_i\bB\gamma^d +\\
    &\qquad\quad(\mu+1)\sum_{i=\ell}^J\bB^T_i\bB\gamma^d - (\mu-1)\sum_{i=\ell}^J\bB^T_i\bB\gamma^u -\\
    &\qquad\quad[(1-\mu)\one^T\bB\gamma^u+(\mu+1)\one^T\bB\gamma^d]\\
    &=(\mu+1)\sum_{i=1}^k\bB^T_i\bB\gamma^u - (\mu-1)\sum_{i=1}^k\bB^T_i\bB\gamma^d +\\
    &\qquad\quad (\mu-1)\sum_{i=1}^{\ell-1}\bB_i^T\bB\gamma^u - (\mu+1)\sum_{i=1}^{\ell-1}\bB_i^T\bB\gamma^d\\
    &=2\mu\sum_{i=1}^{\ell-1}\bB_i^T\bB\gamma^u + (\mu+1)\sum_{i=\ell}^k\bB^T_i\bB\gamma^u - 2\mu\sum_{i=1}^{\ell-1}\bB_i^T\bB\gamma^d - (\mu-1)\sum_{i=\ell}^k\bB^T_i\bB\gamma^d\,,
\end{align*}
then 
\begin{align*}
\begin{bmatrix}
\bB_1^Ty\\
\vdots\\
\bB_{\ell-1}^Ty\\
\sum_{i=\ell}^k\bB_i^Ty\\
\bB_{k+1}^Ty\\
\vdots\\
\bB_J^Ty\\
\end{bmatrix}
&= \begin{bmatrix}
1- \mu & \cdots & 0 & 0 & 0 & \cdots & 0\\
\vdots & \ddots & 0 & 0 & 0 & \cdots & 0\\
0 & \cdots & 1-\mu & 0 & 0 & \cdots & 0\\
2\mu & \cdots & 2\mu & 1+\mu & 0 & \cdots & 0\\
0 & \cdots & 0 & 0 & 1+\mu & \cdots & 0\\
\vdots & \vdots & \vdots & \vdots & \vdots & \ddots & \vdots\\
0 & \cdots & 0 & 0 & 0 & \cdots & 1+\mu
\end{bmatrix}
\bB^T\bB\gamma^u + \notag\\
&\qquad\qquad
\begin{bmatrix}
1+ \mu & \cdots & 0 & 0 & 0 & \cdots & 0\\
\vdots & \ddots & 0 & 0 & 0 & \cdots & 0\\
0 & \cdots & 1+\mu & 0 & 0 & \cdots & 0\\
-2\mu & \cdots & -2\mu & 1-\mu & 0 & \cdots & 0\\
0 & \cdots & 0 & 0 & 1-\mu & \cdots & 0\\
\vdots & \vdots & \vdots & \vdots & \vdots & \ddots & \vdots\\
0 & \cdots & 0 & 0 & 0 & \cdots & 1-\mu
\end{bmatrix}
\bB^T\bB\gamma^d\\
&\triangleq \bA \bB^T\bB\gamma^u + (2\bI-\bA)\bB^T\bB\gamma^d\,.
\end{align*}
If $\ell = k$, then
$$
\bB^Ty = \bA\bB^T\bB\gamma^u + (2\bI-\bA)\bB^T\bB\gamma^d\,.
$$

\end{proof}

\section{Proof of Proposition 4}

\begin{proof}
The fitting vector is
\begin{align}
    \hat \bff &= \bB\hat\gamma^u+\bB\hat\gamma^d\notag\\
    &=\frac{1}{\mu+1}\bB\bG^T(\bG\bB^T\bB\bG^T)^{-1}\bG\bB^T\bfy + \frac{2\mu}{\mu+1}\frac{\one^T_n\hat \bff}{2n}\one_n\notag\\
    &\triangleq k\bA \bfy + \frac 1n(1-k)\one^T_n\hat \bff\one_n\label{eq:proof_md_fitvector}\,,
\end{align}
then
\begin{equation}
\one^T_n\hat \bff = k\one^T_n\bA \bfy + (1-k)\one^T_n\hat \bff\,.\label{eq:proof_md_1f}
\end{equation}
It follows that
$$
\one^T_n\hat \bff = \one^T_n\bA \bfy\,,
$$
then
\begin{equation}
\hat \bff = k\bA \bfy + \frac{1-k}{n}\one^T_n\bA \bfy\one_n = \left[k\bI+\frac{1-k}{n}\one_n\one_n^T\right]\bA \bfy\,.
\label{eq:proof_md_fitvector0}
\end{equation}
Note that $\bG$ depends on $\bfy$, and hence $\bA$ also depends on $\bfy$. 
Take expectation on \eqref{eq:proof_md_fitvector0} and by the law of total expectation, we have
$$
\bbE \hat \bff = \bbE[\bbE[\hat\bff\mid \bG]] = \bbE[k\bA \bff + \frac{1-k}{n}\one^T_n\bA \bff\one_n]\,.
$$
Note that $\bA\bA^T = \bA = \bA^T$ and $\one_n = \bB\one_J = \bG^T\one_g = \bB\bG^T\one_g$, then 
\begin{align*}
    \one^T_n\bA &= \one^T_n\bB\bG^T(\bG\bB^T\bB\bG^T)^{-1}\bG\bB^T\\
    &=\one^T_g\bG\bB^T\bB\bG^T(\bG\bB^T\bB\bG^T)^{-1}\bG\bB^T\\
    &=\one^T_g\bG\bB^T\\
    &=\one^T_n\,,
\end{align*}
and hence $\one_n^T\bA\bA^T\one_n = n$. It follows that
$$
\bbE\hat\bff = k\bbE\bA \bff + \frac{1-k}{n}\one_n^T\bff\one_n\,.
$$
The square of bias is
\begin{align*}
    \Vert \bff-\bbE \hat \bff\Vert^2 &= \Vert \bff - k\bbE\bA \bff - \frac{1-k}{n}\one^T_n \bff\one_n\Vert^2\\
    &=\bff^T(\bI-k\bbE\bA)^2\bff - \frac{2(1-k)}{n}\bff^T(\bI-k\bbE\bA)\one^T_n\bff\one_n + \frac{(1-k)^2}{n^2}(\one^T_n\bff)^2\one^T_n\one_n\\
    &=\bff^T(\bI-k\bbE\bA)^2\bff - \frac{2(1-k)}{n}(\one^T_n\bff) \bff^T(\bI-k\bbE\bA)\one_n + \frac{(1-k)^2}{n}(\one^T_n\bff)^2\\
    &=\bff^T(\bI-k\bbE\bA)^2\bff -\frac{(1-k)^2}{n}(\one^T_n\bff)^2\\
    &=\bff^T\bff-2k\bff^T\bbE\bA \bff+k^2\bff^T[\bbE\bA]^2 \bff-\frac{(1-k)^2}{n}(\one^T_n\bff)^2\,.
\end{align*}
Note that
$$
[\bbE\bA]^2 = \bbE\bA^2 - \Var(\bA)\,,
$$
and $\bA^2 = \bA$, we can write
$$
k^2\bff^T[\bbE\bA]^2\bff = k^2\bff^T[\bbE\bA - \Var(\bA)]\bff\,.
$$
It follows that
\begin{align*}
    \Vert \bff-\bbE\hat\bff\Vert^2 &= \bff^T\bbE\bA \bff - 2k\bff^T\bbE\bA\bff + k^2\bff^T\bbE\bA\bff + \bff^T\bff - \bff^T\bbE\bA\bff - k^2\bff^T\Var(\bA)\bff-\frac{(1-k)^2}{n}(\one^T_n\bff)^2\\
    &=(1-k)^2\bff^T\bbE\bA\bff + \bff^T(\bI-\bbE\bA)\bff - k^2\bff^T\Var(\bA)\bff-\frac{(1-k)^2}{n}(\one^T_n\bff)^2\\
    &=(1-k)^2\left(\bff^T\bbE\bA\bff - \frac{(\one_n^T\bff)^2}{n}\right)+ \bff^T(\bI-\bbE\bA)\bff - k^2\bff^T\Var(\bA)\bff\\
    &\triangleq (1-k)^2C_1 + C_2 -k^2C_3\,.
\end{align*}
Since for each $\bA$, we have $(\bI-\bA)^2=(\bI-\bA)$ and $(\bA - \one\one^T/n)^2=\bA - \one\one^T/n$, then both $\bI-\bA$ and $\bA-\one\one^T$ are idempotent, and hence are positive semi-definite. It follows that both $\bI-\bbE\bA$ and $\bbE \bA -\one\one^T/n$ are also positive semi-definite. Hence $C_1\ge 0$ and $C_2 \ge 0$.

On the other hand, by the law of total variance, we have
\begin{equation}
\Var(\hat\bff) = \bbE[\Var(\hat\bff\mid\bG)] +\Var[\bbE[\hat\bff\mid \bG]] \,.    
\label{eq:var_f}
\end{equation}
For the first term of \eqref{eq:var_f}, we have
\begin{align*}
\Var(\hat \bff\mid\bG) &= \sigma^2\left[k\bI+\frac{1-k}{n}\one_n\one_n^T\right]\bA\bA^T\left[k\bI+\frac{1-k}{n}\one_n\one_n^T\right]    \\
&=\sigma^2\left[k^2\bA\bA^T + \frac{1-k}{n}\one_n\one_n^T\bA\bA^T + \frac{1-k}{n}\bA\bA^T\one_n\one_n^T + \frac{(1-k)^2}{n^2}\one_n\one_n^T\one_n\one_n^T \right]\,.
\end{align*}
Since
\begin{align*}
    \tr(\bA\bA^T) & = \tr(\bA) \\
    &=\tr(\bB\bG^T(\bG\bB^T\bB\bG^T)^{-1}\bG\bB^T)\\
    &=\tr((\bG\bB^T\bB\bG^T)^{-1}\bG\bB^T\bB\bG^T)\\
    &=g\,,
\end{align*}
then
$$
\tr[\Var(\hat \bff\mid\bG)] = \sigma^2\left[k^2g + 2(1-k) + (1-k)^2\right] \,.
$$
For the second term of \eqref{eq:var_f}, we have
\begin{align*}
    \Var[\bbE[\hat\bff\mid \bG]] = \Var[k\bA\bff + \frac{1-k}{n}\one_n^T\bff\one_n] = k^2\Var[\bA]\bff\bff^T\,.
\end{align*}
Thus,
\begin{align*}
    \tr[\Var(\hat\bff)] &= \tr[\bbE[\Var(\hat\bff\mid \bG)]] + \tr[\Var[\bbE[\hat\bff\mid \bG]]]\\
    &=\bbE[\tr[\Var(\hat\bff\mid \bG)]] +\tr[k^2\Var[\bA]\bff\bff^T]\\
    &=\sigma^2[k^2\bbE g+2(1-k)+(1-k)^2] + k^2\bff^T\Var(\bA)\bff\\
    &\triangleq \sigma^2[k^2\bbE g+2(1-k)+(1-k)^2] + k^2C_3\,.
\end{align*}
It follows that the mean square error is
\begin{align*}
    \MSE &= \Vert\Bias\Vert^2 + \tr[\Var(\hat \bff)]\\
    &=(1-k)^2C_1 +C_2 + \sigma^2\left[k^2\bbE g + 2(1-k) + (1-k)^2\right]\,.
\end{align*}

The unconstrained B-spline fitting has
$$
\MSE{}^\ls = J\sigma^2\,.
$$
To have $\MSE < \MSE{}^\ls$,
$$
(1-k)^2C_1 +C_2 + \sigma^2\left[k^2\bbE g + 2(1-k) + (1-k)^2\right] < J\sigma^2\,,
$$
that is
$$
h(k) \triangleq
\left[C_1+(\bbE g+1)\sigma^2\right]k^2 - \left[2C_1+4\sigma^2\right]k+C_1+C_2+3\sigma^2-J\sigma^2 < 0\,.
$$
Note that the minimum is obtained at
$$
k_{\min} = \frac{\left[2C_1+4\sigma^2\right]}{2\left[C_1+(\bbE g+1)\sigma^2\right]} = \frac{C_1+2\sigma^2}{C_1+(\bbE g+1)\sigma^2}\in (0, 1)\,,
$$
since $g\ge 1$, and the minimum value is
\begin{align*}
h_{\min} &= C_1+C_2+3\sigma^2-J\sigma^2 - \frac{\left[2C_1+4\sigma^2\right]^2}{4\left[C_1+(\bbE g+1)\sigma^2\right]}\\
&=\frac{\sigma^4\left[(3-J)(\bbE g+1)-4\right] + \sigma^2\left[-C_1(J-\bbE g)+C_2(\bbE g+1)\right]+C_1C_2}{C_1+(\bbE g+1)\sigma^2}\\
&=\frac{-\sigma^4\left[(J-\bbE g)(\bbE g+1)+(\bbE g-1)^2\right] + \sigma^2\left[-C_1(J-\bbE g)+C_2(\bbE g+1)\right]+C_1C_2}{C_1+(\bbE g+1)\sigma^2}\\
&=\frac{u(\sigma^2)}{C_1+(\bbE g+1)\sigma^2}\,,
\end{align*}
where
$$
u(t) = -t^2\left[(J-\bbE g)(\bbE g+1)+(\bbE g-1)^2\right] + t\left[-C_1(J-\bbE g)+C_2(\bbE g+1)\right]+C_1C_2\,.
$$
Since $C_1, C_2\ge 0$,
\begin{align*}
\Delta & = \left[-C_1(J-\bbE g)+C_2(\bbE g+1)\right]^2+4C_1C_2\left[(J-\bbE g)(\bbE g+1)+(\bbE g-1)^2\right]\ge 0\\
&=\left[C_1(J-\bbE g)+C_2(\bbE g+1)\right]^2+4C_1C_2(\bbE g-1)^2 \ge 0\,,
\end{align*}
then $u(t) < 0$ if 
$$
t > \frac{C_1(J-\bbE g)-C_2(\bbE g+1)-\sqrt{\Delta}}{-2[(J-\bbE g)(\bbE g+1)+(\bbE g-1)^2]} = \frac{-C_1(J-\bbE g)+C_2(\bbE g+1)+\sqrt{\Delta}}{2[(J-\bbE g)(\bbE g+1)+(\bbE g-1)^2]}\,.
$$



\subsection{\texorpdfstring{$\bG=\bI$}{G = I}}

If $\bG=\bI$, that is, no ties in the solution,
$$
\bA \bff = \bA \bB\gamma = \bB(\bB^T\bB)^{-1}\bB^T\bB\gamma = \bB\gamma = \bff\,.
$$
\begin{align*}
    \Vert \bff-\bbE \hat \bff\Vert^2 &= \Vert \bff - k\bff - \frac{1-k}{n}\one^T \bff\one\Vert^2\\
    &=(1-k)^2\Vert \bff - \frac 1n\one^T \bff\one \Vert^2\\
    &=(1-k)^2\left[\Vert \bff\Vert^2 - \frac 1n (\one^T_n\bff)^2\right]
\end{align*}

If $\bG=\bI$, $J=g, C_2=0$, then
$$
h_{\min} = \frac{-\sigma^4(g-1)^2}{C_1+(g+1)\sigma^2} < 0\,.
$$

\end{proof}

\section{Proof of Proposition 5}

\subsection{(i)}\label{sec:md_ss_const}

\begin{proof}

Since if $\gamma^u+\gamma^d = \tilde\gamma^u +\tilde\gamma^d$, then the roughness penalty does not change, 
$$
(\gamma^u+\gamma^d)^T\bOmega(\gamma^u+\gamma^d) = (\tilde\gamma^u+\tilde\gamma^d)^T\bOmega(\tilde\gamma^u+\tilde\gamma^d)\,.
$$
Then we can repeat the procedure in Section \ref{sec:md_cs_constant}.

\end{proof}

\subsection{(ii)}

\begin{proof}
A special case when $\bG=\bI$ in the following result.
\end{proof}

\subsection{(iii)} \label{sec:proof_md_ss_increase2}

\begin{proof}
Take the derivatives on the Lagrange form, 
$$
-2\bZ^T(\bfy-\bZ\gamma) + 2\bH^T\nu + 2\mu\bW\gamma + 2\lambda \begin{bmatrix}
\bOmega & \bOmega\\
\bOmega & \bOmega
\end{bmatrix}\gamma = 0
$$
Then
\begin{align}
-\bB^T_i(\bfy-\bZ\gamma) + \bH_{i}^T\nu + \mu \bB_i^T\bB(\gamma^u-\gamma^d) +\lambda\bOmega_i(\gamma^u+\gamma^d) = 0\\
-\bB^T_i(\bfy-\bZ\gamma) + \bH_{i+J}^T\nu + \mu \bB_i^T\bB(-\gamma^u+\gamma^d)+\lambda\bOmega_i(\gamma^u+\gamma^d) = 0
\end{align}
Rewrite them as
\begin{align}
    \bB^T_i\bfy+\xi_i =  ((1+\mu) \bB_i^T\bB + \lambda\bOmega_i)\gamma^u +((1-\mu)\bB_i^T\bB + \lambda\bOmega_i)\gamma^d\label{eq:ss_sol_1}\\
    \bB^T_i\bfy+\xi_{i+J} =  ((1-\mu) \bB_i^T\bB + \lambda\bOmega_i)\gamma^u +((1+\mu)\bB_i^T\bB + \lambda\bOmega_i)\gamma^d\label{eq:ss_sol_2}\,.
\end{align}
If there are no ties in the solution, i.e.,
$$
\gamma_1^u < \gamma_2^u < \cdots < \gamma_J^u\,,\qquad \gamma_1^d > \gamma_2^d > \cdots > \gamma_J^d\,,
$$
by the KKT condition, $\nu = 0$, and hence $\xi_i=\xi_{i+J}=0$, then \eqref{eq:ss_sol_1} - \eqref{eq:ss_sol_2} yields
$$
0=2\mu\bB_i^T\bB\gamma^u - 2\mu\bB_i^T\bB\gamma^d\,,
$$
that is
$$
0 = 2\mu \bB^T\bB(\gamma^u-\gamma^d)\,,
$$
which implies $\mu = 0$ since $\gamma^u > \gamma^d$. Then it reduces to unconstrained B-spline fitting. This argument proves the first point of Proposition 2.

If $\hat\gamma^u_1 < \cdots < \hat\gamma^u_{k_1}=\cdots=\hat\gamma^u_{k_2}<\cdots <\hat\gamma^u_{k_{2m-1}} =\cdots = \hat\gamma^u_{k_{2m}}< \hat\gamma_J$, then similar to Section \ref{sec:proof_md_cs_sol}, we have
$$
\bG\bB^Ty = ((1+\mu) \bG\bB^T\bB + \lambda\bG\bOmega)\gamma^u +((1-\mu)\bG\bB^T\bB + \lambda\bG\bOmega)\gamma^d\,.
$$

Let $\gamma^u = \bG^T\beta$, where $\beta$ is the sub-vector of $\gamma^u$ constructed by the unique elements. By \ref{sec:md_ss_const}, $\gamma^d=c\one_J$, and note that $\bG^T\one_g = \one_J$, then
$$
\bG\bB^Ty = ((1+\mu) \bG\bB^T\bB\bG^T + \lambda\bG\bOmega\bG^T)\beta +c((1-\mu)\bG\bB^T\bB\bG^T + \lambda\bG\bOmega\bG^T)\one_g\,,
$$
it follows that
\begin{align*}
\beta &= ((1+\mu) \bG\bB^T\bB\bG^T + \lambda\bG\bOmega\bG^T)^{-1}\bG\bB^Ty - \\
&\qquad
c((1+\mu) \bG\bB^T\bB\bG^T + \lambda\bG\bOmega\bG^T)^{-1}((1-\mu)\bG\bB^T\bB\bG^T + \lambda\bG\bOmega\bG^T)\one_g\,,
\end{align*}
and
$$
c = \frac{\one_n^T\bB\gamma^u}{n} = \frac{\one_n^T\bB\bG^T\beta}{n}\,.
$$
\end{proof}

\subsection{Corollary \ref{coro:md_ss}}

Analogously, we can obtain the monotone decomposition with smoothing splines on decreasing functions, as summarized in Corollary \ref{coro:md_ss}.
\begin{corollary}\label{coro:md_ss}
Let $\hat\gamma=(\hat\gamma^u, \hat\gamma^d)$ be the monotone decomposition to problem (11).
Suppose $\hat\gamma^u+\hat\gamma^d$ is decreasing, then
\begin{enumerate}[label=(\roman*)]
    \item $\hat\gamma^u$ is a vector with identical elements; 
    \item if $\hat\gamma^d_1 > \cdots > \hat\gamma^d_{k_1}=\cdots=\hat\gamma^d_{k_2} > \cdots > \hat\gamma^d_{k_{2m-1}} =\cdots = \hat\gamma^d_{k_{2m}}> \hat\gamma_J$, where $1\le k_1 \le k_2\le \cdots\le k_{2m-1}\le k_{2m}\le J$, then 
\begin{align*}
\hat\gamma^d &= \bG^T((1+\mu) \bG\bB^T\bB\bG^T + \lambda\bG\bOmega\bG^T)^{-1}\bG\bB^Ty - \\
&\qquad
c\bG^T((1+\mu) \bG\bB^T\bB\bG^T + \lambda\bG\bOmega\bG^T)^{-1}((1-\mu)\bG\bB^T\bB\bG^T + \lambda\bG\bOmega\bG^T)\one_g\,,\\
\hat\gamma^u &=c\one_J = \frac{\one^T\bB\hat\gamma^d}{n}\one_J\,,
\end{align*}
\end{enumerate}
\end{corollary}
\begin{proof}
Similar to the proof \ref{sec:proof_coro_cs} for Corollary \ref{coro:md_cs_decre}.
\end{proof}

\section{Proof of Proposition 6}

\subsection{Lemmas}

\begin{lemma}[{\cite{magnusMatrixDifferentialCalculus2019}}]\label{eq:trace_cauchy-schwarz}
Let $\bA, \bB$ be two real matrices of the same size, then
$$
[\tr(\bA\bB)]^2\le [\tr(\bA^2)][\tr(\bB^2)]\,.
$$
\end{lemma}

\begin{lemma}\label{eq:trace_sq_ieq}
Let $\bA$ be a real positive semi-definite matrix, then
$$
\tr(\bA^2)\le[\tr(\bA)]^2\,.
$$
\end{lemma}
\begin{proof}
Let $\lambda_i$ be the eigenvalues of $\bA$, then $\lambda_i^2$ are the eigenvalues of $\bA^2$. Note that
$$
\tr[\bA^2]=\sum \lambda_i^2 \le (\sum\lambda_i)^2 = [\tr (\bA)]^2\,.
$$
\end{proof}

\begin{lemma}
The eigenvalues of $\one_n\one_n^T$ is
$$
\lambda_1=n,\lambda_2=\cdots=\lambda_n=0\,.
$$
\end{lemma}
\begin{proof}
Since $\one_n\one^T_n$ is a rank-1 matrix, which implies that it has $n-1$ eigenvalues which are zero. Denote the eigenvectors as $\xi_i,i=1,\ldots,n$. Suppose the nonzero eigenvalues is $\lambda_1$ with associated eigenvectors $\xi_1$, then
$$
\one_n\one^T_n\xi_1=\lambda_1\xi_1\,,
$$
left multiplying $\one^T_n$ yields
$$
n\one_n^T\xi_1 = \lambda_1\one^T\xi_1\,,
$$
which implies that $\lambda_1 =n$. 
\end{proof}

\subsection{\texorpdfstring{$\bG = \bI$}{G = I} }
\begin{proof}
If $\bG=\bI$, the solution is
\begin{align*}
\hat\gamma^u &= \frac{1}{1+\mu} \left[\bB^T\bB+\frac{\lambda}{1+\mu}\bOmega\right]^{-1}\bB^T\bfy - c ((1+\mu)\bB^T\bB+\lambda\bOmega)^{-1}((1-\mu)\bG\bB^T\bB+\lambda\bOmega)\one_J\,,\\
&\triangleq \frac{1}{1+\mu} \left[\bB^T\bB+\lambda_0\bOmega\right]^{-1}\bB^T\bfy - c\bK \one_J\,,\\
\hat\gamma^d &= c\one_J = \frac{\one_n^T \hat \bff}{2n}\one_J\,,
\end{align*}
then the fitting vector is
\begin{align*}
    \hat \bff &= \bB\hat\gamma^u + \bB\hat\gamma^d\\
    &=\frac{1}{1+\mu}\hat \bff^\sspl + c\bB(-\bK+\bI)\one_J\\
    &=\frac{1}{1+\mu}\hat \bff^\sspl +\frac{\one^T_n\hat \bff}{2n}\bB\left[(1+\mu)\bB^T\bB+\lambda\bOmega\right]^{-1}2\mu\bB^T\bB\one_J\\
    &=\frac{1}{1+\mu}\hat \bff^\sspl + \frac{\mu}{1+\mu}\frac{\one^T_n\hat \bff}{n}\bB\left[\bB^T\bB+\frac{\lambda}{1+\mu}\bOmega\right]^{-1}\bB^T\one_n\\
    &\triangleq k\hat \bff^\sspl + (1-k)\frac{\one^T_n\hat \bff}{n}\bQ\one_n\,,
\end{align*}
where $\bQ = \bB\left[\bB^T\bB+\lambda_0\bOmega\right]^{-1}\bB^T$. 
In practice, $\lambda_0$ is given as a constant, so $\bQ$ does not depend on $k$. Left multiplying $\one_n^T$ yields
$$
\one^T_n\hat \bff = k\one^T_n\hat \bff^\sspl + (1-k)\one^T_n\hat \bff\frac{\one^T_n\bQ\one_n}{n}\,,
$$
that is,
$$
\one^T_n\hat \bff = \frac{k\one^T_n\hat \bff^\sspl}{1-\frac{\one^T_n\bQ\one_n}{n}(1-k)}\triangleq \frac{k\one^T_n\hat \bff^\sspl}{1-\eta(1-k)}\,.
$$
It follows that
$$
\hat \bff = \left[k\bI+ \frac{1-k}{n}\frac{k}{1-\eta(1-k)}\bQ\one_n\one_n^T\right]\hat \bff^\sspl\triangleq (k\bI+\alpha\bQ\one_n\one_n^T)\hat \bff^\sspl\,.
$$
Then the squared bias is
\begin{align*}
    &\Vert \bff-\bbE\hat \bff\Vert^2_2 = \Vert \bff - (k\bI+\alpha\bQ\one_n\one_n^T)\bbE \hat \bff^\sspl\Vert^2_2\\
    &=\Vert (k\bI+\alpha\bQ\one_n\one_n^T)(\bff - \bbE\hat \bff^\sspl) + ((1-k)\bI - \alpha\bQ\one_n\one_n^T) \bff\Vert^2\\
    &=(\bff - \bbE\hat \bff^\sspl)^T(k\bI+\alpha\one_n\one_n^T\bQ)(k\bI+\alpha\bQ\one_n\one_n^T)(\bff - \bbE\hat \bff^\sspl) + \\
    &\qquad 2(\bff-\bbE \hat \bff^\sspl)^T(k\bI+\alpha\bQ\one_n\one_n^T)^T((1-k)\bI - \alpha\bQ\one_n\one_n^T) \bff + \\
    &\qquad \bff^T((1-k)\bI - \alpha\bQ\one_n\one_n^T)^T((1-k)\bI - \alpha\bQ\one_n\one_n^T)\bff\\
    &\triangleq C_1+2C_2+C_3\,.
\end{align*}
First for term $C_1$, we have
\begin{align*}
    C_1&=(\bff - \bbE\hat \bff^\sspl)^T(k\bI+\alpha\one_n\one_n^T\bQ)(k\bI+\alpha\bQ\one_n\one_n^T)(\bff - \bbE\hat \bff^\sspl)\\
    &=k^2\Vert \bff -\bbE\hat \bff^\sspl\Vert^2 + 2k\alpha (\bff - \bbE\hat \bff^\sspl)^T\one_n\one_n^T\bQ(\bff - \bbE\hat \bff^\sspl) + \\
    &\qquad\qquad \alpha^2(\bff - \bbE\hat \bff^\sspl)^T\one_n\one_n^T\bQ^2\one_n\one_n^T(\bff - \bbE\hat \bff^\sspl)\,.
\end{align*}
Take the derivative with respect to $k$, 
\begin{align*}
    \frac{\partial C_1}{\partial k} &= 2k \Vert \bff -\bbE\hat \bff^\sspl\Vert^2 + 2\left(k\frac{\partial \alpha}{\partial k} + \alpha\right)(\bff - \bbE\hat \bff^\sspl)^T\one_n\one_n^T\bQ(\bff - \bbE\hat \bff^\sspl) \\
    &\qquad \qquad + 2\alpha\frac{\partial \alpha}{\partial k}(\bff - \bbE\hat \bff^\sspl)^T\one_n\one_n^T\bQ^2\one_n\one_n^T(\bff - \bbE\hat \bff^\sspl)\,.
\end{align*}
Note that $\frac{\partial \alpha}{\partial k}\mid_{k=1}=-\frac{1}{n}$ and $\alpha(1)=0$,
then
$$
\frac{\partial C_1}{\partial k}\mid_{k=1} = 2\Vert \bff -\bbE\hat \bff^\sspl\Vert^2 -\frac 2n (\bff - \bbE\hat \bff^\sspl)^T\one_n\one_n^T\bQ(\bff - \bbE\hat \bff^\sspl)\,.
$$
Similarly, for term $C_2$ and $C_3$, we have
\begin{align*}
    C_2 &= (\bff-\bbE \hat \bff^\sspl)^T(k\bI+\alpha\bQ\one_n\one_n^T)^T((1-k)\bI - \alpha\bQ\one_n\one_n^T) \bff \\
    &= \bff^T(\bI-\bQ)^T(k\bI+\alpha\bQ\one_n\one_n^T)^T((1-k)\bI - \alpha\bQ\one_n\one_n^T) \bff\\
    &= k(1-k)\bff^T(\bI-\bQ) \bff +(1-k)\alpha \bff^T(\bI-\bQ)\one_n\one_n^T\bQ \bff-k\alpha \bff^T(\bI-\bQ)\bQ\one_n\one_n^T \bff + \\
    &\qquad\qquad
    \alpha^2\bff^T(\bI-\bQ)\one_n\one_n^T\bQ^2\one_n\one_n^T \bff\,,\\
    C_3 &= \bff^T((1-k)\bI - \alpha\bQ\one_n\one_n^T)^T((1-k)\bI - \alpha\bQ\one_n\one_n^T)\bff\\
    &=(1-k)^2\bff^T\bff - 2\alpha(1-k)\bff^T\one_n\one_n^T\bQ \bff + \alpha^2 \bff^T\one_n\one_n^T\bQ^2\one_n\one_n^T \bff\,,
\end{align*}
and then take their derivatives with respect to $k$,
\begin{align*}
    \frac{\partial C_2}{\partial k} &=(1-2k)\bff^T(\bI-\bQ) \bff + \left( (1-k)\frac{\partial \alpha}{\partial k} -\alpha\right)\bff^T(\bI-\bQ)\one_n\one_n^T\bQ \bff -\\
    &\qquad \qquad \left(\frac{\partial \alpha}{\partial k} + \alpha\right)\bff^T(\bI-\bQ)\bQ\one_n\one_n^T \bff + 2\alpha \frac{\partial \alpha}{\partial k}\bff^T(\bI-\bQ)\one_n\one_n^T\bQ^2\one_n\one_n^T \bff\,,\\
    \frac{\partial C_3}{\partial k} &=-2(1-k)\bff^T\bff-2\left( (1-k)\frac{\partial \alpha}{\partial k} -\alpha\right)\bff^T\one_n\one_n^T\bQ \bff + 2\alpha \frac{\partial \alpha}{\partial k}\bff^T\one_n\one_n^T\bQ^2\one_n\one_n^T \bff\,,
\end{align*}
it follows that
\begin{align*}
    \frac{\partial C_2}{\partial k}\mid_{k=1} &= -\bff^T(\bI-\bQ) \bff +\frac 1n \bff^T(\bI-\bQ)\bQ\one_n\one_n^T \bff\,,\\
    \frac{\partial C_3}{\partial k}\mid_{k=1} &=0\,.
\end{align*}

On the other hand, the variance is
\begin{align*}
    \Var(\hat \bff) &=(k\bI+\alpha\bQ\one_n\one_n^T)\Var(\hat \bff^\sspl)(k\bI+\alpha\bQ\one_n\one_n^T)^T\\
    &=k^2\Var(\hat \bff^\sspl) + k\alpha\bQ\one_n\one_n^T\Var(\hat \bff^\sspl) + k\alpha\Var(\hat \bff^\sspl)\one_n\one_n^T\bQ +\alpha^2\bQ\one_n\one_n^T\Var(\hat \bff^\sspl)\one_n\one_n^T\bQ\,,
\end{align*}
and
$$
\tr[\Var(\hat \bff)] = k^2\tr[\Var(\hat \bff^\sspl)] + 2k\alpha \tr[\bQ\one_n\one_n^T\Var(\hat \bff^\sspl)] + \alpha^2\tr[\bQ\one_n\one_n^T\Var(\hat \bff^\sspl)\one_n\one_n^T\bQ]\,,
$$
then its derivative with respect to $k$ is
\begin{align*}
    \frac{\partial \tr[\Var(\hat \bff)]}{\partial k} &= 2k\tr[\Var(\hat \bff^\sspl)] + 2\left(k\frac{\partial \alpha}{\partial k}+\alpha\right)\tr[\bQ\one_n\one_n^T\Var(\hat \bff^\sspl)] +\\
    &\qquad\qquad 2\alpha\frac{\partial \alpha}{\partial k}\bQ\one_n\one_n^T\Var(\hat \bff^\sspl)\one_n\one_n^T\bQ\,.
\end{align*}
Evaluate at $k=1$,
$$
\frac{\partial \tr[\Var(\hat \bff)]}{\partial k} \mid_{k=1} = 2\tr[\Var(\hat \bff^\sspl)]-\frac{2}{n}\tr[\bQ\one_n\one_n^T\Var(\hat \bff^\sspl)]\,.
$$
Thus
\begin{align}
    \frac{\partial\MSE(k)}{\partial k}\mid_{k=1} &= 2\Vert \bff -\bbE\hat \bff^\sspl\Vert^2 -\frac 2n (\bff - \bbE\hat \bff^\sspl)^T\one_n\one_n^T\bQ(\bff - \bbE\hat \bff^\sspl) +\notag\\
    &\qquad\qquad 2\left[-\bff^T(\bI-\bQ) \bff +\frac 1n \bff^T(\bI-\bQ)\bQ\one_n\one_n^T \bff\right] + \notag\\
    &\qquad\qquad 2\tr[\Var(\hat \bff^\sspl)]-\frac{2}{n}\tr[\bQ\one_n\one_n^T\Var(\hat \bff^\sspl)]\notag\\
    &=2\bff^T\bQ\frac{\one_n\one_n^T\bQ}{n}(\bI-\bQ)f - 2\bff^T\bQ(\bI-\bQ)\bff + \notag\\
    &\qquad\qquad 2\tr[\Var(\hat \bff^\sspl)]-\frac{2}{n}\tr[\bQ\one_n\one_n^T\Var(\hat \bff^\sspl)]\notag\\
    &=2\left[\tr[\Var(\hat \bff^\sspl)]-\frac{1}{n}\tr[\bQ\one_n\one_n^T\Var(\hat \bff^\sspl)] - \bff^T\bQ(\bI-\frac{\one_n\one_n^T\bQ}{n})(\bI-\bQ)\bff\right]\label{eq:mse_deriv}\\
    &=2\left[\sigma^2\tr[(\bI-\frac{\one_n\one_n^T\bQ}{n})\bQ^2]-\tr[(\bI-\frac{\one_n\one_n^T\bQ}{n})(\bI-\bQ)\bff\bff^T\bQ]\right]\,.\label{eq:mse_deriv2}
\end{align}
Next, we will prove the first term on the right-hand side of \eqref{eq:mse_deriv2} is positive. By the Cauchy-Schwarz inequality for trace in Lemma \ref{eq:trace_cauchy-schwarz},
\begin{align*}
    \tr[\bQ\one_n\one_n^T\Var(\hat \bff^\sspl)] &\le \sqrt{\tr[(\bQ\one_n\one_n^T)^2]\tr[ (\Var(\hat \bff^\sspl))^2]}\\
    &=\one_n^T\bQ\one_n \sqrt{\tr[ (\Var(\hat \bff^\sspl))^2]}\,,
\end{align*}
and since $\Var(\hat \bff^\sspl)=\sigma^2\bQ^2$ is a positive semi-definite matrix, then by Lemma \ref{eq:trace_sq_ieq},
$$
\tr[(\Var(\hat \bff^\sspl))^2]\le [\tr(\Var(\hat \bff^\sspl))]^2\,.
$$
Note that
\begin{align*}
    \eta & = \frac{\one^T_n\bQ\one_n}{n} = \frac{\one^T_n\bQ\one_n}{\one^T_n\one_n}\le \lambda_{\max}(\bQ)= \lambda_{\max}(\bB(\bB^T\bB+\lambda_0\bOmega)^{-1}\bB^T)\,.
\end{align*}
Perform singular value decomposition (SVD) on $\bB$, $\bB=\bU\bD\bV^T$, where $\bU$ and $\bV$ are $n\times J$ and $J\times J$ orthogonal matrices, and $\bD$ is a $J\times J$ diagonal matrix, with diagonal entries $d_1\ge d_2\ge\cdots d_p\ge 0$. Then
$$
\bQ = \bU\bD\bV^T (\bV\bD^2\bV^T+\lambda_0\bOmega)^{-1}\bV\bD\bU^T = \bU(\bI + \lambda_0\bD^{-1}\bV^T\bOmega\bV\bD^{-1})^{-1}\bU^T\,,
$$
it follows that
\begin{align*}
    \eta & \le \lambda_{\max}(\bU(\bI + \lambda_0\bD^{-1}\bV^T\bOmega\bV\bD^{-1})^{-1}\bU^T)\\
    &=\lambda_{\max}((\bI + \lambda_0\bD^{-1}\bV^T\bOmega\bV\bD^{-1})^{-1})\\
    &=\frac{1}{\lambda_{\min}(\bI + \lambda_0\bD^{-1}\bV^T\bOmega\bV\bD^{-1}) }\\
    &=\frac{1}{1+\lambda_{\min}(\lambda_0\bD^{-1}\bV^T\bOmega\bV\bD^{-1})} < 1\,,
\end{align*}
thus
$$
\frac{1}{n}\tr[\bQ\one_n\one_n^T\Var(\hat \bff^\sspl)] < \tr[\Var(\hat \bff^\sspl)]\,.
$$
Note that if $\frac{\partial \MSE(k)}{\partial k} > 0$, that is,
$$
\sigma^2 > \frac{\bff^T\bQ(\bI-\frac{\one_n\one_n^T\bQ}{n})(\bI-\bQ)\bff}{\tr[(\bI-\frac{\one_n\one_n^T\bQ}{n})\bQ^2]}\,,
$$
then there exists $k\in(0, 1)$ such that $\MSE(k) < \MSE(1)$.

If $\lambda_0=0$, then $(\bI-\bQ)\bff = \zero$, and
$$
\one_n\one_n^T\bQ=\one_n(\bB\one_J)^T\bB(\bB^T\bB)^{-1}\bB^T=\one_n\one_J^T\bB^T=\one_n\one_n^T\,,
$$
$$
\bI-\frac{\one_n\one_n^T\bQ}{n} = \bI - \frac{\one_n(\bB\one_J)^T\bB(\bB^T\bB)^{-1}\bB^T}{n} = \bI - \frac{\one_n\one_J^T\bB^T}{n} = \bI - \frac{\one_n\one_n^T}{n}\,,
$$
then
$$
\tr[(\bI-\frac{\one_n\one_n^T\bQ}{n})\bQ^2] = \tr[(\bI - \frac{\one_n\one_n^T}{n})\bQ] = J - \frac{1}{n}\tr[\one_n\one_n^T\bQ]=J-1\,,
$$
and hence $\frac{\partial \MSE(k)}{\partial k} > 0$ if $\sigma^2 > 0$, which always holds.

\end{proof}

\end{document}